\documentclass[11pt,a4paper]{article}
\pdfoutput=1

\usepackage{jcappub}
\usepackage{graphicx}%
\usepackage{multirow}%
\usepackage{amsmath,amssymb,amsfonts}%
\usepackage{amsthm}%
\usepackage{mathrsfs}%
\usepackage[title]{appendix}%
\usepackage{xcolor}%
\usepackage{textcomp}%
\usepackage{manyfoot}%
\usepackage{booktabs}%
\usepackage{algorithm}%
\usepackage{algorithmicx}%
\usepackage{algpseudocode}%
\usepackage{listings}%
\usepackage{float}

\usepackage{tikz}
\usetikzlibrary{calc}
\usetikzlibrary{positioning}
\usepackage{array} 
\usepackage{aas_macros} 

\newcommand{\Gaia}{\textit{Gaia}}
\newcommand{\MWP}{{\rm MWP14}}

\newcommand{\appropto}{\mathrel{\vcenter{
  \offinterlineskip\halign{\hfil$##$\cr
    \propto\cr\noalign{\kern2pt}\sim\cr\noalign{\kern-2pt}}}}}

\newcommand{\approximatelylessthan}{\mathrel{\vcenter{
  \offinterlineskip\halign{\hfil$##$\cr
    <\cr\noalign{\kern1pt}\sim\cr\noalign{\kern-1pt}}}}}

\newcommand{\solaraccX}{-6.56} 
\newcommand{\solaraccXerr}{0.04} 
\newcommand{\solaraccY}{0.31} 
\newcommand{\solaraccYerr}{0.03} 
\newcommand{\solaraccZ}{-0.12} 
\newcommand{\solaraccZerr}{0.03} 

\newcommand{\localdensityradial}{0.84} 
\newcommand{\localdensityradialerr}{0.08} 
\newcommand{\localdensityradialGeV}{0.32} 
\newcommand{\localdensityradialerrGeV}{0.03} 

\newcommand{\localdensityazim}{0.63} 
\newcommand{\localdensityazimerr}{1.25} 

\newcommand{\localdensitypoint}{-1.71} 
\newcommand{\localdensitypointerr}{1.43} 

\newcommand{\NFWchisq}{1.69} 
\newcommand{\NFWrho}{0.89} 
\newcommand{\NFWrhoUpperErr}{0.06} 
\newcommand{\NFWrhoLowerErr}{0.06} 
\newcommand{\NFWrScale}{3.1} 
\newcommand{\NFWrScaleUpperErr}{2.2} 
\newcommand{\NFWrScaleLowerErr}{1.5} 

  \newcommand{\gNFWchisq}{1.70} 
\newcommand{\gNFWrScale}{1.9} 
\newcommand{\gNFWrScaleUpperErr}{1.3} 
\newcommand{\gNFWrScaleLowerErr}{1.0} 
\newcommand{\gNFWbeta}{0.0} 
\newcommand{\gNFWbetaUpperErr}{1.4} 
\newcommand{\gNFWbetaLowerErr}{0.0} 

\newcommand{\triaxialchisq}{1.59} 
\newcommand{\triaxialrho}{1.14} 
\newcommand{\triaxialrhoErr}{0.07} 
\newcommand{\triaxialrScale}{2.6} 
\newcommand{\triaxialrScaleUpperErr}{1.6} 
\newcommand{\triaxialrScaleLowerErr}{1.5} 
\newcommand{\triaxialxiOne}{0.89} 
\newcommand{\triaxialxiOneErr}{0.09} 
\newcommand{\triaxialxiTwo}{0.33} 
\newcommand{\triaxialxiTwoErr}{0.06} 
\newcommand{\triaxialtheta}{52} 
\newcommand{\triaxialthetaErr}{11} 
\newcommand{\triaxialpsi}{-89} 
\newcommand{\triaxialpsiUpperErr}{2} 
\newcommand{\triaxialpsiLowerErr}{1} 
\newcommand{\triaxialT}{0.21} 
\newcommand{\triaxialTUpperErr}{0.17} 
\newcommand{\triaxialTLowerErr}{0.18} 

\title{ClearPotential: Revealing Local Dark Matter in Three Dimensions}

\author[1]{Eric Putney,}

\author[1]{David Shih,}

\author[2,1]{Sung Hak Lim,}

\author[1]{Matthew R.~Buckley}

\affiliation[1]{NHETC, Department of Physics and Astronomy, Rutgers, the State University of New Jersey,126 Frelinghuysen Road, Piscataway, NJ 08854 USA}

\affiliation[2]{Particle Theory and Cosmology Group, Center for Theoretical Physics of the Universe, Institute for Basic Science (IBS), 55 Expo-ro, Yuseong-gu, Daejeon, 34126, Republic of Korea}

\emailAdd{eputney@physics.rutgers.edu}
\emailAdd{shih@physics.rutgers.edu}
\emailAdd{sunghak.lim@ibs.re.kr}
\emailAdd{sunghak.lim@ibs.re.kr}

\abstract{
We present \texttt{ClearPotential}, a data-driven, three-dimensional measurement of the gravitational potential of the local Milky Way using unsupervised machine learning, without the symmetry assumptions, specific functional forms, and binning required in previous work.
The potential is modeled as a neural network, optimized to solve the equilibrium collisionless Boltzmann equation for the observed phase space density of \Gaia{} DR3 Red Clump stars within 4~kpc of the Sun. This density is obtained from data using normalizing flows, and our unsupervised solution to the Boltzmann equation automatically corrects for selection effects from crowding and the dust-driven extinction of starlight.
Our fully-differentiable model of the gravitational potential allows us to map the acceleration and mass density of the Galaxy in the volume around the Sun, including in the dust-obscured disk towards the Galactic Center.
We determine the dark matter density at the Solar radius to be $(\localdensityradial\pm\localdensityradialerr)\times 10^{-2}\,{M}_\odot/{\rm pc}^3$, and analyze the structure of the dark matter halo. We find strong evidence for a tilted oblate halo, weak preference for a cored inner profile, and the strongest constraints to date on a possible dark matter disk. We place a bound on the timescale of disequilibrium in the local Milky Way, and find mild evidence for disequilibrium using independent acceleration measurements from timings of binary pulsar systems. This work provides the clearest map of the local Galactic potential to date and marks an important step in the era of data-driven astrometry.
}


\usepackage{fancyhdr}
\fancypagestyle{titlepage}{
  \fancyhf{}
  \fancyhead[R]{\raggedleft CTPU-PTC-25-33}  
}
\fancypagestyle{otherpage}{
    \fancyhf{}
    \fancyhead[R]{}
    \fancyhead[L]{}
    \fancyhead[C]{}
}

\begin{document}
\maketitle

\section{Introduction}\label{sec1}

Despite long-standing evidence for dark matter in astrophysics and cosmology, its fundamental nature remains unknown.
So far, our knowledge of dark matter's particle properties comes solely from its gravitational imprint on the visible structures of the Universe: constraints on dark matter's self-interactions, dissipative processes, couplings to the Standard Model, and primordial phase space distribution all arise from mapping the distribution of dark matter in cosmological structures across the lifespan of the Universe
\cite{Salucci:2018hqu,2011ARA&A..49..409A,1933AcHPh...6..110Z,Planck:2018vyg,Clowe:2003tk,Buckley:2017ijx, 2000ApJ...534L.143B, 2000ApJ...543..514K, 2000ApJ...544L..87Y, 2001ApJ...547..574D, 2002ApJ...581..777C, 2011MNRAS.415.1125K, 2012MNRAS.423.3740V, 2013MNRAS.430...81R, 2013MNRAS.430..105P, 2013MNRAS.431L..20Z, 2018ApJ...853..109E}.

Within a galaxy, the stellar phase space density evolves in response to all visible matter (stars, gas) and dark matter components. Early measurements \cite{1939LicOB..19...41B,1970ApJ...159..379R,1979ARA&A..17..135F,1980ApJ...238..471R} of dark matter halos in distant galaxies relied on rotation curves: the observed circular velocities (extrapolated from line-of-sight velocity measurements) of visible matter as a function of galactocentric radius. In our own Milky Way Galaxy, where stellar velocities and positions can be measured in three dimensions, the collisionless Boltzmann Equation (CBE) links the stellar phase space density $f$ to the gravitational potential $\Phi$ and acceleration $\vec{a} = -\vec{\nabla}\Phi$ under the assumption of equilibrium:
\begin{equation} \label{eq:CBE}
\frac{\partial f}{\partial t} = -\vec{v} \cdot \vec{\nabla} f + \vec{\nabla}\Phi \cdot \frac{\partial f}{\partial \vec{v}} = 0.
\end{equation}
From the potential, the total mass density $\rho$ can be calculated using the Poisson Equation.

The trove of high-quality stellar kinematics measured by the \Gaia{} space observatory \cite{2016Gaia}, which recorded the angular positions, parallax, and 3D velocities of billions of stars, provides a new window into the Galactic gravitational potential. 
However, even with such a rich dataset, traditional techniques (such as solving the velocity moments of the CBE -- the Jeans equation, see e.g.,~Refs.~\cite{2018MNRAS.478.1677S,2020A&A...643A..75S,2020MNRAS.494.6001N,2021ApJ...916..112N,2020MNRAS.495.4828G,2018A&A...615A..99H}) require binning stellar data under assumptions of axisymmetry, as the high-dimensionality of the phase space density has historically made directly solving the CBE intractable. Deep learning techniques can overcome the curse of dimensionality, with new algorithms
(known as normalizing flows \cite{papamakarios2021normalizing, 9089305}) that allow for accurate, flexible and differentiable data-driven models of the phase space density. These flows can subsequently be used to find solutions for the potential, acceleration field and mass density from the CBE under the assumption of equilibrium.

These new deep-learning techniques were first applied to simulated stellar distributions drawn from both analytic and non-analytic potentials in Refs.~\cite{2023ApJ...942...26G,2021MNRAS.506.5721A, 10.1093/mnras/stac153,2023MNRAS.521.5100B, 2024MNRAS.52712284K}. In Ref.~\cite{lim2023mapping_journal}, we presented the first application of these ideas to real data, producing a model-free map of the acceleration and mass density in select locations within 4~kpc of the Sun using Masked Autoregressive Flows (MAFs, see Ref.~\cite{papamakarios2018masked}) trained on Red Clump (RC) and Red Giant Branch (RGB) stars in the \Gaia{} $3^{\rm rd}$ data release (DR3) \cite{2021Gaia,2022arXiv220800211G}. More recently, Ref.~\cite{2025arXiv250703742K} presented a similar analysis using \Gaia{} data, restricted to main sequence stars within 1~kpc around the Solar location.

In this paper, we go beyond previous works to create the first-ever high-resolution, data-driven, fully 3D map of the gravitational potential, acceleration, and mass density within 4~kpc of the Solar location in the Milky Way. 
In order to achieve this, we solve three problems that limited all prior results. 
First, the observed stellar populations from \Gaia{} are incomplete due to dust extinction. Stars dimmed below the magnitude threshold of \Gaia{}'s spectrometer will not be observed, and thus dusty regions of the sky appear as dark ``voids" with few stars. The observed phase space density reflects the dust-driven position-dependent suppression of the data, and does not satisfy the CBE. Second, the derived accelerations 
were not continuous functions of position, and did not satisfy the curl-free and positive mass density physical constraints. Third, the mass density calculation required averaging over a kernel and was extremely computationally expensive, limiting the resulting spatial resolution of the density map.

We overcome all three of these limitations by introducing two new position-dependent functions, parametrized as trainable neural networks. The first of these functions is the potential $\Phi(\vec{x})$. The second is a position-dependent efficiency function $\epsilon(\vec{x})$. This second function relates a ``corrected'' phase space density $f_{\rm corr}$ (which satisfies the CBE under the assumption of equilibrium) to the observed probability density of stars $f_{\rm obs}$:
\begin{equation}
f_{\rm obs}(\vec{x},\vec{v}) \equiv \epsilon(\vec{x}) f_{\rm corr}(\vec{x},\vec{v}).
\end{equation}
This efficiency primarily represents the extinction of stars by interstellar dust; as was verified in Ref.~\cite{dustpaperI}, it is independent of the stellar velocity. As would be expected from dust extinction due to dust clouds physically located between us and the star, the efficiency only depends on the star's position.
In terms of $f_{\rm obs}$ and the two functions $\epsilon$ and $\Phi$, the CBE takes the form
\begin{equation}
\label{eq:CBE_eps}
\vec{v}\cdot\vec{\nabla} \ln f_{\rm obs} - \vec{v}\cdot\vec{\nabla} \ln \epsilon (\vec{x}) - \vec{\nabla}\Phi(\vec{x})\cdot \frac{\partial \ln f_{\rm obs}}{\partial \vec{v}} = 0.
\end{equation}

To learn $f_{\rm obs}$, we train normalizing flows on the complete sample of RC/RGB stars from \Gaia{} used in Ref.~\cite{lim2023mapping_journal} (with minor updates to the training procedure). Once $f_{\rm obs}$ is trained, we demonstrated in Ref.~\cite{dustpaperI} that we can learn both $\epsilon$ and $\Phi$ simultaneously by minimizing Eq.~\eqref{eq:CBE_eps} as a mean-squared error (MSE) loss function. We call the combined algorithm -- encompassing all three sets of neural networks -- \texttt{ClearPotential}.

By encoding the potential $\Phi$ as a neural network, the curl-free condition is automatically satisfied, and appropriate regularizer terms added to Eq.~\eqref{eq:CBE_eps} can be used to enforce the positive mass density condition. 
This allows for continuous maps of the potential, acceleration, and mass density to be quickly calculated across the sphere of observed stars. Furthermore, we develop a robust pipeline for modeling the correlation scale length (or resolution) of the learned density map, which is itself a nontrivial function of position based on the model flexibility of $\Phi$ and the varying sparsity of stellar data.

To evaluate the consistency of our results, we compare the measured potential with a simple mass model of the Milky Way (\texttt{MilkyWayPotential2014}) \cite{2015ApJS..216...29B}, finding excellent agreement. Our resulting accelerations also largely display the expected symmetries, and our method allows a direct measurement of the vertical acceleration profile, a key input into a standard Jeans analysis. We obtain a precise measurement of the Galactic acceleration at the Solar location. We investigate deviations from the expected axisymmetric acceleration fields; such deviations can be connected to disequilibrium in phase space, which is known to exist to some extent in the Milky Way \cite{2006ApJ...643..881L,2009MNRAS.396L..56M,2009MNRAS.399L.118C,2011Natur.477..301P,2012ApJ...750L..41W,2016ARA&A..54..529B,2017AstRv..13..113V,2018Natur.561..360A,2018Natur.563...85H,2019ApJ...886...67C,2020ApJ...900..186E,2020RAA....20..159S,2025NewAR.10001721H}. 
To test the limits of the equilibrium assumption, we compare our acceleration maps with the measured line-of-sight accelerations of nearby pulsar binaries \cite{donlon, moran}, and construct an estimate of the disequilibrium timescale or ``non-stationarity" at the location of each pulsar binary. Furthermore, we compare spatial distribution of disequilibrium assuming various acceleration fields across our 4~kpc region of interest.

From our differentiable model for the potential, we obtain a fully three-dimensional map of the mass density within a 4~kpc sphere centered on the Sun. Subtracting an analytic model for the baryonic density results in a map of the dark matter distribution within the local volume of space. Because the point-wise measurement of the local dark matter density has large errors, we make the assumption of spherical symmetry and obtain a measurement of the local dark matter density of $\rho_\odot(r_\odot)=(\localdensityradial\pm\localdensityradialerr)\times 10^{-2}\:{M}_\odot/{\rm pc}^3 =(\localdensityradialGeV\pm\localdensityradialerrGeV)\,{\rm GeV/cm^3}$. We fit to the Navarro-Frenk-White (NFW), generalized NFW, and triaxial global models of the dark matter distribution, finding a preference for short scale radii in all models, as well as evidence for a highly tilted and oblate halo with a flat inner slope. Finally, we set the strongest limits yet on the presence of a dark matter disk aligned with the baryonic disk. In our analysis of the dark matter density, the uncertainties in the baryonic model are significant, and demonstrate the need for more accurate understanding of the mass distribution of the visible components within the Milky Way.

In Section~\ref{sec:data}, we describe the \Gaia{} DR3 dataset from which stars are selected for our analysis. The architecture of the neural networks used to solve the CBE are described in Section~\ref{sec:methods}, along with our methods of uncertainty estimation. We discuss our results in Section~\ref{sec:results}, and conclude in Section~\ref{sec:conclusion}.

\section{Data}\label{sec:data}

We adopt right-handed Galactocentric coordinates with the positive $y$ direction oriented opposite to the rotation of the stellar disk, with the Galactic Center at the origin. The stellar disk is centered on the Galactic midplane at $z=0$~kpc. The Sun is located at $(x,y,z)=(8.122,0,0.0208)$~kpc \cite{2018A&A...615L..15G, 2019MNRAS.482.1417B} in this coordinate system.

Our dataset is constructed from stars in \Gaia{} DR3. This dataset is the same as used previously in Refs.~\cite{2023MNRAS.521.5100B,dustpaperI}. We select a sample of $\sim25$ million stars in DR3 with measured positions, proper motions, radial velocities, and measured photometric magnitudes with parallax distances less than 4~kpc from the Sun. We further require small parallax errors $\Delta\varpi/\varpi < 1/3$. The \Gaia{} DR3 catalog includes an estimate of the standard deviation of the Gaussian error for each measured quantity for each star; we retain and use these measurement errors in our error propagation Section~\ref{sec:uncertainties}.

We apply a completeness condition, requiring that each star's absolute magnitude $M_G$ (inferred from the RVS spectrometer magnitude $G_{\rm RVS}$ and the parallax-measured distance) is sufficiently small so that the star would have been visible in the \Gaia{} spectrometer (with a magnitude threshold of 14) if located at a distance of 4~kpc:
\begin{equation}
M_G + \mu(4~{\rm kpc}) < 14.
\end{equation}
where $\mu(4~{\rm kpc}) = 13.010$ is the distance modulus at 4~kpc.

There are 5,811,956 bright stars which survive these selection criteria, the majority ($\sim 69\%$) belong to the RC and RGB. The RC stars in particular are expected to be an older and equilibrated stellar population (barring outside perturbers on the Galaxy) \cite{2016ARA&A..54...95G, 2014JPhG...41f3101R}, and have been used in previous Jeans analyses of the Milky Way \cite{2020A&A...643A..75S, 2024arXiv240608158B}.

\section{Methods} \label{sec:methods}
\subsection{Neural Networks}\label{sec:networks}

We model the phase space density of the complete sample of well-measured \Gaia{} DR3 stars within 4~kpc using normalizing flows. Specifically, we use Masked Autoregressive Flows (MAFs) \cite{papamakarios2018masked}, which learn an invertible and differentiable map from a simple latent base distribution (in our case, a multidimensional Gaussian) to the probability density of data. This map is a composition of autoregressive transformations, ensuring a tractable Jacobian. This allows simple calculation of the probability distribution with respect to the input parameters.

Before training the MAFs, the input position and velocity data are preprocessed and standardized. We adopt a slightly updated procedure from the one adopted in Ref.~\cite{2023MNRAS.521.5100B}. Our position-space data is confined to a sphere centered on the Solar location, with a hard cutoff at $r_{\rm max}=4$~kpc. First, we center and scale the data to a unit ball with the mapping
\begin{equation}
    \vec{x}\rightarrow\frac{\vec{x}-\vec{x}_\odot}{r_{\rm max}\cdot (1+c)}
\end{equation}
where $c$ is a small boundary multiplier that creates a buffer between the transformed dataset and $r_{\rm max}$. Next, since the discontinuity at $r_{\rm max}$ cannot be modeled by MAFs using continuous base distributions, we map the unit sphere to $\mathbb{R}^3$ with the logit transformation:
\begin{equation}
    \vec{x}\rightarrow\frac{\vec{x}}{|\vec{x}|}\tanh^{-1}|\vec{x}|.
\end{equation}
Afterwards, each Cartesian dimension is re-centered and standardized to unit norm via
\begin{equation}
    x_i\rightarrow\frac{x_i-\langle x_i \rangle}{\sigma_{x_i}}.
\end{equation}
Finally, as an additional precaution to avoid autoregressive bias from the fixed ordering of dimensions in the MAF, we rotate the frame by a random angle around a random axis. Our velocity space data is continuous, so we need only apply the standardization $v_i\rightarrow\frac{v_i-\langle v_i \rangle}{\sigma_{v_i}}$ and rotation steps.

We factorize the observed phase space density of stars into $f_{\rm obs}(\vec{x},\vec{v})=n_{\rm obs}(\vec{x})p_{\rm obs}(\vec{v}|\vec{x})$, where $n_{\rm obs}(\vec{x})$ and $p_{\rm obs}(\vec{v}|\vec{x})$ are the observed number density and conditional velocity distribution of stars, respectively.
This factorization is useful for two reasons. First, the velocity distribution of a collisionless ``gas" of stars likely resembles a Gaussian, which can be easily mapped to the Gaussian base distribution of the MAF.
Second, and most importantly, this factorization allows us to sample multiple velocities at a fixed position when later solving the CBE. Both MAFs are trained to minimize the negative log-likelihood (NLL) over the stars in our dataset:
\begin{equation}\label{eq:NLL}
\begin{aligned}
    & \mathcal{L}=-{\mathbb{E}}_{\vec{x}\sim {\rm data}}\log n_{\rm obs}(\vec{x}) \\
    & \mathcal{L}=-{\mathbb{E}}_{\vec{x},\vec{v}\sim {\rm data}}\log p_{\rm obs}(\vec{v}|\vec{x}).
\end{aligned}
\end{equation}

After training the MAFs, we factorize $f_{\rm obs}$ into $f_{\rm corr}$ and the dust efficiency $\epsilon$, and parameterize $\Phi$ and $\epsilon$ with fully connected neural networks described by parameters $\boldsymbol{\theta}$ and $\boldsymbol{\vartheta}$: $\Phi_{\boldsymbol{\vartheta}}(\vec{x})$ and $\epsilon_{\boldsymbol{\theta}}(\vec{x})$. We solve for these networks by minimizing a regularized form of the MSE of Eq.~\eqref{eq:CBE_eps} summed over pairs of $\{\vec{x},\vec{v}\}\sim f_{\rm obs}(\vec{x},\vec{v})$ with respect to the neural network parameters $\boldsymbol{\theta}$ and $\boldsymbol{\vartheta}$: 

\begin{equation}\label{eq:Lmseloss_reg}
\begin{aligned}
\mathcal{L}_{\boldsymbol{\theta} , \boldsymbol{\vartheta}} = & {\mathbb{E}}_{\genfrac{}{}{0pt}{}{\vec{x}\sim n_{\rm obs}(\vec x)}{\vec{v}\sim p_{\rm obs}(\vec v|\vec x)} } \Biggl( \bigg| \vec{v}\,\cdot\vec{\nabla}\ln f_{\rm obs} -\vec{\nabla}\Phi_{\boldsymbol{\vartheta}}(\vec{x})\cdot \frac{\partial \ln f_{\rm obs}}{\partial \vec{v}\,}\\ & - \vec{v}\,\cdot\vec{\nabla}\ln \epsilon_{\boldsymbol{\theta}}(\vec{x}) \bigg|^2  + \mathcal{L}_{\rm reg}(\vec{x})\Bigg)
\end{aligned}
\end{equation}
The expectation value involves a double sum $\vec x$ sampled from the learned number density, and $\vec v$ sampled from the learned conditional velocity distribution. This is necessary because at a single point $\vec x$, one must sample enough velocities to constrain all six degrees of freedom of $\vec{\nabla}\ln \epsilon_{\boldsymbol{\theta}}(\vec{x})$ and $\vec{\nabla}\ln \Phi_{\boldsymbol{\vartheta}}(\vec{x})$. We realize this by first sampling $N_x=2^{22}$ positions from the number density flow $\vec{x}\sim n_{\rm obs}(\vec{x})$. Sampling $\vec{x}$ directly from the flows (instead of a uniform grid) allows us to best constrain $\Phi(\vec{x})$ in regions of space of high data availability. Next, for each sampled $\vec{x}$, we overconstrain the equilibrium CBE by sampling $N_v=16$ unique velocities from the conditional velocity flow at that location $\vec{v}\sim p_{\rm obs}(\vec{v}|\vec{x})$.

The regularization term $\mathcal{L}_{\rm reg}$ in Eq.~\eqref{eq:Lmseloss_reg} is
\begin{equation}
\mathcal{L}_{\rm reg}(\vec{x}) = \lambda_\epsilon \left|\ln{\epsilon}(\vec{x})\right|^2+\lambda_\Phi \, \textrm{max}\left(0,-\nabla^2 \Phi(\vec{x})\right).
\end{equation}
The first term of the regularizer imposes an absolute scale for the $\epsilon$ dust correction, whereas the second term penalizes negative mass densities encountered at the sampled $\vec{x}$ used for training. Ideally, $\lambda_\Phi$ should be infinitely large so as to rigidly enforce the positive-mass requirement. However, computing $-\nabla^2\Phi(\vec{x})$ during training can introduce severe training instabilities. This restricts $\lambda_\Phi$ to be large, but finite. We set fiducial scales for these terms as $\lambda_\epsilon=10^{-1}$ and $\lambda_\Phi=10$. 

\subsection{Uncertainty Estimation}\label{sec:uncertainties}

As in Refs.~\cite{2023MNRAS.521.5100B,lim2023mapping_journal,dustpaperI}, we account for three sources of error in our analysis: instrumental measurement error, uncertainty from finite statistics, and error from training variance. The first two are intrinsic to the dataset and cannot be reduced from any downstream analysis, while the third is reducible by ensembling. To derive $1\sigma$ uncertainty brackets for $\Phi$, acceleration $\vec{a}$, and mass density $\rho$, we repeat the training regimen described above over variations of the training data as well as random initializations for the neural networks, as described below.

First, to compute measurement uncertainty in $\Phi$, we generate $10$~realizations of the \Gaia{} astrometric data (right ascension $\alpha$, declination $\delta$, parallax $\omega$, proper motions $\mu$, and radial velocities $v_r$) perturbed by their measurement uncertainties (assuming Gaussian statistics). We repeat the entire training procedure on all $10$ of these perturbed datasets, and use the resulting variance of estimates for $\Phi$ as $1\sigma$ measurement errors. 

Next, to compute statistical uncertainty in $\Phi$ from sparse regions of space, we create $10$~resampled versions of the original \Gaia{} dataset via the non-parametric bootstrap, and again repeat the training regimen and take the variation of estimates as a $1\sigma$ statistical error. 

Our central estimates and training uncertainties are produced by repeating the $\Phi$ training procedure on the original dataset with $100$ different random seeds for the neural network initialization. We ensemble average the $100$ realizations for our central estimates, and divide the variation of the $100$ realizations by $\sqrt{N}$ as an estimate for the training uncertainty on the mean. As a result, this source of uncertainty is generally subdominant to the other two. 

\section{Results}\label{sec:results}

\subsection{Galactic Potential}
After fitting flows to the {\it Gaia} data to learn $f_{\rm obs}$~\cite{lim2023mapping_journal,dustpaperI},
we substitute into Eq.~\eqref{eq:CBE_eps} and solve for the gravitational potential $\Phi$ and dust efficiency map $\epsilon$ across a sphere of radius 4~kpc centered on the Solar location.
The resulting dust efficiency map $\epsilon(\vec{x})$ is discussed in detail in Ref.~\cite{dustpaperI}; here we note that it is largely successful in matching known extinction features and returns a smooth $f_{\rm corr}(\vec{x},\vec{v})$ across the sky. Some residual artifacts remain in regions of high dust extinction (i.e., towards the Galactic Center within the disk). In these regions, we should expect $\Phi$ and its derived quantities (acceleration and mass density) to also have increasing systematic errors which are not captured in our error budget.

\begin{figure}[t]
  \centering
  \begin{tikzpicture}
    \node[anchor=south west, inner sep=0] at (0,0)
      {\includegraphics[width=\columnwidth]{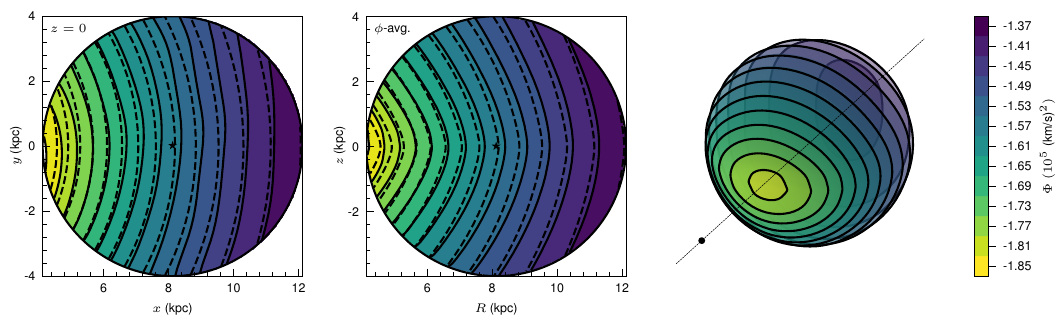}};
    \node[anchor=south west, inner sep=0] at (0,0)
      {\includegraphics[width=\columnwidth]{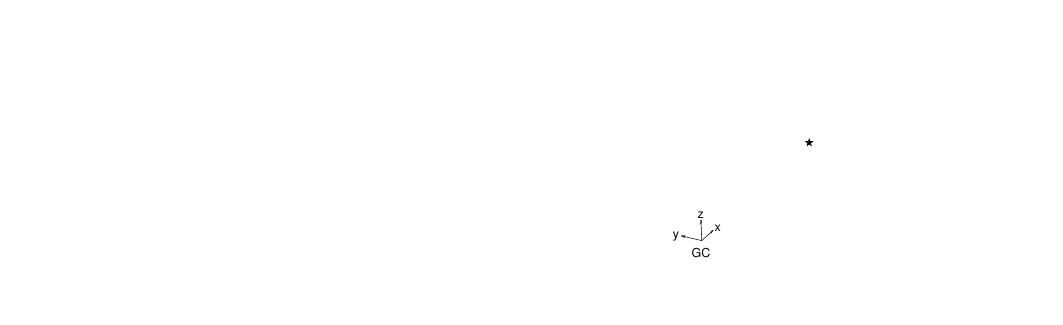}};
  \end{tikzpicture}
  \caption{Left: Contours of the total gravitational potential $\Phi$ estimated in this work (solid) compared to \MWP{} (dashed) in the midplane ($z=0$). Middle: Contours of the azimuthally averaged $\Phi$ in the $R-z$ plane. Right: 3-dimensional isocontours of $\Phi$ within $3.8$~kpc of the Solar location ($\star$). 
  }
  \label{fig:potential_contours}
\end{figure}

We compare our $\Phi$ function to an analytic axisymmetric model of the Milky Way's gravitational potential (\texttt{MilkyWayPotential2014}, henceforth referred to as \MWP{}) which was fit to pre-\Gaia{} stellar kinematics \cite{galpy}. In Figure~\ref{fig:potential_contours}, we show $\Phi$ contours from our model and \MWP{} in the midplane of the disk and over the $R-z$ plane averaged over azimuthal angles. 
As can be seen, our data-driven potential demonstrates a remarkable level of axisymmetry and morphological similarity to the analytic expectation from \MWP{}. This is without any enforced prior on the structure and symmetries of the potential. However, deviations from axisymmetry are visible by eye. These will become more prominent in the acceleration and mass density, as those rely on derivatives of the potential. 

\subsection{Galactic Accelerations}

The Galactic acceleration field can be straightforwardly computed from the gradient of the $\Phi$ neural network.
We obtain a precise estimate of the acceleration at the Solar location of $\vec{a}_\odot=(\solaraccX\pm\solaraccXerr, \solaraccY\pm\solaraccYerr, \solaraccZ\pm\solaraccZerr)~\textrm{mm/s/yr}$ in Galactocentric coordinates. This is consistent with the results from our previous analysis Ref.~\cite{lim2023mapping_journal}, with much smaller uncertainties. We are also in closer agreement with (and more precise than) the \Gaia{} measurement of Ref.~\cite{2021A&A...649A...9G} than previously. These improvements are likely attributable to $\vec{a}$ being computed from a consistent potential, obeying the curl-free condition. A full description of the uncertainties in this analysis in provided in Section~\ref{sec:uncertainties}.

\begin{figure}[t]
    \centering
    \includegraphics[width=\columnwidth]{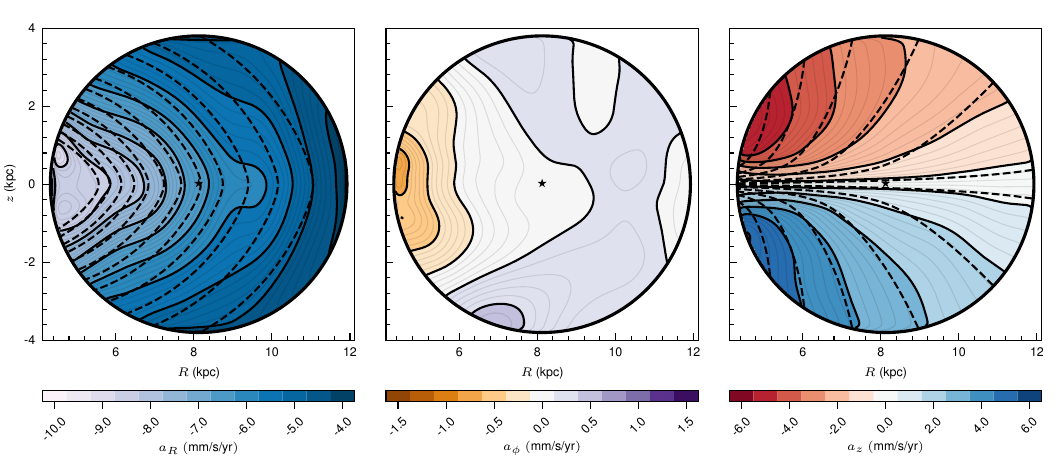}
    \caption{Azimuthally averaged radial $R$ (left), aziumuthal $\phi$ (center), and vertical $z$ (right) accelerations in the $R-z$ plane, compared to \MWP{} (dashed). \MWP{} predicts $a_\phi=0$ due to the assumption of azimuthal symmetry.}
    \label{fig:RZ_accels}
\end{figure}

In Figure~\ref{fig:RZ_accels}, we show the resulting acceleration field, converted to Galactocentric cylindrical components $(a_R,a_\phi,a_z)$ azimuthally averaged in the $R-z$ plane. We are in agreement with the expectations from symmetry and equilibrium across much of the observational volume.
In particular, the acceleration map near the Solar location and the regions directly above and below the Sun (relative to the Galactic disk) align with expectations. This suggests that the local measures of acceleration (and hence mass density) may be less affected by these sources of disequilibrium compared to directions towards the Galactic center and anti-center.

The vertical acceleration profile $a_z$ (Figure~\ref{fig:RZ_accels}, right panel) is one of the most important quantities in Jeans analyses that is typically indirectly estimated through velocity dispersions and assumed functional forms.  The simple vertical acceleration profile assumed in \MWP{} appears to be in good agreement with our direct data-driven measurement across the full 4~kpc range.

Some deviations from the expected equilibrium acceleration field are visible in Figure~\ref{fig:RZ_accels}. First, the radial acceleration (left panel) appears to flatten near the Galactic Center. 
Secondly, there is an unexpected ``shelf-like'' feature located between $9<R<10$~kpc and $|z|<2$~kpc. 
Finally, the axial acceleration $a_\phi$ (center panel of Figure~\ref{fig:RZ_accels}) is expected to be zero in the equilibrium solution, but we find small, nonzero $a_\phi$ across most of the observation window. These non-zero accelerations ($\lesssim 10\%$ across the entire observational volume) are significant at the $3-5\sigma$ level in the disk (most prominently towards the Galactic Center), and $1-2\sigma$ in the halo. 
The origins of these features are not fully understood, and could be attributable to either disequilibrium or a failure of the dust-correction model.

\begin{figure}[t]
    \centering
    \includegraphics[width=\columnwidth]{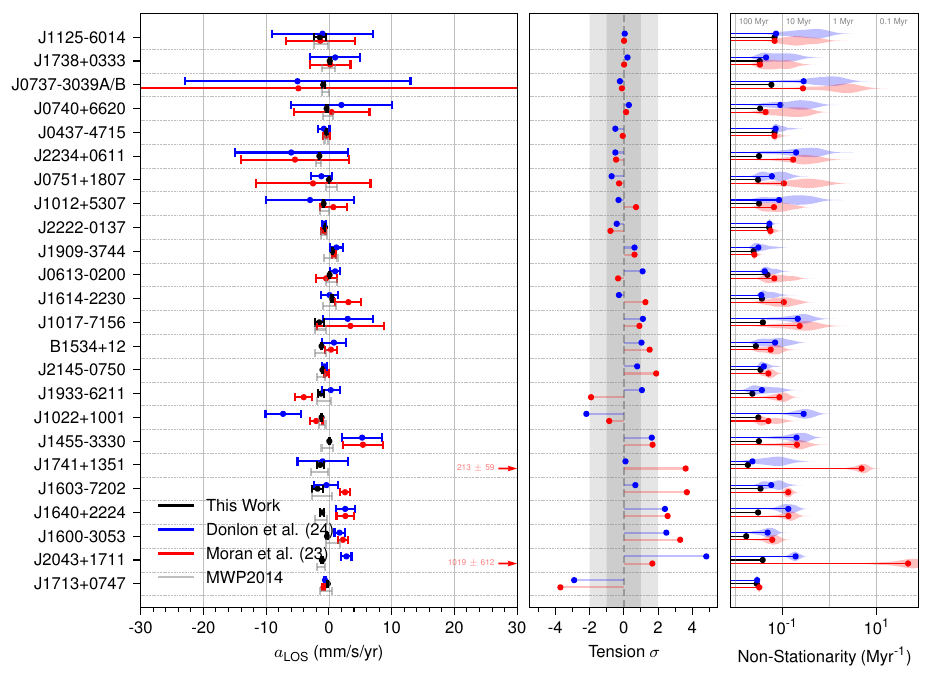}
    \caption{Left column: Measured relative line-of-sight (LOS) accelerations of nearby binary pulsar systems as measured by this work (black), Donlon et al.\ (blue, Ref.~\cite{donlon}), Moran et al.\ (red, Ref.~\cite{moran}), and as predicted by \MWP{} (grey). Center column: $\sigma$-tension between the LOS pulsar acceleration measurements and this work. Ordering of pulsars is sorted from left to right by lowest to highest level of mutual agreement with this work. Right column: Non-stationarity metric (inverse dynamic timescale of disequilibrium) given the LOS accelerations estimated in this work, Donlon, or Moran. Transparent contours denote the uncertainty in non-stationarity given the LOS acceleration uncertainties.}
    \label{fig:pulsar_LOS_comparison}
\end{figure}

\begin{figure}[t]
    \centering
    \includegraphics[width=\columnwidth]{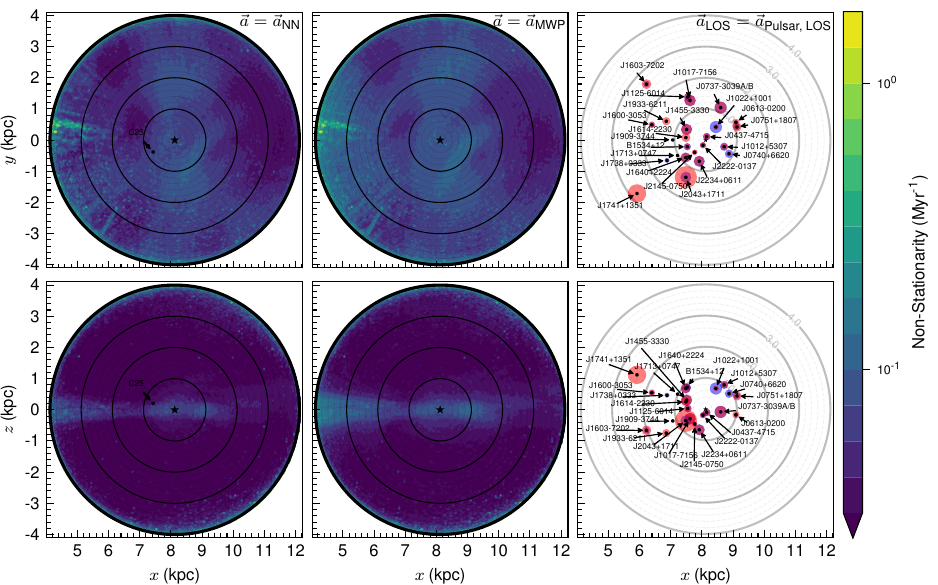}
    \caption{Left column: Non-stationarity (NS) map in the $x-y$ (top) and $x-z$ (bottom) plane, using our flow-based gradients of $f$ and assuming the neural-network-based acceleration $\vec{a}$ satisfies the CBE. Center column: NS map assuming the accelerations in \MWP{} satisfy the CBE. 
    Right column: Locations of the pulsars in Galactocentric coordinates, with overlapping blue (Donlon) and red (Moran) circles of area proportional to the increase in NS assuming $\vec{a}_{\rm LOS}=\vec{a}_{\rm Pulsar,\:LOS}$. A black dot denotes the location of the proposed subhalo (labeled C25) in Ref.~\cite{2025arXiv250716932C}.}
    \label{fig:dlnf_dt_comparisons}
\end{figure}

Our smooth three-dimensional acceleration field, derived under the assumption of equilibrium, can be compared to measurements of the relative line-of-sight (LOS) accelerations from binary pulsar system timings, which are independent this assumption. In Figure~\ref{fig:pulsar_LOS_comparison}, we compare the LOS component of our acceleration measurements to two separate measurements of $24$ binary pulsar systems \cite{donlon, moran}. Overall we tend to agree much better with the results from Ref.~\cite{donlon} than from Ref.~\cite{moran}. A more detailed breakdown shows that our measurements (which have smaller uncertainties\footnote{In addition to the standard error model discussed in Section~\ref{sec:uncertainties}, we include uncertainty in the distance to the binary pulsar systems for the evaluation of our LOS accelerations.}) are consistent (within $2\sigma$) with both studies for $17$ binary pulsar systems. Of the remaining seven binary pulsars, 
four of our measurements are consistent with one study but not the other. Finally, there are three binary pulsar systems (J1640+2224, J1600-3053, and J1713+0747) which are in $>2\sigma$ tension with both studies.

Contrasting independent equilibrium and non-equilibrium acceleration fields allows us to probe (and potentially map) disequilibrium within the Milky Way. To quantify the disequilibrium interpretation of this tension,  we use the ``non-stationarity" (NS) metric \cite{lim2023mapping_journal,2024MNRAS.52712284K, 2025arXiv250703742K}. 
This metric computes the inverse of an effective disequilibrium timescale, derived from the dispersion of the residual of the equilibrium CBE for a given acceleration vector:\footnote{
In practice, the mean of the residuals is much smaller than the dispersion for the choices of $\vec a$ we will take below. This justifies using the dispersion as the basis for the NS metric.} 
\begin{equation}
    {\rm NS}(\vec{x};\vec a)^2={\rm Var}_{\vec{v}\sim f(\vec{x},\vec{v})}\left[\vec{v} \cdot \vec{\nabla} \ln f -\vec a \cdot \frac{\partial \ln f}{\partial \vec{v}}\right]. 
\end{equation}
A longer disequilibrium timescale at a location $\vec x$ (corresponding to a smaller value of the NS metric) suggests less disequilibrium at $\vec x$. We visualize the distribution of NS across the $4$~kpc observation volume in Figure~\ref{fig:dlnf_dt_comparisons}, assuming that either the neural-network-derived $\vec{a}$ presented in this work (left column) or the $\vec{a}$ derived from \MWP{} solves the CBE (center column). As $\vec{a}_{\rm NN}$ was explicitly derived as the best solution to the CBE, wherever $\vec{a}_{\rm MWP}\neq\vec{a}_{\rm NN}$ the MWP-based NS will be larger than the NN-derived NS. It is immediately apparent in the left and center columns of Figure~\ref{fig:dlnf_dt_comparisons} that the majority of the disequilibrium (as identified by the NS metric) is located within the disk. The disequilibrium towards the Galactic Center is possibly attributable to mismodeling of the dust extinction, our method also identifies larger-than typical NS values both near the Solar radius and further from the Galactic Center, in regions where dust expected to be less significant. Further work will be required to understand whether these NS values are a result of true dynamical disequilibrium or some other source.

In the right column of Figure~\ref{fig:pulsar_LOS_comparison}, we compute the NS metric at the location of each binary pulsar system using the transverse accelerations obtained from our method and three choices for the LOS accelerations: the measurements 
provided by Refs.~\cite{donlon, moran}, and our own equilibrium-based estimate. 
Using our own equilibrium-based estimate of the LOS acceleration, we find an average disequilibrium timescale of $28.9~{\rm Myr}$ across the locations of all $24$ binary pulsar systems. We interpret this as a baseline expectation for the longest timescale of disequilibrium our method is sensitive to. 
Using the LOS acceleration components measured by pulsar timing, when marginalizing over measurement uncertainties, yields average disequilibrium timescales of $9.9~{\rm Myr}$ and $6.1~{\rm Myr}$ for Refs.~\cite{donlon, moran}, respectively. For some pulsars, this increase is attributable to the large uncertainties in the LOS accelerations. For others, this difference may be a meaningful indicator for a larger-than-expected scale of disequilibrium.  

In the right column of Figure~\ref{fig:dlnf_dt_comparisons}, we visualize the increase of NS for these different choices of acceleration at the location of each system. Two binary pulsar systems \textrm{J0437-4715} and \textrm{J2222-0137} near the Solar location show no increase in the baseline NS due to the consistency between LOS acceleration measurements, providing no indication of local disequilibrium.
Two other binary pulsar systems (\textrm{J1741+1351} and \textrm{J2043+1711}) have extremely large 
LOS accelerations reported by Ref.~\cite{moran}, but not by Ref.~\cite{donlon} or by our method. 
None of the surrounding binary pulsar systems appear to exhibit similarly large NS, so we do not view these two pulsars as providing strong evidence for disequilibrium. Furthermore, the anomalous acceleration of \textrm{J2043+1711} may be explained by a close stellar encounter \cite{2025ApJ...983...62D}.

One source of disequilibrium expected in the Milky Way comes from compact dark matter subhalos, which perturb the orbits of nearby stars. While we expect a population of subhalos distributed across the Galaxy, there has not yet been a definitive detection of an individual subhalo.
Recently, Ref.~\cite{2025arXiv250716932C} postulated the existence of a $10^7~M_\odot$ dark matter subhalo at $(x,y,z)=(7.43,-0.38,0.21)$~kpc (shown in the right column of Figure~\ref{fig:pulsar_LOS_comparison}) based on the anomalous accelerations\footnote{Anomalous relative to a smooth disk and halo potential fit by Ref.~\cite{2025arXiv250716932C} to rotation curves and $27$ binary pulsar systems.}
of $19$ binary pulsar systems, in the vicinity of \textrm{J1640+2224} and \textrm{J1713+0747}. 

The measured LOS accelerations of these two binary pulsar systems are in $>2\sigma$ tension with the equilibrium-based accelerations estimated in this work. For \textrm{J1713+0747}, this does not lead to an increase in the NS metric, because the statistical tension is driven entirely by the very small uncertainties in the pulsar measurements, and the magnitude of the difference in LOS accelerations is so small that it does not affect the dispersion of the CBE residuals that source the NS metric. Meanwhile 
we do observe a moderate increase in NS at the other binary pulsar system \textrm{J1640+2224}, but this is in line with increases in disequilibrium elsewhere in our analysis volume. 
We do observe an increased level of non-stationarity within $7<R<9$~kpc i.e.~around the Solar radius, but as seen in Figure~\ref{fig:dlnf_dt_comparisons} we do not observe any substantial localized increase in the proximity of the proposed subhalo.
We conclude that if there is a subhalo near the suggested location sourcing these accelerations, it is either inducing disequilibrium over longer timescales (that is, has a smaller $\partial \ln f/\partial t$) than the maximum detectable timescale of $\sim30~{\rm Myr}$ in this work, or it is one subhalo of many inducing the average observed increase in NS across the bulk.

\subsection{Galactic Dark Matter Mass Density}
\begin{figure}[t]
  \centering
  \begin{minipage}[t]{0.9\textwidth}\vspace{0pt}%
    \centering
    \begin{tikzpicture}[baseline=(current bounding box.north)]
      \node[inner sep=0] (main) {\includegraphics[width=\linewidth]{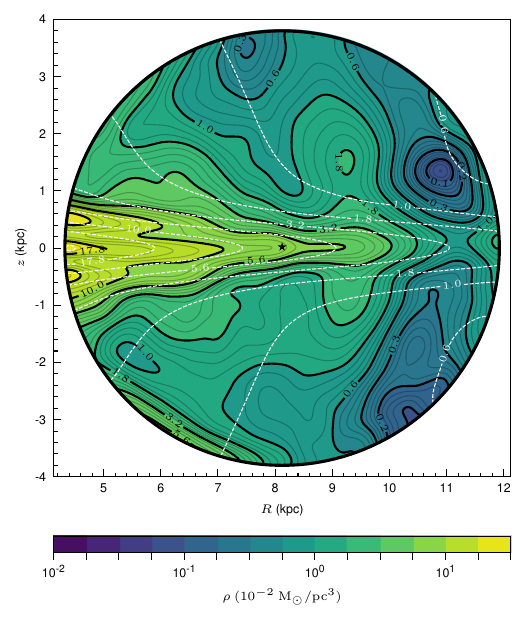}};
      \node[anchor=north east, xshift=30, yshift=45] at (main.north east)
        {\includegraphics[width=0.33\linewidth]{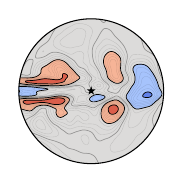}};
    \end{tikzpicture}
  \end{minipage}\hfill
  
  \caption{The inferred azimuthally-averaged total Galactic mass density field $\rho(\vec{x})$ (black) in the $R$–$z$ plane, compared to \MWP{} (white dashed). The inset shows the average pull relative to \MWP{} (light/dark red for $2\sigma$–$5\sigma$/$>5\sigma$ higher, light/dark blue for lower; grey indicates $\sigma<2$ agreement).
  }
  \label{fig:2D_mass_density}
\end{figure}

Finally, we calculate the mass density using the Poisson Equation
\begin{equation}\label{eq:Poisson}
    \nabla^2\Phi=4\pi G \rho,
\end{equation}
once again leveraging the calculable gradients in the $\Phi$ network. This results in a fully three-dimensional map of the local (total) mass density of the Milky Way within the $4$~kpc observation window. We visualize the total density in the $R-z$ plane in Figure~\ref{fig:2D_mass_density}, averaging over the azimuthal angle $\phi$. A complete discussion of this averaging procedure is provided in Appendix~\ref{app:correlations}, in particular how the averaging handles spatial correlations. Though $\Phi$ (and thus $\rho$) is a continuous function that can be evaluated at any point in the volume in which the data has support, the function values are correlated over length scales varying from $\sim 50$~pc (in the disk, where the stellar density is high) to $\sim 300$~pc (in the low-density halo).

In most regions of the observation volume, we find $<2\sigma$ agreement between our density model and \MWP{}. 
We observe larger discrepancies in two regions in the disk: towards the Galactic Center and at $R \sim 9-11.5$~kpc. Both of these correspond to locations where we had previously noted signs of disequilibrium in the acceleration field. 
Next, we subtract a model of the baryonic density -- convolved by the correlation function described in Appendix~\ref{app:correlations} -- from our measurement of the total mass density. The baryonic model is described in Appendix~\ref{app:baryons}.
This results in an inferred distribution of dark matter across the entire the 4~kpc sphere centered on the Sun. 

The dark matter density at the Solar location $\rho_{\rm DM, \odot}$ is an important input for direct detection experiments. At the exact Solar location we find $\rho_\odot(\vec{x}_\odot)=(\localdensitypoint\pm\localdensitypointerr) \times 10^{-2}\:{ M}_{\odot}/{\rm pc}^3$, which places a $98\%$ confidence upper limit of $\rho_\odot(\vec{x}_\odot)<1.15 \times 10^{-2}\:{ M}_{\odot}/{\rm pc}^3$. 
The error on the dark matter estimate everywhere in the disk (including the Solar location) is dominated by the uncertainties in the baryonic mass model of the Milky Way. 
As we enter the precision era for dark matter astrometry, reducing the errors on the baryonic model is an important priority for future work.

\begin{figure}[t]
  \centering
  
    \centering
    \includegraphics[width=0.9\linewidth]{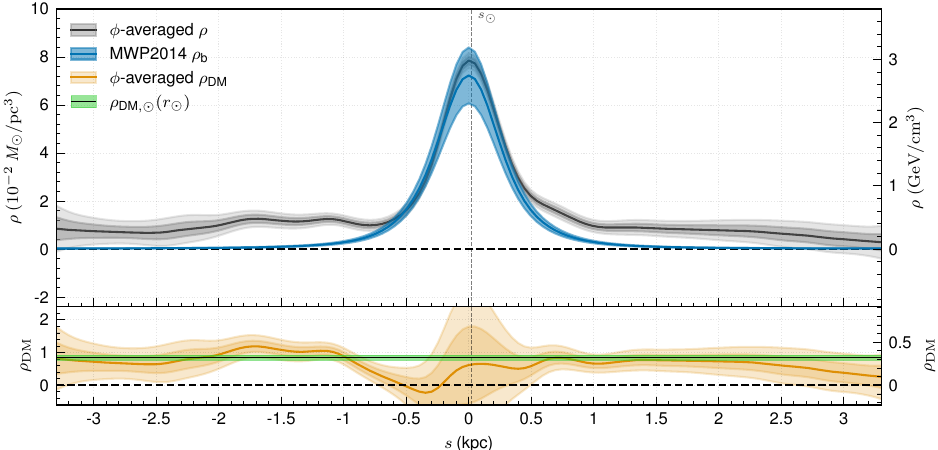}
    
  \caption{ Top: Weighted azimuthal average total (black) and baryonic (blue) mass density along $s\equiv r_\odot(\pi/2-\theta)$ with $1\sigma$ ($2\sigma$) uncertainties in the dark (light) bands. Bottom: Subtracted weighted azimuthal average dark matter density.
  }
  \label{fig:s_mass_density}
\end{figure}

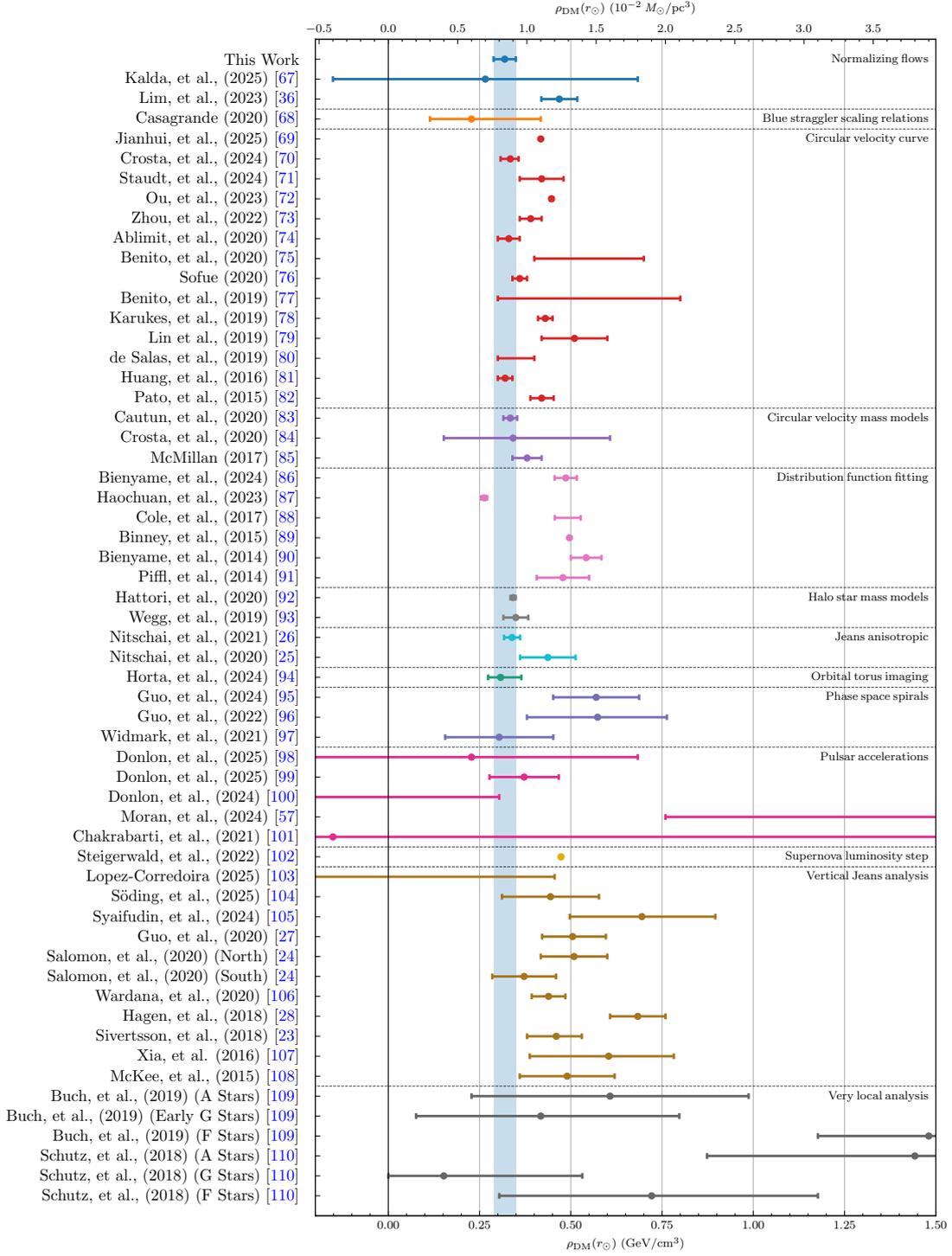
\begin{figure*}[ht!]
    \centering
    \resizebox{!}{0.87\textheight}{
    \begin{tikzpicture}[every node/.style={inner sep=0,outer sep=0}]
        \node [anchor=south west] at (0,0) {\includegraphics{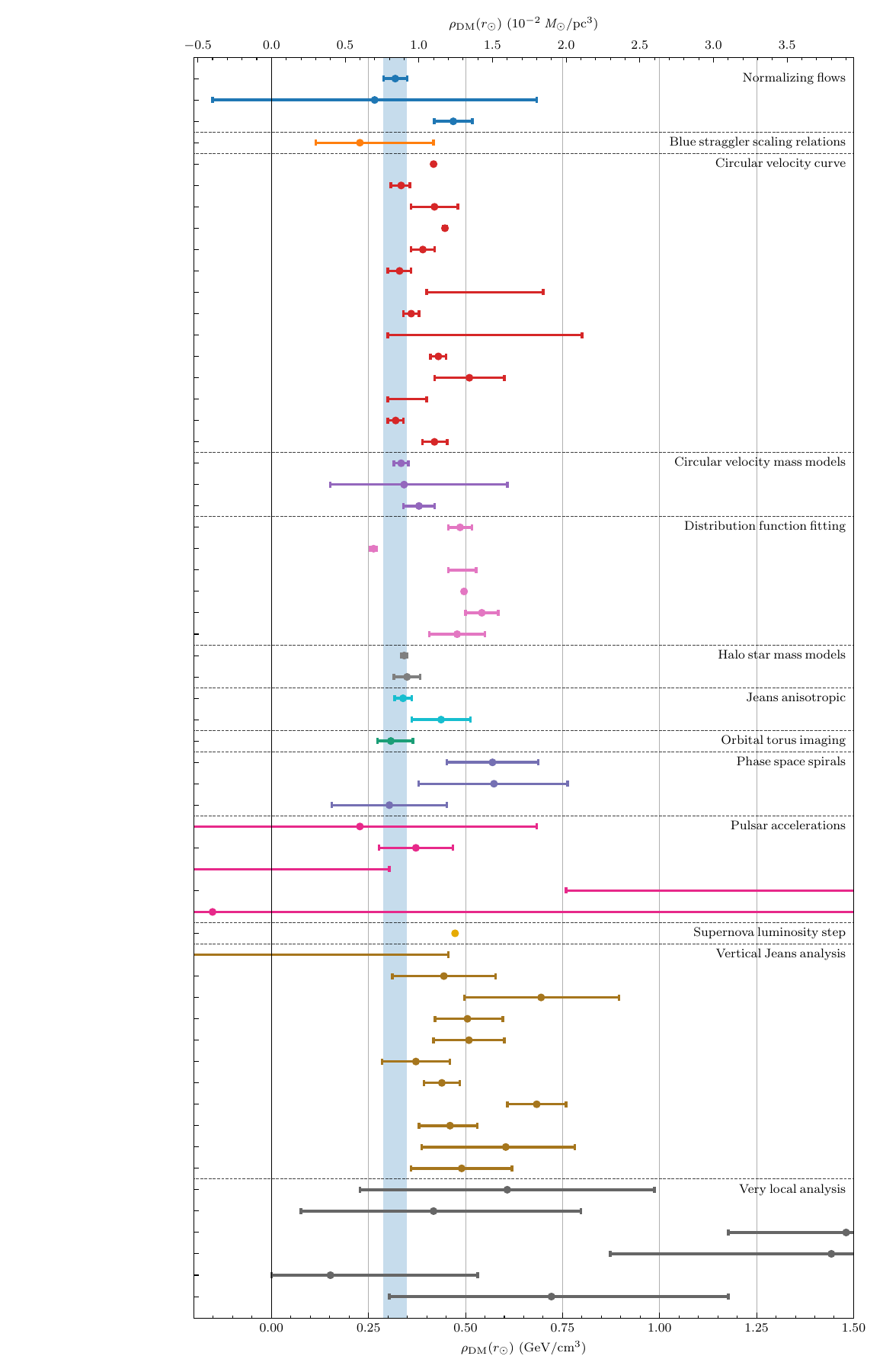}};
        \draw [white] (0in, 0) -- (0in, 8.5 in);
\node [draw=none, anchor=east, align=right] at ($(1.6in,  11.312in) - 0*(0in, 0.187in)$) {
            This Work
        };
\node [draw=none, anchor=east, align=right] at ($(1.6in,  11.312in) - 1*(0in, 0.187in)$) {
            Kalda, et al., (2025) \cite{2025ApJ...992...84K}
        };
\node [draw=none, anchor=east, align=right] at ($(1.6in,  11.312in) - 2*(0in, 0.187in)$) {
            Lim, et al., (2023) \cite{lim2023mapping_journal}
        };
\node [draw=none, anchor=east, align=right] at ($(1.6in,  11.312in) - 3*(0in, 0.187in)$) {
            Casagrande (2020) \cite{2020ApJ...896...26C}
        };
\node [draw=none, anchor=east, align=right] at ($(1.6in,  11.312in) - 4*(0in, 0.187in)$) {
            Jianhui, et al., (2025) \cite{2025ApJ...990L..37L}
        };
\node [draw=none, anchor=east, align=right] at ($(1.6in,  11.312in) - 5*(0in, 0.187in)$) {
            Crosta, et al., (2024) \cite{2024MNRAS.529.4681B}
        };
\node [draw=none, anchor=east, align=right] at ($(1.6in,  11.312in) - 6*(0in, 0.187in)$) {
            Staudt, et al., (2024) \cite{2024JCAP...08..022S}
        };
\node [draw=none, anchor=east, align=right] at ($(1.6in,  11.312in) - 7*(0in, 0.187in)$) {
            Ou, et al., (2023) \cite{2024MNRAS.528..693O}
        };
\node [draw=none, anchor=east, align=right] at ($(1.6in,  11.312in) - 8*(0in, 0.187in)$) {
            Zhou, et al., (2022) \cite{2023ApJ...946...73Z}
        };
\node [draw=none, anchor=east, align=right] at ($(1.6in,  11.312in) - 9*(0in, 0.187in)$) {
            Ablimit, et al., (2020) \cite{2020ApJ...895L..12A}
        };
\node [draw=none, anchor=east, align=right] at ($(1.6in,  11.312in) - 10*(0in, 0.187in)$) {
            Benito, et al., (2020) \cite{2021PDU....3200826B}
        };
\node [draw=none, anchor=east, align=right] at ($(1.6in,  11.312in) - 11*(0in, 0.187in)$) {
            Sofue (2020) \cite{2020Galax...8...37S}
        };
\node [draw=none, anchor=east, align=right] at ($(1.6in,  11.312in) - 12*(0in, 0.187in)$) {
            Benito, et al., (2019) \cite{2019JCAP...03..033B}
        };
\node [draw=none, anchor=east, align=right] at ($(1.6in,  11.312in) - 13*(0in, 0.187in)$) {
            Karukes, et al., (2019) \cite{2019JCAP...09..046K}
        };
\node [draw=none, anchor=east, align=right] at ($(1.6in,  11.312in) - 14*(0in, 0.187in)$) {
            Lin et al., (2019) \cite{2019MNRAS.487.5679L}
        };
\node [draw=none, anchor=east, align=right] at ($(1.6in,  11.312in) - 15*(0in, 0.187in)$) {
            de Salas, et al., (2019) \cite{2019JCAP...10..037D}
        };
\node [draw=none, anchor=east, align=right] at ($(1.6in,  11.312in) - 16*(0in, 0.187in)$) {
            Huang, et al., (2016) \cite{2016MNRAS.463.2623H}
        };
\node [draw=none, anchor=east, align=right] at ($(1.6in,  11.312in) - 17*(0in, 0.187in)$) {
            Pato, et al., (2015) \cite{2015JCAP...12..001P}
        };
\node [draw=none, anchor=east, align=right] at ($(1.6in,  11.312in) - 18*(0in, 0.187in)$) {
            Cautun, et al., (2020) \cite{2020MNRAS.494.4291C}
        };
\node [draw=none, anchor=east, align=right] at ($(1.6in,  11.312in) - 19*(0in, 0.187in)$) {
            Crosta, et al., (2020) \cite{2020MNRAS.496.2107C}
        };
\node [draw=none, anchor=east, align=right] at ($(1.6in,  11.312in) - 20*(0in, 0.187in)$) {
            McMillan (2017) \cite{2017MNRAS.465...76M}
        };
\node [draw=none, anchor=east, align=right] at ($(1.6in,  11.312in) - 21*(0in, 0.187in)$) {
            Bienyame, et al., (2024) \cite{2024A&A...689A.280B}
        };
\node [draw=none, anchor=east, align=right] at ($(1.6in,  11.312in) - 22*(0in, 0.187in)$) {
            Haochuan, et al., (2023) \cite{2023MNRAS.520.3329L}
        };
\node [draw=none, anchor=east, align=right] at ($(1.6in,  11.312in) - 23*(0in, 0.187in)$) {
            Cole, et al., (2017) \cite{2017MNRAS.465..798C}
        };
\node [draw=none, anchor=east, align=right] at ($(1.6in,  11.312in) - 24*(0in, 0.187in)$) {
            Binney, et al., (2015) \cite{2015MNRAS.454.3653B}
        };
\node [draw=none, anchor=east, align=right] at ($(1.6in,  11.312in) - 25*(0in, 0.187in)$) {
            Bienyame, et al., (2014) \cite{2014A&A...571A..92B}
        };
\node [draw=none, anchor=east, align=right] at ($(1.6in,  11.312in) - 26*(0in, 0.187in)$) {
            Piffl, et al., (2014) \cite{2014MNRAS.445.3133P}
        };
\node [draw=none, anchor=east, align=right] at ($(1.6in,  11.312in) - 27*(0in, 0.187in)$) {
            Hattori, et al., (2020) \cite{2021MNRAS.508.5468H}
        };
\node [draw=none, anchor=east, align=right] at ($(1.6in,  11.312in) - 28*(0in, 0.187in)$) {
            Wegg, et al., (2019) \cite{2019MNRAS.485.3296W}
        };
\node [draw=none, anchor=east, align=right] at ($(1.6in,  11.312in) - 29*(0in, 0.187in)$) {
            Nitschai, et al., (2021) \cite{2021ApJ...916..112N}
        };
\node [draw=none, anchor=east, align=right] at ($(1.6in,  11.312in) - 30*(0in, 0.187in)$) {
            Nitschai, et al., (2020) \cite{2020MNRAS.494.6001N}
        };
\node [draw=none, anchor=east, align=right] at ($(1.6in,  11.312in) - 31*(0in, 0.187in)$) {
            Horta, et al., (2024) \cite{2024ApJ...962..165H}
        };
\node [draw=none, anchor=east, align=right] at ($(1.6in,  11.312in) - 32*(0in, 0.187in)$) {
            Guo, et al., (2024) \cite{2024ApJ...960..133G}
        };
\node [draw=none, anchor=east, align=right] at ($(1.6in,  11.312in) - 33*(0in, 0.187in)$) {
            Guo, et al., (2022) \cite{2022ApJ...936..103G}
        };
\node [draw=none, anchor=east, align=right] at ($(1.6in,  11.312in) - 34*(0in, 0.187in)$) {
            Widmark, et al., (2021) \cite{2021A&A...653A..86W}
        };
\node [draw=none, anchor=east, align=right] at ($(1.6in,  11.312in) - 35*(0in, 0.187in)$) {
            Donlon, et al., (2025) \cite{2025arXiv251115865D}
        };
\node [draw=none, anchor=east, align=right] at ($(1.6in,  11.312in) - 36*(0in, 0.187in)$) {
            Donlon, et al., (2025) \cite{2025PhRvD.111j3036D}
        };
\node [draw=none, anchor=east, align=right] at ($(1.6in,  11.312in) - 37*(0in, 0.187in)$) {
            Donlon, et al., (2024) \cite{2024PhRvD.110b3026D}
        };
\node [draw=none, anchor=east, align=right] at ($(1.6in,  11.312in) - 38*(0in, 0.187in)$) {
            Moran, et al., (2024) \cite{moran}
        };
\node [draw=none, anchor=east, align=right] at ($(1.6in,  11.312in) - 39*(0in, 0.187in)$) {
            Chakrabarti, et al., (2021) \cite{2021ApJ...907L..26C}
        };
\node [draw=none, anchor=east, align=right] at ($(1.6in,  11.312in) - 40*(0in, 0.187in)$) {
            Steigerwald, et al., (2022) \cite{2022MNRAS.510.4779S}
        };
\node [draw=none, anchor=east, align=right] at ($(1.6in,  11.312in) - 41*(0in, 0.187in)$) {
            Lopez-Corredoira (2025) \cite{2025ApJ...978...45L}
        };
\node [draw=none, anchor=east, align=right] at ($(1.6in,  11.312in) - 42*(0in, 0.187in)$) {
            S\"oding, et al., (2025) \cite{2025MNRAS.542.2987S}
        };
\node [draw=none, anchor=east, align=right] at ($(1.6in,  11.312in) - 43*(0in, 0.187in)$) {
            Syaifudin, et al., (2024) \cite{2024MNRAS.534.3387S}
        };
\node [draw=none, anchor=east, align=right] at ($(1.6in,  11.312in) - 44*(0in, 0.187in)$) {
            Guo, et al., (2020) \cite{2020MNRAS.495.4828G}
        };
\node [draw=none, anchor=east, align=right] at ($(1.6in,  11.312in) - 45*(0in, 0.187in)$) {
            Salomon, et al., (2020) (North) \cite{2020A&A...643A..75S}
        };
\node [draw=none, anchor=east, align=right] at ($(1.6in,  11.312in) - 46*(0in, 0.187in)$) {
            Salomon, et al., (2020) (South) \cite{2020A&A...643A..75S}
        };
\node [draw=none, anchor=east, align=right] at ($(1.6in,  11.312in) - 47*(0in, 0.187in)$) {
            Wardana, et al., (2020) \cite{2020EPJWC.24004002W}
        };
\node [draw=none, anchor=east, align=right] at ($(1.6in,  11.312in) - 48*(0in, 0.187in)$) {
            Hagen, et al., (2018) \cite{2018A&A...615A..99H}
        };
\node [draw=none, anchor=east, align=right] at ($(1.6in,  11.312in) - 49*(0in, 0.187in)$) {
            Sivertsson, et al., (2018) \cite{2018MNRAS.478.1677S}
        };
\node [draw=none, anchor=east, align=right] at ($(1.6in,  11.312in) - 50*(0in, 0.187in)$) {
            Xia, et al. (2016) \cite{2016MNRAS.458.3839X}
        };
\node [draw=none, anchor=east, align=right] at ($(1.6in,  11.312in) - 51*(0in, 0.187in)$) {
            McKee, et al., (2015) \cite{2015ApJ...814...13M}
        };
\node [draw=none, anchor=east, align=right] at ($(1.6in,  11.312in) - 52*(0in, 0.187in)$) {
            Buch, et al., (2019) (A Stars) \cite{2019JCAP...04..026B}
        };
\node [draw=none, anchor=east, align=right] at ($(1.6in,  11.312in) - 53*(0in, 0.187in)$) {
            Buch, et al., (2019) (Early G Stars) \cite{2019JCAP...04..026B}
        };
\node [draw=none, anchor=east, align=right] at ($(1.6in,  11.312in) - 54*(0in, 0.187in)$) {
            Buch, et al., (2019) (F Stars) \cite{2019JCAP...04..026B}
        };
\node [draw=none, anchor=east, align=right] at ($(1.6in,  11.312in) - 55*(0in, 0.187in)$) {
            Schutz, et al., (2018) (A Stars) \cite{2018PhRvL.121h1101S}
        };
\node [draw=none, anchor=east, align=right] at ($(1.6in,  11.312in) - 56*(0in, 0.187in)$) {
            Schutz, et al., (2018) (G Stars) \cite{2018PhRvL.121h1101S}
        };
\node [draw=none, anchor=east, align=right] at ($(1.6in,  11.312in) - 57*(0in, 0.187in)$) {
            Schutz, et al., (2018) (F Stars) \cite{2018PhRvL.121h1101S}
        };
    \end{tikzpicture}
    }
    \caption{Our averaged measurement of the dark matter density at the Solar radius (top line), compared to recent measurements of the dark matter density at or near the Solar location.
    }
    \label{fig:darkmatter_densities}
\end{figure*}

These pointwise error estimates can be reduced by averaging over multiple points at the cost of introducing new symmetry assumptions.
In Figure~\ref{fig:s_mass_density} we show the total azimuthally-averaged mass density evaluated at $r=r_\odot$. The coordinate $s$ represents the arclength distance from the disk $(s=0)$ at constant Galactic radius $r=r_\odot$, defined as $s=r_\odot(\pi/2-\theta)$ where $\theta$ is the Galactocentric polar angle.
From this azimuthally-averaged mass density, we subtract the \MWP{} baryonic density model. The resulting rippling in the dark matter density within $|s|<0.5$ (seen in the left panel of Figure~\ref{fig:s_mass_density}) is the result of a mismatch between the disk scale height observed this work and in \MWP{}. 
Under the assumption of azimuthal symmetry, the local dark matter density at the (cylindrical) Solar radius is $\rho_\odot(R_\odot)=(\localdensityazim \pm \localdensityazimerr)\times{10^{-2}\:{M}}_\odot/{\rm pc}^3$.
Outside the disk, we observe a near-constant dark matter density as a function of $s$, indicating approximate spherical symmetry in the dark matter halo. If we assume that the dark matter is spherically symmetric (and thus has the same mass density at all values of $s$), we find an average dark matter density at the Solar radius of $\rho_\odot(r_\odot)=(\localdensityradial\pm\localdensityradialerr)\times 10^{-2}\:{M}_\odot/{\rm pc}^3 =(\localdensityradialGeV\pm\localdensityradialerrGeV)\,{\rm GeV/cm^3}$. For comparison, we provide an updated summary of all recent measurements of the local dark matter density in Figure~\ref{fig:darkmatter_densities}. 

This measurement of the spherically-averaged local dark matter density is notably smaller than the one provided in our previous work Ref.~\cite{lim2023mapping_journal}, which found $\rho_\odot(r_\odot)=(0.47\pm0.05)\,{\rm GeV/cm^3}$ -- corresponding to nearly a $3\sigma$ tension with the present result. Barring the significant difference in the number of samples used for these averages -- $15$ (independent) samples in the prior work and $215$ (mostly independent) samples in the present analysis -- there are two effects which are primarily responsible for this change. First is the implementation of dust correction. The previous average avoided the disk, and $14$ out of the $15$ points at which the density was evaluated were selected to fall within lines of sight we judged likely to be ``dust-free." However, in our present work we find that our dust efficiency function results in a $\sim10\%$ correction across a majority of these points, challenging our previous assumption that these lines of sight were dust-free. As part of the dust-avoidance in our previous analysis, the Solar location was the only point within the disk at which we measured density. Now, we include several dozens of samples within the disk. Second, and perhaps more importantly, our measured density field is derived from a continuous potential that enforces the curl-free acceleration condition. These improvements in our analysis explain the differences between our current \texttt{ClearPotential} measurement and the previous result.

\begin{figure}[t]
  \centering
    \centering

    \includegraphics[width=0.9\linewidth]{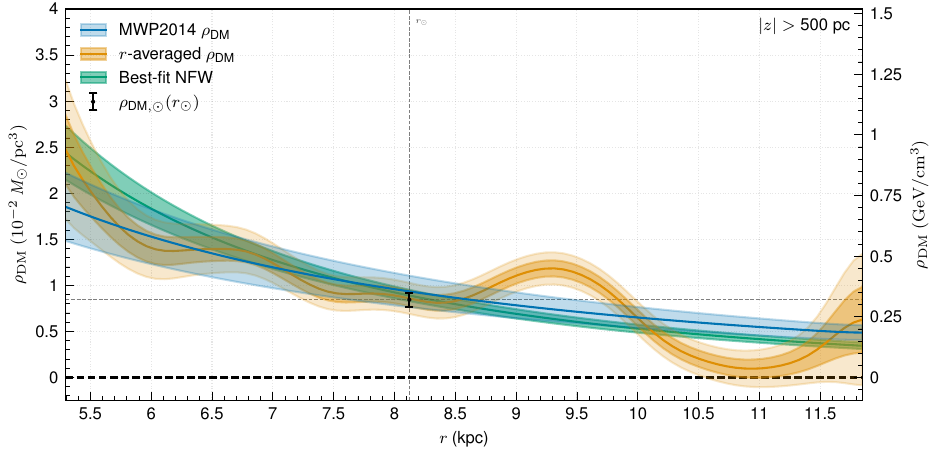}

  \caption{Weighted spherical average dark matter density (orange) compared to the \MWP{} halo model (blue); the spherically averaged local dark matter density (black error bar at $r=8.122$~kpc) and best-fit NFW profile with $1\sigma$ intervals (green).
  }
  \label{fig:r_mass_density}
\end{figure}

In Figure~\ref{fig:r_mass_density} we extend the assumption of spherical symmetry to the entire observational volume, presenting the angular-averaged dark matter density as a function of radius. At each radius, we evaluate the dark matter density over the three-dimensional bulk (the sampling procedure is outlined in Section~\ref{app:spatial_sampling}) and compute a correlation-weighted azimuthal average.  
The previously identified discrepant regions of mass density at both low and high radii are apparent, as the oscillations in $\rho_{\rm DM}$ seen in Figure~\ref{fig:2D_mass_density} at $6-7$~kpc and $9-10$~kpc. We again note that these regions displayed anomalies in the acceleration solutions. However such anomalies are not immediately apparent near the Solar location (or above and below the disk near the Solar radius), providing a level of confidence that the local density of dark matter found through our method is not being systematically shifted by unknown sources of disequilibrium or failures in the dust-correction model. Even with these oscillations, the overall trend of the dark matter radial density profile tracks the general expectation from well-motivated models such as the Navarro-Frenk-White (NFW) potential.

\subsection{Dark Matter Density Profiles}\label{sec:profiles}

Beyond basic symmetry assumptions, we can estimate the local dark matter density by modeling the global structure of the dark matter halo, including its radial profile and non-spherical morphology. We fit points sampled across the halo ($|z|>500$~pc) to three global models of the halo -- a standard NFW profile \cite{1996ApJ...462..563N}, a generalized NFW profile (gNFW) \cite{1996MNRAS.278..488Z}, and a tilted Lee-Suto \cite{2003ApJ...585..151L} triaxial NFW profile.
(For details of the fitting procedure, see Section~\ref{app:correlations}.) Fitting to the standard and generalized NFW profiles provides two measurements of the local density under the assumption of spherical symmetry, while fitting to the triaxial profile allows us to relax the assumption of spherical symmetry and characterize the potentially complex, non-spherical structure of the Milky Way's dark matter halo. We parameterize each density profile as:
\begin{equation}
    \rho(\chi)=\frac{\rho_0}{\chi^\beta(1+\chi)^{3-\beta}}=\rho_\odot\frac{\chi_\odot^\beta(1+\chi_\odot)^{3-\beta}}{\chi^\beta(1+\chi)^{3-\beta}},
\end{equation}
where $\rho_\odot$ is the dark matter halo density at the Solar location,
$\beta$ is the inner power law, and $\chi$ is a dimensionless radius. The spherically symmetric NFW and gNFW profiles use $\chi= r/r_s$, where $r_s$ is the scale radius of the break between the inner and outer power laws. The tilted triaxial potential uses
\begin{equation}
    \chi^2\equiv\left({x'}/{r_s}\right)^2+\left({y'}/{\xi_1 r_s}\right)^2+\left({z'}/{\xi_2 r_s}\right)^2,
\end{equation}
where $x', y', z'$ are the major, intermediate, and minor axes rotated from the standard Cartesian coordinates first by yaw (an angle $\psi$ about the $z$-axis) and then pitch (an angle $\theta$ about the $y'$-axis) angles. $\xi_1$ and $\xi_2$ are the major-to-intermediate and major-to-minor principal axis ratios, respectively. For both the NFW profile and the triaxial profile, the inner power law slope is fixed to $\beta=1$. The gNFW profile floats this power-law slope as a fit parameter.

\begin{figure}[t]
  \centering
  \includegraphics[width=0.5\columnwidth]{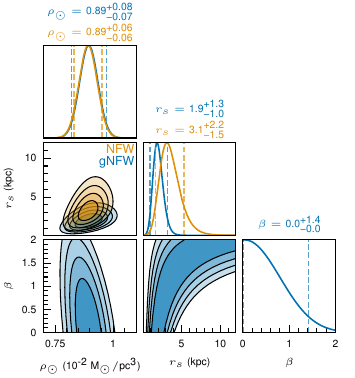}

  \caption{The $0.5\sigma$, $1\sigma$, $1.5\sigma$, and $2\sigma$ likelihood contours of the NFW and gNFW dark matter halo models over scale density $\rho_\odot$, scale length $r_s$, and power law index $\beta$
  }
  \label{fig:NFW_model_fits}
\end{figure}

\begin{figure}[t]
  \centering
  
  \begin{minipage}[t]{0.9\columnwidth}\vspace{0pt}%
    \centering
    \begin{tikzpicture}[baseline=(current bounding box.north)]
      \path (0,0) coordinate (gbb_nw);             
      \path (\linewidth,0) coordinate (gbb_ne);     

      \node[inner sep=0, anchor=north west] (top) at (gbb_nw)
        {\includegraphics[width=\columnwidth]{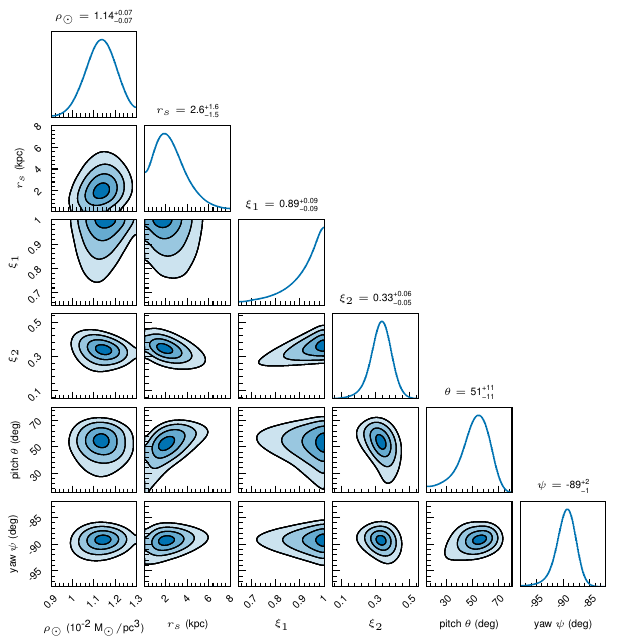}};

      \node[anchor=north east, xshift=0, yshift=0] (sideview) at (gbb_ne)
        {\includegraphics{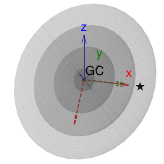}};

      \node[anchor=north, inner sep=0] at (sideview.south)
        {\includegraphics{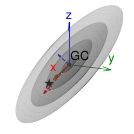}};
    \end{tikzpicture}
  \end{minipage}

  \caption{%
  (Left): The $0.5\sigma$, $1\sigma$, $1.5\sigma$, and $2\sigma$ likelihood contours of the tilted triaxial NFW model over scale density and radius, as well as shape parameters such as the axis ratios $\xi_1$, $\xi_2$ and pitch/yaw angles $\theta$, $\psi$.
  (Upper Right): A visualization of the best-fit triaxial model is shown, with principal and rotated axes centered on the Galactic center (GC) shown with respect to the Solar location ($\odot$).
  }
  \label{fig:triaxial_model_fits}
\end{figure}

The local dark matter density estimates resulting from the fits to the NFW and gNFW profiles are shown in Figure~\ref{fig:NFW_model_fits}. The best fitting NFW model has a reduced $\chi^2$ per degree of freedom of $\chi^2_\nu=\NFWchisq$, the best-fitting gNFW yields $\chi^2_\nu=\gNFWchisq$. The resulting fit to the triaxial model are shown in Figure~\ref{fig:triaxial_model_fits}. and the best-fitting triaxial NFW yields $\chi^2_\nu=\triaxialchisq$. 
Both the spherically-symmetric models predict a local dark matter density of $\NFWrho^{+\NFWrhoUpperErr}_{-\NFWrhoLowerErr}\,M_\odot/{\rm pc}^3$ (the gNFW fit results in slightly larger errors with the same central value). This is consistent (within $1\sigma$) with our spherical average, with a slight gain in precision. The triaxial model (the only non-spherically symmetric model) predicts $\triaxialrho\pm\triaxialrhoErr\, M_\odot/{\rm pc}^3$, which is higher than (and in slight tension with) the density found by the spherically-symmetric profiles.

Across all of these model fits, we find that small values of the scale radius $r_s$ are preferred. In the spherical and triaxial NFW profile, we find $r_s=\NFWrScale_{-\NFWrScaleLowerErr}^{+\NFWrScaleUpperErr}$~kpc and $r_s=\triaxialrScale_{-\triaxialrScaleLowerErr}^{+\triaxialrScaleUpperErr}$~kpc, respectively. When relaxing the power-law slope in the gNFW profile, we observe a clear preference for a much shorter scale radius of $r_s=\gNFWrScale_{-\gNFWrScaleLowerErr}^{+\gNFWrScaleUpperErr}$~kpc with a flat $\beta=\gNFWbeta_{-\gNFWbetaLowerErr}^{+\gNFWbetaUpperErr}$ inner power law slope; possibly indicative of a cored inner profile. It is interesting that the preference for a very short scale radius and flat inner power law slope was also observed in Ref.~\cite{2023arXiv230312838O}. 

Our fit to a tilted triaxial NFW potential allows us to characterize the non-spherical structure of the halo \cite{2022AJ....164..249H, 2022ApJ...934...14H, 2024arXiv240612969H}. We evaluate the six-dimensional likelihood with the Monte-Carlo code \textsc{emcee} \cite{emcee}, halting on convergence.
In addition to a local dark matter density which is $28\%$ larger than found in our spherical fits,
we find evidence for a tilted and non-spherical halo. 
The best fit solution predicts axis ratios of $\xi_1=\triaxialxiOne\pm\triaxialxiOneErr$ and $\xi_2=\triaxialxiTwo\pm\triaxialxiTwoErr$, a pitch of $(\triaxialtheta\pm\triaxialthetaErr)^\circ$, and rotated by a yaw of $\left({\triaxialpsi}_{-\triaxialpsiLowerErr}^{+\triaxialpsiUpperErr}\right)^\circ$. These axis ratios correspond to a triaxiality parameter of $T\equiv(1-\xi_1^2)/(1-\xi_2^2)=\triaxialT_{-\triaxialTLowerErr}^{+\triaxialTUpperErr}$. This result is in excellent agreement with the shape parameters and orientations found in one of the two modes presented by Ref.~\cite{2025ApJ...985L..22N}, which measured the triaxial halo using the accelerations of the GD-1 stellar stream and found a pitch angle of $56^\circ$, a triaxiality parameter of $0.17$, and a yaw of $-97^\circ$. Interestingly, it is unclear whether such small values of $T$ are expected based on $N$-body simulations of dark matter halos \cite{2006MNRAS.367.1781A, 2012JCAP...05..030S, 2025ApJ...985L..22N, 2023ApJ...957L..24H}. 

As a final demonstration of our method's ability to map the density field of the Milky Way, we perform a search for an additional disk-like component of dark matter. It has long been understood that dark matter models with self-interactions (for example, mirror sector dark matter \cite{1991SvA....35...21K, 1996PhLB..375...26B, 2000PhRvD..62f3506M, 2002PhRvD..66f3002M, 2004IJMPD..13.2161F, 2008arXiv0808.2595S, 2013JCAP...08..031H} and atomic dark matter \cite{2010JCAP...05..021K, 2011JCAP...10..011K, 2012PhRvD..85j1302C}) can possess dissipative cooling processes similar to the baryonic sector. The component of dark matter within the Galaxy with such dissipative interactions would cool and collapse into a rotationally-supported disk, with a thickness $h_{\rm DD}$ set by the masses and coupling between the dark matter and dark radiation \cite{Fan:2013yva}. 
Thicker dark disks have been postulated in the absence of dark sector interactions \cite{1989AJ.....98.1554L,2008MNRAS.389.1041R,2009MNRAS.397...44R,2017MNRAS.472.3722G}. While studies of stellar populations suggest that a dark disk is not present in the Milky Way \cite{2015MNRAS.450.2874R}, accreted structures have been identified \cite{2020NatAs...4.1078N} that would create thick disks. We concentrate on the thin disk parameter space, though our analysis applies to disks up to $\sim500$~pc scale heights. In this work, we follow the common assumption that the dark disk collapses in alignment with the baryonic disk.

We adopt the parameterization of the dark disk used in Refs.~\cite{Fan:2013tia, Fan:2013yva, 2021A&A...653A..86W} in terms of the surface density at the Solar radius $\Sigma_{\rm DD}$ and the scale height of the disk $h_{\rm DD}$. In analogy with the baryonic disk, we include an additional radial exponent with scale radius $R_{\rm DD}$. Previous works only covered short ranges (within $<1$~kpc) away from the Solar radius and did not include explicit radial dependence, equivalent to an effective scale radius of $R_{\rm DD}\gtrsim1$~kpc. Our assumed functional form is
\begin{equation}\label{eq:dark_disk}
    \rho_{\rm DD}=\frac{\Sigma_{\rm DD}}{4h_{\rm DD}}e^{-(R-R_\odot)/R_{\rm DD}}\cosh^{-2}\left({\frac{z}{2h_{\rm DD}}}\right).
\end{equation}

\begin{figure}[t]
    \centering
    \includegraphics[width=0.7\columnwidth]{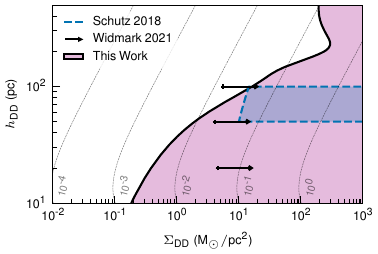}

  \caption{%
  $95\%$ confidence exclusion interval (pink) of a dark disk component in terms of local surface density $\Sigma_{\rm DD}$ and scale height $h_{\rm DD}$, compared to prior works by Refs.~\cite{2018PhRvL.121h1101S, 2021A&A...653A..86W}. Dashed gray lines indicate the density of the dark disk at the solar location in $M_\odot/{\rm pc}^3$.
  }
  \label{fig:darkdisk_fits}
\end{figure}

We model the total dark matter density distribution as our best-fit NFW halo plus the dark disk component, and set limits on the dark disk $(\Sigma_{\rm DD},~h_{\rm DD})$ parameter space using our measured dark matter density field (Figure~\ref{fig:darkdisk_fits}).
In our initial studies, we find 
stronger constraints on $(\Sigma_{\rm DD},~h_{\rm DD})$ for smaller $R_{\rm DD}$, where $\rho_{\rm DD}$ increases rapidly towards the Galactic Center. The constraints gradually weaken until $R_{\rm DD}\gtrsim 3$~kpc, after which the variation of $\rho_{\rm DD}$ across our analysis volume is small enough that it no longer has a strong effect on the constraints. Here we fix $R_{\rm DD}=3$~kpc to present the most conservative constraints on the dark disk in the $(\Sigma_{\rm DD},~h_{\rm DD})$ plane. 

We find that we are able to place substantially improved ($95\%$ confidence) null observation constraints on this form of a dark disk compared to previous literature, especially for small values of $h_{\rm DD}$. For $h_{\rm DD}=20$~pc we exclude $\Sigma_{\rm DD}>0.38\,{M}_\odot/{\rm pc}^2$, strengthening to $\Sigma_{\rm DD}>1.94\,{M}_\odot/{\rm pc}^2$ at $h_{\rm DD}=50$~pc, and $\Sigma_{\rm DD}>16.31\,{M}_\odot/{\rm pc}^2$ for $h_{\rm DD}=100$~pc. At the smallest $h_{\rm DD}$, this is an order-of-magnitude improvement on the constraint compared to Ref.~\cite{2021A&A...653A..86W}, and an improvement by over a factor of two for disks of intermediate thickness. As the scale height of the dark disk increases and approaches that of the baryonic disk, our constraints weaken, as would be expected given their morphological similarities.
At $h_{\rm DD}=100$~pc (approximately the scale height of the baryonic disk), we are consistent with the constraints obtained by Ref.~\cite{2018PhRvL.121h1101S}, but weaker than those found in Ref.~\cite{2021A&A...653A..86W}. We note that correlations in the large systematic uncertainties from the baryonic model are taken into account in our analysis (which generally weaken the bounds), whereas these correlations are apparently neglected in Ref.~\cite{2021A&A...653A..86W}.

\section{Conclusions}\label{sec:conclusion}

In this work, we have presented the first-ever fully data-driven, three-dimensional measurement of the gravitational potential of the local Milky Way. Our measurement technique leverages modern machine learning applied to the exquisite data from the \Gaia{} Space Telescope. By correcting for the effects of dust extinction without reliance on external maps, our machine-learning algorithm \texttt{ClearPotential} enables measurements closer to the Galactic Center than previously possible. The neural network encoding the potential is completely differentiable, and as a result, the gravitational accelerations and total mass density can be directly inferred within the entire 4~kpc observational volume.

While the potential itself largely reflects the expected axisymmetry of the Milky Way, small deviations can be observed. These deviations are more apparent when considering the acceleration and mass density results, where we find evidence for nonzero azimuthal acceleration at the Solar location and unexpected structures in the density distribution.
By comparing our inferred accelerations to independently measured LOS accelerations of 24 nearby binary pulsar systems, we are able to quantitatively test the equilibrium assumptions made in this analysis. We are in agreement with the independent measurements at all but three of the 24 pulsar systems, providing strong evidence for the overall validity of the equilibrium assumption. Interestingly, two of the three pulsars that we are in tension with were also highlighted in a recent work claiming that they provide evidence for a nearby DM subhalo. However, the level of disequilibrium we detect at these locations is similar to that found elsewhere in the observation volume.

Further differentiating the potential provides a flexible data-driven model of the mass density, correlated over length scales from $\sim 50$~pc (in the disk) to $\sim 300$~pc (in the halo). Subtracting a model of the baryonic density provides an experimental determination of the dark matter density across our 4~kpc observational volume. Though the point-wise measurement of the dark matter at the Solar location has significant errors originating in the baryonic model, these can be reduced by additional symmetry assumptions which allow averaging over uncorrelated points. With spherical symmetry, we find a density at the Solar radius of $(\localdensityradial \pm \localdensityradialerr)\times 10^{-2}\,M_\odot/{\rm pc}^3$. We further fit to a variety of dark matter profiles, finding preference for a short scale radius, a possible cored inner halo profile, and significant departure from spherical symmetry. While this  analysis is restricted to our local Milky Way neighborhood, it is striking that our best-fit model of a tilted triaxial halo agrees with a recent analysis of the GD-1 stream \cite{2025ApJ...985L..22N} performed on the other side of the Galaxy.

In the disk, we place the most competitive constraints to date on the existence of an additional disk component of dark matter aligned with the stellar disk, spanning scale heights of $10-500$~pc. This was achieved in spite of the conservative systematic uncertainties and correlation structure we adopted for the baryonic model of the disk, and was successful because of our ability to directly search for the dark disk across radial scales in $3$D. 

Looking to the future, the large uncertainties on the baryonic components of the MW gravitational potential emerged as the primary limiting factor across all aspects of this study. It is notable that the advancement in analysis techniques and control of dust extinction made possible by machine learning
have created a new demand for high resolution and accurate $3$D mass models of the Milky Way's stellar and gas components. Improving our understanding of the visible components of the Milky Way should be a clear priority for future research into Galactic dynamics.

Future {\it Gaia} data releases (DR4 and DR5) are expected to reduce the measurement errors intrinsic to the analysis. In these data releases, full 6D phase space will be available for fainter stars, possibly allowing for measurements of the gravitational potential over a larger volume of the Milky Way. In addition, it is possible that stellar phase space density (currently modeled using normalizing flows) could be more accurately represented using even more expressive density estimators, such as continuous normalizing flows trained with conditional flow matching~\cite{lipman2023flowmatchinggenerativemodeling,ot_cfm_2, fm_4, fm_5, fm_6}. Improvements of the phase space model of the dataset may increase our ability to resolve of the presence of disequilibrium, opening new windows on the structure and history of the distribution of mass within the Milky Way.

\section*{Acknowledgments}

This work was supported by the DOE under Award Number DOE-SC0010008. 
The work of SHL was also partly supported by IBS under the project code, IBS-R018-D1. 
This work was also performed in part at Aspen Center for Physics, which is supported by National Science Foundation grant PHY-2210452. 
The authors acknowledge the Office of Advanced Research Computing (OARC) at Rutgers, The State University of New Jersey for providing access to the Amarel cluster and associated research computing resources that have contributed to the results reported here. URL: \url{https://oarc.rutgers.edu}.
This research used resources of the National Energy Research Scientific Computing Center, a DOE Office of Science User Facility supported by the Office of Science of the U.S.~Department of Energy under Contract No.~DE-AC02-05CH11231 using NERSC award HEP-ERCAP0027491.

\appendix

\section{Model Likelihoods}\label{app:correlations}
We evaluate the global model fit likelihoods of the dark matter mass density field in Section~\ref{sec:results} using a $\chi^2$ parameter defined as:
\begin{equation}
    \chi^2 = (\boldsymbol{\rho}_{\mathrm{obs}} - \boldsymbol{\rho}_{\mathrm{pred}})^{\mathsf{T}}
             \, \Sigma_{\mathrm{dm}}^{-1} \,
             (\boldsymbol{\rho}_{\mathrm{obs}} - \boldsymbol{\rho}_{\mathrm{pred}}),
\end{equation}
and use a covariance-weighted average for spatial averaging:
\begin{equation}
    \overline{\rho}
    = \frac{\mathbf{1}^{\mathsf{T}} \, \Sigma_{\mathrm{dm}}^{-1} \, \boldsymbol{\rho}_{\rm obs}}
           {\mathbf{1}^{\mathsf{T}} \, \Sigma_{\mathrm{dm}}^{-1} \, \mathbf{1}}.
\end{equation}
Here $\boldsymbol{\rho}_{\mathrm{obs}}=(\rho_{\rm obs}(\vec x_1),\dots,\rho_{\rm obs}(\vec x_n))$ and $\boldsymbol{\rho}_{\mathrm{pred}}=(\rho_{\rm pred}(\vec x_1),\dots,\rho_{\rm pred}(\vec x_n))$ are vectors (at different points in space) of measured and model-predicted dark matter densities respectively, $\mathbf{1}=(1,\dots,1)$ serves to contract the indices, and the dark matter covariance matrix $\Sigma_{\rm dm}$ is the sum of the total and baryonic mass density covariance matrices:
\begin{equation}
    \Sigma_{\rm dm}=\Sigma_{\rm tot}+\Sigma_{\rm b}.
\end{equation}
We describe the method with which we estimate $\Sigma_{\rm b}$ in Section~\ref{app:baryons}.
We treat the intrinsic errors -- discussed in Section~\ref{sec:uncertainties} -- that contribute to the estimate of the total density covariance $\Sigma_{\rm tot}$ as spatially uncorrelated. However, our machine learning algorithms introduce spatial correlations in these errors, as the networks modeling the phase space density $f_{\rm obs}(\vec{x},\vec{v})$, the potential $\Phi(\vec{x})$, and efficiency factor $\epsilon(\vec{x})$ are flexible, but not infinitely so. The total density field correlations in $\Sigma_{\rm tot}$ must be correctly accounted for in order to compute the full $\Sigma_{\rm dm}$ for fits to the dark matter mass density field.

To construct a model of the correlations of the total density field, we assume that the measured field $\rho(\vec{x})$ represents a smoothed version of an inaccessible ``truth-level" $\hat\rho(\vec{x})$ that contains no spatial correlations, convolved with a spatially-varying kernel $K$:
\begin{equation}
\rho(\vec{x})=(\hat\rho*K)(\vec{x}) = \int d^3y\, \hat\rho(\vec x + \vec y)K(\vec y|\vec x).
\end{equation}
For every realization of modeling pipeline for $\Phi$ (as discussed in Section~\ref{sec:uncertainties}), we measure a single instance of the function $\rho(\vec{x})$ perturbed by some spatially-correlated noise $\delta\rho(\vec{x})$. We measure many instances of the sum of the smeared $\rho$ and $\delta\rho$, which are related to the ``perfectly resolved" fields by:
\begin{equation}(\hat\rho+\delta\hat\rho)*K=\rho+\delta\rho
\end{equation}
where $\delta\hat\rho$ represents the intrinsic error that appears as an uncorrelated random noise added to $\hat\rho$. Under these assumptions, the covariance of the total density is derived from the two-point correlation function by
\begin{equation}\label{eq:two-point-correlation}
\begin{aligned}
\Sigma_{\rm tot}(\vec{x}_1, \vec{x}_2) &= \langle\delta\rho(\vec x_1)\delta\rho(\vec x_2)\rangle \\ &=\int d^3y K(\vec y|\vec x_1)K(\vec y+\vec x_1-\vec x_2|\vec x_2){\mathcal N}(\vec{x}_1+\vec{y}).
\end{aligned}
\end{equation}
where ${\mathcal N}(\vec x)$ is a normalization factor from the two point function of $\delta\hat\rho$.

Our goal is to reconstruct the kernel $K$ with a smooth model so we can produce stable estimates of the two-point correlation function of the errors of our measured mass density field.
Stability of the two-point correlation function is critical for well-behaved covariance matrices. In principle, we could use the direct estimates of these correlations from our $10$ bootstrapped and error-smoothed realizations of $\rho$. However, this introduces significant noise in the covariance matrix, making it ill-conditioned.
We adopt the ansatz that $K$ is a Gaussian kernel of spatially-varying bandwidth $s(\vec{x})$:
\begin{equation}\label{eq:Kansatz}
    K(\vec{y}|\vec{x})=
    \left(2\pi s^2(\vec{x})\right)^{-3/2}
    \exp\left(-\frac{||\vec{y}||^2}{2s^2(\vec{x})}\right).
\end{equation}
This is a reasonable approach as this Gaussian ``blur" only introduces local correlations, with a length scale of locality $s(\vec{x})$ that is unknown {\em a priori} but is directly interpretable from data. 

After substituting Eq.~\eqref{eq:Kansatz} into Eq.~\eqref{eq:two-point-correlation} and taking the lowest-order saddle point approximation, the resulting covariance matrix takes the form:
\begin{equation}\label{eq:double_sigma_corr}
   \Sigma_{{\rm tot}}(\vec{x}_1, \vec{x}_2)
   =\mathcal{N}(\vec x_1,\vec x_2)\sigma_1\sigma_2\exp\left(-\frac{||\vec{x}_1-\vec{x}_2||^2}{2(s^2_1+s^2_2)}\right),
\end{equation}
where $\sigma_1$ and $\sigma_2$ are the uncertainties in the mass density at $\vec{x}_1$ and $\vec{x}_2$, respectively; and $\mathcal{N}(\vec x_1,\vec x_2)$ is a normalization factor that goes to one as $\vec x_2\to \vec x_1$. We ignore this normalization factor in the following, as we find that the fits to determine $s(\vec x)$ are largely insensitive to it.
This ansatz simplifies the problem of determining the correlation model to only measuring the position-dependent correlation length $s(\vec{x})$.

We model $\ln s(\vec{x})$ as a neural network that is trained using the (MSE) residual of Eq.~\eqref{eq:double_sigma_corr} and measurements of the two-point correlations anchored at $4096$ points uniformly distributed within the $4$~kpc sphere, normalized by uncertainty. Around each anchor point, we sample $32$ ``secondary" points on concentric shells of increasing radius $r\in[0,2]$~kpc. We then compute the normalized two-point correlations between each anchor point and all of its secondary points from each source of error described in Section~\ref{sec:uncertainties} (measurement (m) and statistical (s)). We combine the correlations from each of these independent sources of error as the sum of covariances:
\begin{equation}
    \Sigma_{{\rm tot}}(\vec{x}_1, \vec{x}_2) = \sum_{a={\rm m,s}} \Sigma_{{\rm a}}(\vec{x}_1, \vec{x}_2).
\end{equation}

\begin{figure}
    \centering
    \includegraphics{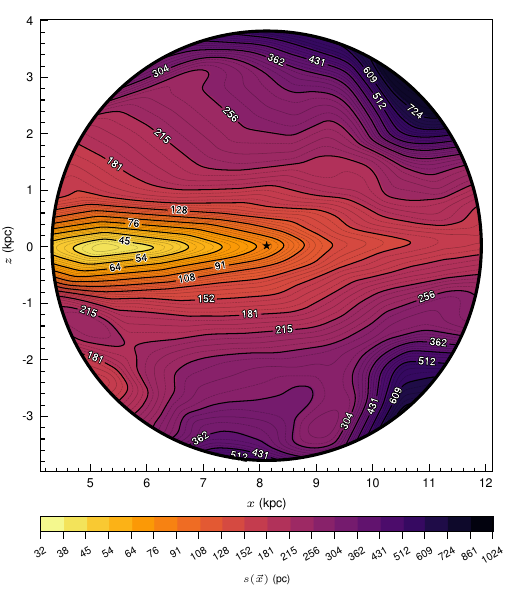}
    \caption{Correlation scale length and spatial resolution $s(\vec{x})$ in $\log_2$ scale of the mass density field in the $x-z$ plane, modeled by a neural network. Smaller and larger scale lengths indicate higher and lower spatial resolutions, respectively. We are able to resolve the disk to very short length scales of $\sim50$~pc due to the high density of available data in this region. The halo is less resolved, typically over length scales of $\sim300$~pc.}
    \label{fig:density_scale_length}
\end{figure}

Training is done with an $80\%/20\%$ training validation split, where batches of the correlations between $64$ anchor points and their accompanying secondaries are provided. Training halts when the validation loss does not improve for 5 epochs in a fast phase with a large learning rate ($10^{-2}$), and after 10 epochs in a slower ``refinement" phase with a small learning rate ($10^{-3}$). We obtain a minimum MSE loss of $0.036$ corresponding to a precision of $\sim20\%$. We conclude that the initial ansatz for $K$ was accurate, and accordingly that our neural network provides an effective parameterization of the correlation structure of $\rho$. We visualize the trained $s(\vec{x})$ function across the $x-z$ plane in Figure~\ref{fig:density_scale_length}.

\section{Baryonic Model}\label{app:baryons}

In order to characterize the distribution of nearby dark matter from the total mass density, we must quantify the distribution of baryonic matter.

In Ref.~\cite{lim2023mapping_journal}, we drew from the compilation of mass models in Ref.~\cite{2015ApJ...814...13M} and used Ref.~\cite{2023arXiv230312838O} to construct an approximate radial profile for each component. By constructing our own mass model, it was convenient to clearly characterize uncertainties based on the individual measurement uncertainties of these component mass models. However, the derived radial profile was somewhat crude. Additionally, it proved to be numerically difficult to compute a consistent acceleration and potential sourced from this baryonic mass model. To this end, throughout this work we have adopted the baryonic components of the \MWP{} model \cite{2015ApJS..216...29B} from the \textsc{gala} code \cite{gala}, which has efficient methods for computing the potential, acceleration, and density of this model.

We note that there is substantial disagreement between different baryonic mass models of the local Milky Way. The baryonic component of the mass model used in Ref.~\cite{lim2023mapping_journal} as well as the ``2022" variant of \MWP{} in the $\textsc{gala}$ code both differ from \MWP{}, with the latter having much higher densities in the midplane. Given the lack of uncertainties provided for either \MWP{} or its 2022 variant, it is difficult to determine the statistical extent to which these models disagree. 
For the purposes of this work, we adopt $10\%$ uncertainties on all model parameters of \MWP{}, which covers the range of density values predicted by the 2022 variant and another mass model (Ref.~\cite{2015ApJ...814...13M}) used in our previous work.

The total mass density field derived from data has an intrinsic resolution, given by the correlation length calculated in Section~\ref{app:correlations}. In order to consistently subtract the baryonic mass density from the total, the \MWP{} baryonic model must be convolved by the same correlation length scale. Finally, we compute two-point correlations across $1000$ random realizations of the baryonic profile, and include this covariance in all spatially-averaged measurements of the dark matter density.

\section{Spatial Sampling}\label{app:spatial_sampling}

When computing spatial averages of a smooth function (e.g.,~$\rho$), we must carefully consider the method used for sampling space to avoid numerical instabilities in the inversion of the covariance matrix. In general we sample spatial points such that no points are placed within two total density correlation lengths (as defined in Section~\ref{app:correlations}). This allows for sampled points to be approximately statistically independent, weighted primarily by their relative uncertainties.

A similar but simpler approach is taken for fits to the dark matter halo profile. First of all, we choose to avoid sampling the disk region $|z|<500$~pc, to avoid contamination by the poorly-modeled and highly-correlated baryonic disk. In the region above the disk, we sample uniformly and quasi-randomly.
After sampling, we drop out points in the halo spaced too closely together until an average low correlation threshold is satisfied. Finally, outlier points are identified in the NFW profile fit (approximately $15\%$ of points) which are removed from all other fits. This leaves us with $215$ statistically independent points across the halo. We repeat this procedure $10$ times to account for random sample variation, and ensemble the results of the model fits.

An inverse approach is taken for the dark disk exclusion analysis. The correlations of the underlying baryonic model are a critical input for the error model, so we sample the disk $|z|<500$~pc as densely as possible to probe the shortest length scales before numerical instabilities dominate. We do not drop out heavily correlated points. We sample 3,200 points within the disk, and repeat this procedure $10$ times to account for random sampling variation, and ensemble average the likelihoods obtained for each set of samples.

\bibliography{refs}

@misc{ot_cfm_2,
      title={Improving and generalizing flow-based generative models with minibatch optimal transport}, 
      author={Alexander Tong and Kilian Fatras and Nikolay Malkin and Guillaume Huguet and Yanlei Zhang and Jarrid Rector-Brooks and Guy Wolf and Yoshua Bengio},
      year={2024},
      eprint={2302.00482},
      archivePrefix={arXiv},
      primaryClass={cs.LG},
      url={https://arxiv.org/abs/2302.00482}, 
}

@misc{fm_4,
      title={Multisample Flow Matching: Straightening Flows with Minibatch Couplings}, 
      author={Aram-Alexandre Pooladian and Heli Ben-Hamu and Carles Domingo-Enrich and Brandon Amos and Yaron Lipman and Ricky T. Q. Chen},
      year={2023},
      eprint={2304.14772},
      archivePrefix={arXiv},
      primaryClass={cs.LG},
      url={https://arxiv.org/abs/2304.14772}, 
}

@misc{fm_5,
      title={Flow Straight and Fast: Learning to Generate and Transfer Data with Rectified Flow}, 
      author={Xingchao Liu and Chengyue Gong and Qiang Liu},
      year={2022},
      eprint={2209.03003},
      archivePrefix={arXiv},
      primaryClass={cs.LG},
      url={https://arxiv.org/abs/2209.03003}, 
}

@misc{fm_6,
      title={Building Normalizing Flows with Stochastic Interpolants}, 
      author={Michael S. Albergo and Eric Vanden-Eijnden},
      year={2023},
      eprint={2209.15571},
      archivePrefix={arXiv},
      primaryClass={cs.LG},
      url={https://arxiv.org/abs/2209.15571}, 
}

@misc{lipman2023flowmatchinggenerativemodeling,
      title={Flow Matching for Generative Modeling}, 
      author={Yaron Lipman and Ricky T. Q. Chen and Heli Ben-Hamu and Maximilian Nickel and Matt Le},
      year={2023},
      eprint={2210.02747},
      archivePrefix={arXiv},
      primaryClass={cs.LG},
      url={https://arxiv.org/abs/2210.02747}, 
}

@article{2021ApJ...907L..26C,
	adsurl = {https://ui.adsabs.harvard.edu/abs/2021ApJ...907L..26C},
	archiveprefix = {arXiv},
	author = {{Chakrabarti}, Sukanya and {Chang}, Philip and {Lam}, Michael T. and {Vigeland}, Sarah J. and {Quillen}, Alice C.},
	doi = {10.3847/2041-8213/abd635},
	eid = {L26},
	eprint = {2010.04018},
	journal = {Astrophys. J. Letters},
	month = feb,
	number = {2},
	pages = {L26},
	primaryclass = {astro-ph.GA},
	title = {{A Measurement of the Galactic Plane Mass Density from Binary Pulsar Accelerations}},
	volume = {907},
	year = 2021}

@article{9089305,
	author = {Kobyzev, Ivan and Prince, Simon J.D. and Brubaker, Marcus A.},
	doi = {10.1109/TPAMI.2020.2992934},
	journal = {IEEE Transactions on Pattern Analysis and Machine Intelligence},
	number = {11},
	pages = {3964-3979},
	title = {Normalizing Flows: An Introduction and Review of Current Methods},
	volume = {43},
	year = {2021}}

@article{2018PhRvL.121h1101S,
	adsurl = {https://ui.adsabs.harvard.edu/abs/2018PhRvL.121h1101S},
	archiveprefix = {arXiv},
	author = {{Schutz}, Katelin and {Lin}, Tongyan and {Safdi}, Benjamin R. and {Wu}, Chih-Liang},
	doi = {10.1103/PhysRevLett.121.081101},
	eid = {081101},
	eprint = {1711.03103},
	journal = {Phys. Rev. Lett.},
	month = aug,
	number = {8},
	pages = {081101},
	primaryclass = {astro-ph.GA},
	title = {{Constraining a Thin Dark Matter Disk with G a i a}},
	volume = {121},
	year = 2018}

@article{2018A&A...615A..99H,
	adsurl = {https://ui.adsabs.harvard.edu/abs/2018A&A...615A..99H},
	archiveprefix = {arXiv},
	author = {{Hagen}, Jorrit H.~J. and {Helmi}, Amina},
	doi = {10.1051/0004-6361/201832903},
	eid = {A99},
	eprint = {1802.09291},
	journal = {Astronomy \& Astrophysics},
	month = jul,
	pages = {A99},
	primaryclass = {astro-ph.GA},
	title = {{The vertical force in the solar neighbourhood using red clump stars in TGAS and RAVE. Constraints on the local dark matter density}},
	volume = {615},
	year = 2018}

@article{2018MNRAS.478.1677S,
	adsurl = {https://ui.adsabs.harvard.edu/abs/2018MNRAS.478.1677S},
	archiveprefix = {arXiv},
	author = {{Sivertsson}, S. and {Silverwood}, H. and {Read}, J.~I. and {Bertone}, G. and {Steger}, P.},
	doi = {10.1093/mnras/sty977},
	eprint = {1708.07836},
	journal = {Monthly Notices of the Royal Astronomical Society},
	month = aug,
	number = {2},
	pages = {1677-1693},
	primaryclass = {astro-ph.GA},
	title = {{The local dark matter density from SDSS-SEGUE G-dwarfs}},
	volume = {478},
	year = 2018
}

@article{2020MNRAS.495.4828G,
	adsurl = {https://ui.adsabs.harvard.edu/abs/2020MNRAS.495.4828G},
	archiveprefix = {arXiv},
	author = {{Guo}, Rui and {Liu}, Chao and {Mao}, Shude and {Xue}, Xiang-Xiang and {Long}, R.~J. and {Zhang}, Lan},
	doi = {10.1093/mnras/staa1483},
	eprint = {2005.12018},
	journal = {Monthly Notices of the Royal Astronomical Society},
	month = jul,
	number = {4},
	pages = {4828-4844},
	primaryclass = {astro-ph.GA},
	title = {{Measuring the local dark matter density with LAMOST DR5 and Gaia DR2}},
	volume = {495},
	year = 2020}

@article{2020A&A...643A..75S,
	adsurl = {https://ui.adsabs.harvard.edu/abs/2020A&A...643A..75S},
	archiveprefix = {arXiv},
	author = {{Salomon}, Jean-Baptiste and {Bienaym{\'e}}, Olivier and {Reyl{\'e}}, C{\'e}line and {Robin}, Annie C. and {Famaey}, Benoit},
	doi = {10.1051/0004-6361/202038535},
	eid = {A75},
	eprint = {2009.04495},
	journal = {Astronomy \& Astrophysics},
	month = nov,
	pages = {A75},
	primaryclass = {astro-ph.GA},
	title = {{Kinematics and dynamics of Gaia red clump stars. Revisiting north-south asymmetries and dark matter density at large heights}},
	volume = {643},
	year = 2020}

@article{2020MNRAS.494.6001N,
	adsurl = {https://ui.adsabs.harvard.edu/abs/2020MNRAS.494.6001N},
	archiveprefix = {arXiv},
	author = {{Nitschai}, Maria Selina and {Cappellari}, Michele and {Neumayer}, Nadine},
	doi = {10.1093/mnras/staa1128},
	eprint = {1909.05269},
	journal = {Monthly Notices of the Royal Astronomical Society},
	month = jun,
	number = {4},
	pages = {6001-6011},
	primaryclass = {astro-ph.GA},
	title = {{First Gaia dynamical model of the Milky Way disc with six phase space coordinates: a test for galaxy dynamics}},
	volume = {494},
	year = 2020}

@article{2021ApJ...916..112N,
	adsurl = {https://ui.adsabs.harvard.edu/abs/2021ApJ...916..112N},
	archiveprefix = {arXiv},
	author = {{Nitschai}, Maria Selina and {Eilers}, Anna-Christina and {Neumayer}, Nadine and {Cappellari}, Michele and {Rix}, Hans-Walter},
	doi = {10.3847/1538-4357/ac04b5},
	eid = {112},
	eprint = {2106.05286},
	journal = {Astrophys. J.},
	month = aug,
	number = {2},
	pages = {112},
	primaryclass = {astro-ph.GA},
	title = {{Dynamical Model of the Milky Way Using APOGEE and Gaia Data}},
	volume = {916},
	year = 2021}

@article{2021MNRAS.508.5468H,
	adsurl = {https://ui.adsabs.harvard.edu/abs/2021MNRAS.508.5468H},
	archiveprefix = {arXiv},
	author = {{Hattori}, Kohei and {Valluri}, Monica and {Vasiliev}, Eugene},
	doi = {10.1093/mnras/stab2898},
	eprint = {2012.03908},
	journal = {Monthly Notices of the Royal Astronomical Society},
	month = dec,
	number = {4},
	pages = {5468-5492},
	primaryclass = {astro-ph.GA},
	title = {{Action-based distribution function modelling for constraining the shape of the Galactic dark matter halo}},
	volume = {508},
	year = 2021}

@article{2019MNRAS.485.3296W,
	adsurl = {https://ui.adsabs.harvard.edu/abs/2019MNRAS.485.3296W},
	archiveprefix = {arXiv},
	author = {{Wegg}, Christopher and {Gerhard}, Ortwin and {Bieth}, Marie},
	doi = {10.1093/mnras/stz572},
	eprint = {1806.09635},
	journal = {Monthly Notices of the Royal Astronomical Society},
	month = may,
	number = {3},
	pages = {3296-3316},
	primaryclass = {astro-ph.GA},
	title = {{The gravitational force field of the Galaxy measured from the kinematics of RR Lyrae in Gaia}},
	volume = {485},
	year = 2019}

@article{papamakarios2018masked,
 author = {Papamakarios, George and Pavlakou, Theo and Murray, Iain},
 journal = {Advances in Neural Information Processing Systems (NIPS)},
 editor = {I. Guyon and U. Von Luxburg and S. Bengio and H. Wallach and R. Fergus and S. Vishwanathan and R. Garnett},
 pages = {},
 publisher = {Curran Associates, Inc.},
 title = {Masked Autoregressive Flow for Density Estimation},
 url = {https://proceedings.neurips.cc/paper_files/paper/2017/hash/6c1da886822c67822bcf3679d04369fa-Abstract.html},
 volume = {30},
 year = {2017},
	archiveprefix = {arXiv},
	eprint = {1705.07057},
	primaryclass = {stat.ML},
}

@article{2016Gaia,
	author = {Prusti, T. and de Bruijne, J. H. J. and Brown, A. G. A. and Vallenari, A. and Babusiaux, C. and Bailer-Jones, C. A. L. and Bastian, U. and Biermann, M. and Evans, D. W. and Eyer, L. and Jansen, F. and Jordi, C. and et al.},
	doi = {10.1051/0004-6361/201629272},
	issn = {1432-0746},
	journal = {Astronomy \& Astrophysics},
	month = {Nov},
	pages = {A1},
	publisher = {EDP Sciences},
	title = {TheGaiamission},
	url = {http://dx.doi.org/10.1051/0004-6361/201629272},
	volume = {595},
	year = {2016}}

@article{2021Gaia,
	author = {Lindegren, L. and Klioner, S. A. and Hern{\'a}ndez, J. and Bombrun, A. and Ramos-Lerate, M. and Steidelm{\"u}ller, H. and Bastian, U. and Biermann, M. and de Torres, A. and Gerlach, E. and Geyer, R. and Hilger, T. and Hobbs, D. and et al.},
	doi = {10.1051/0004-6361/202039709},
	issn = {1432-0746},
	journal = {Astronomy \& Astrophysics},
	month = {Apr},
	pages = {A2},
	publisher = {EDP Sciences},
	title = {Gaia Early Data Release 3},
	url = {http://dx.doi.org/10.1051/0004-6361/202039709},
	volume = {649},
	year = {2021}}

@ARTICLE{1996ApJ...462..563N,
       author = {{Navarro}, Julio F. and {Frenk}, Carlos S. and {White}, Simon D.~M.},
        title = "{The Structure of Cold Dark Matter Halos}",
      journal = {\apj},
         year = 1996,
        month = may,
       volume = {462},
        pages = {563},
          doi = {10.1086/177173},
archivePrefix = {arXiv},
       eprint = {astro-ph/9508025},
 primaryClass = {astro-ph},
       adsurl = {https://ui.adsabs.harvard.edu/abs/1996ApJ...462..563N}
}

@article{Fan:2013tia,
	archiveprefix = {arXiv},
	author = {Fan, JiJi and Katz, Andrey and Randall, Lisa and Reece, Matthew},
	doi = {10.1103/PhysRevLett.110.211302},
	eprint = {1303.3271},
	journal = {Phys. Rev. Lett.},
	number = {21},
	pages = {211302},
	primaryclass = {hep-ph},
	title = {{Dark-Disk Universe}},
	volume = {110},
	year = {2013}}

@article{Fan:2013yva,
	archiveprefix = {arXiv},
	author = {Fan, JiJi and Katz, Andrey and Randall, Lisa and Reece, Matthew},
	doi = {10.1016/j.dark.2013.07.001},
	eprint = {1303.1521},
	journal = {Phys. Dark Univ.},
	pages = {139--156},
	primaryclass = {astro-ph.CO},
	title = {{Double-Disk Dark Matter}},
	volume = {2},
	year = {2013}}

@article{Buckley:2017ijx,
	archiveprefix = {arXiv},
	author = {Buckley, Matthew R. and Peter, Annika H. G.},
	doi = {10.1016/j.physrep.2018.07.003},
	eprint = {1712.06615},
	journal = {Phys. Rept.},
	pages = {1--60},
	primaryclass = {astro-ph.CO},
	title = {{Gravitational probes of dark matter physics}},
	volume = {761},
	year = {2018}}

@article{Clowe:2003tk,
	archiveprefix = {arXiv},
	author = {Clowe, Douglas and Gonzalez, Anthony and Markevitch, Maxim},
	doi = {10.1086/381970},
	eprint = {astro-ph/0312273},
	journal = {Astrophys. J.},
	pages = {596--603},
	title = {{Weak lensing mass reconstruction of the interacting cluster 1E0657-558: Direct evidence for the existence of dark matter}},
	volume = {604},
	year = {2004}}

@article{Planck:2018vyg,
	archiveprefix = {arXiv},
	author = {Aghanim, N. and others},
	collaboration = {Planck},
	doi = {10.1051/0004-6361/201833910},
	eprint = {1807.06209},
	journal = {Astronomy \& Astrophysics},
	note = {[Erratum: Astron.Astrophys. 652, C4 (2021)]},
	pages = {A6},
	primaryclass = {astro-ph.CO},
	title = {{Planck 2018 results. VI. Cosmological parameters}},
	volume = {641},
	year = {2020}}

@article{Salucci:2018hqu,
	archiveprefix = {arXiv},
	author = {Salucci, Paolo},
	doi = {10.1007/s00159-018-0113-1},
	eprint = {1811.08843},
	journal = {Astron. Astrophys. Rev.},
	number = {1},
	pages = {2},
	primaryclass = {astro-ph.GA},
	title = {{The distribution of dark matter in galaxies}},
	volume = {27},
	year = {2019}}

@article{2011ARA&A..49..409A,
	adsurl = {https://ui.adsabs.harvard.edu/abs/2011ARA&A..49..409A},
	archiveprefix = {arXiv},
	author = {{Allen}, Steven W. and {Evrard}, August E. and {Mantz}, Adam B.},
	doi = {10.1146/annurev-astro-081710-102514},
	eprint = {1103.4829},
	journal = {Annual Review of Astronomy and Astrophysics},
	month = sep,
	number = {1},
	pages = {409-470},
	primaryclass = {astro-ph.CO},
	title = {{Cosmological Parameters from Observations of Galaxy Clusters}},
	volume = {49},
	year = 2011}

@article{1933AcHPh...6..110Z,
	adsurl = {https://ui.adsabs.harvard.edu/abs/1933AcHPh...6..110Z},
	author = {{Zwicky}, F.},
	journal = {Helvetica Physica Acta},
	month = jan,
	pages = {110-127},
	title = {{Die Rotverschiebung von extragalaktischen Nebeln}},
	volume = {6},
	year = 1933}

@article{1980ApJ...238..471R,
	adsurl = {https://ui.adsabs.harvard.edu/abs/1980ApJ...238..471R},
	author = {{Rubin}, V.~C. and {Ford}, W.~K., Jr. and {Thonnard}, N.},
	doi = {10.1086/158003},
	journal = {Astrophys. J.},
	month = jun,
	pages = {471-487},
	title = {{Rotational properties of 21 SC galaxies with a large range of luminosities and radii, from NGC 4605 (R=4kpc) to UGC 2885 (R=122kpc).}},
	volume = {238},
	year = 1980}

@article{1939LicOB..19...41B,
	adsurl = {https://ui.adsabs.harvard.edu/abs/1939LicOB..19...41B},
	author = {{Babcock}, Horace W.},
	doi = {10.5479/ADS/bib/1939LicOB.19.41B},
	journal = {Lick Observatory Bulletin},
	month = jan,
	pages = {41-51},
	title = {{The rotation of the Andromeda Nebula}},
	volume = {498},
	year = 1939}

@article{2018Natur.561..360A,
	adsurl = {https://ui.adsabs.harvard.edu/abs/2018Natur.561..360A},
	archiveprefix = {arXiv},
	author = {{Antoja}, T. and {Helmi}, A. and {Romero-G{\'o}mez}, M. and {Katz}, D. and {Babusiaux}, C. and {Drimmel}, R. and {Evans}, D.~W. and {Figueras}, F. and {Poggio}, E. and {Reyl{\'e}}, C. and {Robin}, A.~C. and {Seabroke}, G. and {Soubiran}, C.},
	doi = {10.1038/s41586-018-0510-7},
	eprint = {1804.10196},
	journal = {Nature},
	month = sep,
	number = {7723},
	pages = {360-362},
	primaryclass = {astro-ph.GA},
	title = {{A dynamically young and perturbed Milky Way disk}},
	volume = {561},
	year = 2018}

@article{2014JPhG...41f3101R,
	adsurl = {https://ui.adsabs.harvard.edu/abs/2014JPhG...41f3101R},
	archiveprefix = {arXiv},
	author = {{Read}, J.~I.},
	doi = {10.1088/0954-3899/41/6/063101},
	eid = {063101},
	eprint = {1404.1938},
	journal = {Journal of Physics G Nuclear Physics},
	month = jun,
	number = {6},
	pages = {063101},
	primaryclass = {astro-ph.GA},
	title = {{The local dark matter density}},
	volume = {41},
	year = 2014}

@article{2021MNRAS.506.5721A,
	adsurl = {https://ui.adsabs.harvard.edu/abs/2021MNRAS.506.5721A},
	archiveprefix = {arXiv},
	author = {{An}, J. and {Naik}, A.~P. and {Evans}, N.~W. and {Burrage}, C.},
	doi = {10.1093/mnras/stab2049},
	eprint = {2106.05981},
	journal = {Monthly Notices of the Royal Astronomical Society},
	month = oct,
	number = {4},
	pages = {5721-5730},
	primaryclass = {astro-ph.GA},
	title = {{Charting galactic accelerations: when and how to extract a unique potential from the distribution function}},
	volume = {506},
	year = 2021}

@inproceedings{2020EPJWC.24004002W,
	adsurl = {https://ui.adsabs.harvard.edu/abs/2020EPJWC.24004002W},
	author = {{Wardana}, M. Dafa and {Wulandari}, Hesti and {Sulistiyowati} and {Khatami}, Akbar H.},
	booktitle = {European Physical Journal Web of Conferences},
	doi = {10.1051/epjconf/202024004002},
	eid = {04002},
	month = dec,
	pages = {04002},
	series = {European Physical Journal Web of Conferences},
	title = {{Determination of the local dark matter density using K-dwarfs from Gaia DR2}},
	volume = {240},
	year = 2020}

@ARTICLE{2021A&A...653A..86W,
       author = {{Widmark}, A. and {Laporte}, C.~F.~P. and {de Salas}, P.~F. and {Monari}, G.},
        title = "{Weighing the Galactic disk using phase-space spirals. II. Most stringent constraints on a thin dark disk using Gaia EDR3}",
      journal = {\aap},
         year = 2021,
        month = sep,
       volume = {653},
          eid = {A86},
        pages = {A86},
          doi = {10.1051/0004-6361/202141466},
archivePrefix = {arXiv},
       eprint = {2105.14030},
 primaryClass = {astro-ph.GA},
       adsurl = {https://ui.adsabs.harvard.edu/abs/2021A&A...653A..86W}
}

@ARTICLE{2019JCAP...04..026B,
       author = {{Buch}, Jatan and {Leung}, John Shing Chau and {Fan}, JiJi},
        title = "{Using Gaia DR2 to constrain local dark matter density and thin dark disk}",
      journal = {\jcap},
         year = 2019,
        month = apr,
       volume = {2019},
       number = {4},
          eid = {026},
        pages = {026},
          doi = {10.1088/1475-7516/2019/04/026},
archivePrefix = {arXiv},
       eprint = {1808.05603},
 primaryClass = {astro-ph.GA},
       adsurl = {https://ui.adsabs.harvard.edu/abs/2019JCAP...04..026B}
}

@ARTICLE{2015ApJ...814...13M,
       author = {{McKee}, Christopher F. and {Parravano}, Antonio and {Hollenbach}, David J.},
        title = "{Stars, Gas, and Dark Matter in the Solar Neighborhood}",
      journal = {\apj},
         year = 2015,
        month = nov,
       volume = {814},
       number = {1},
          eid = {13},
        pages = {13},
          doi = {10.1088/0004-637X/814/1/13},
archivePrefix = {arXiv},
       eprint = {1509.05334},
 primaryClass = {astro-ph.GA},
       adsurl = {https://ui.adsabs.harvard.edu/abs/2015ApJ...814...13M}
}

@ARTICLE{2016ARA&A..54...95G,
       author = {{Girardi}, L{\'e}o},
        title = "{Red Clump Stars}",
      journal = {\araa},
         year = 2016,
        month = sep,
       volume = {54},
        pages = {95-133},
          doi = {10.1146/annurev-astro-081915-023354},
       adsurl = {https://ui.adsabs.harvard.edu/abs/2016ARA&A..54...95G}
}

@ARTICLE{2022arXiv220800211G,
       author = {{Gaia Collaboration} and {Vallenari}, A. and {Brown}, A.~G.~A. and {Prusti}, T. and {de Bruijne}, J.~H.~J. et al.},
        title = "{Gaia Data Release 3: Summary of the content and survey properties}",
      journal = {arXiv e-prints},
         year = 2022,
        month = jul,
          eid = {arXiv:2208.00211},
        pages = {arXiv:2208.00211},
          doi = {10.48550/arXiv.2208.00211},
archivePrefix = {arXiv},
       eprint = {2208.00211},
 primaryClass = {astro-ph.GA},
       adsurl = {https://ui.adsabs.harvard.edu/abs/2022arXiv220800211G}
}

@ARTICLE{2019MNRAS.482.1417B,
       author = {{Bennett}, Morgan and {Bovy}, Jo},
        title = "{Vertical waves in the solar neighbourhood in Gaia DR2}",
      journal = {\mnras},
         year = 2019,
        month = jan,
       volume = {482},
       number = {1},
        pages = {1417-1425},
          doi = {10.1093/mnras/sty2813},
archivePrefix = {arXiv},
       eprint = {1809.03507},
 primaryClass = {astro-ph.GA},
       adsurl = {https://ui.adsabs.harvard.edu/abs/2019MNRAS.482.1417B}
}

@ARTICLE{2018A&A...615L..15G,
       author = {{GRAVITY Collaboration} and {Abuter}, R. and {Amorim}, A. and {Anugu}, N. and {Baub{\"o}ck}, M. et al.},
        title = "{Detection of the gravitational redshift in the orbit of the star S2 near the Galactic centre massive black hole}",
      journal = {\aap},
         year = 2018,
        month = jul,
       volume = {615},
          eid = {L15},
        pages = {L15},
          doi = {10.1051/0004-6361/201833718},
archivePrefix = {arXiv},
       eprint = {1807.09409},
 primaryClass = {astro-ph.GA},
       adsurl = {https://ui.adsabs.harvard.edu/abs/2018A&A...615L..15G}
}

@article{10.1093/mnras/stac153,
    author = {Naik, A P and An, J and Burrage, C and Evans, N W},
    title = "{Charting galactic accelerations – II. How to ‘learn’ accelerations in the solar neighbourhood}",
    journal = {Monthly Notices of the Royal Astronomical Society},
    volume = {511},
    number = {2},
    pages = {1609-1621},
    year = {2022},
    month = {01},
    issn = {0035-8711},
    doi = {10.1093/mnras/stac153},
    url = {https://doi.org/10.1093/mnras/stac153},
    eprint = {https://academic.oup.com/mnras/article-pdf/511/2/1609/48413075/stac153.pdf},
}

@ARTICLE{2015MNRAS.454.3653B,
       author = {{Binney}, J. and {Piffl}, T.},
        title = "{The distribution function of the Galaxy's dark halo}",
      journal = {\mnras},
         year = 2015,
        month = dec,
       volume = {454},
       number = {4},
        pages = {3653-3663},
          doi = {10.1093/mnras/stv2225},
archivePrefix = {arXiv},
       eprint = {1509.06877},
 primaryClass = {astro-ph.GA},
       adsurl = {https://ui.adsabs.harvard.edu/abs/2015MNRAS.454.3653B}
}

@ARTICLE{2017MNRAS.465..798C,
       author = {{Cole}, D.~R. and {Binney}, James},
        title = "{A centrally heated dark halo for our Galaxy}",
      journal = {\mnras},
         year = 2017,
        month = feb,
       volume = {465},
       number = {1},
        pages = {798-810},
          doi = {10.1093/mnras/stw2775},
archivePrefix = {arXiv},
       eprint = {1610.07818},
 primaryClass = {astro-ph.GA},
       adsurl = {https://ui.adsabs.harvard.edu/abs/2017MNRAS.465..798C}
}

@ARTICLE{2017MNRAS.465...76M,
       author = {{McMillan}, Paul J.},
        title = "{The mass distribution and gravitational potential of the Milky Way}",
      journal = {\mnras},
         year = 2017,
        month = feb,
       volume = {465},
       number = {1},
        pages = {76-94},
          doi = {10.1093/mnras/stw2759},
archivePrefix = {arXiv},
       eprint = {1608.00971},
 primaryClass = {astro-ph.GA},
       adsurl = {https://ui.adsabs.harvard.edu/abs/2017MNRAS.465...76M}
}

@ARTICLE{2020MNRAS.494.4291C,
       author = {{Cautun}, Marius and {Ben{\'\i}tez-Llambay}, Alejandro and {Deason}, Alis J. and {Frenk}, Carlos S. and {Fattahi}, Azadeh and {G{\'o}mez}, Facundo A. and {Grand}, Robert J.~J. and {Oman}, Kyle A. and {Navarro}, Julio F. and {Simpson}, Christine M.},
        title = "{The milky way total mass profile as inferred from Gaia DR2}",
      journal = {\mnras},
         year = 2020,
        month = may,
       volume = {494},
       number = {3},
        pages = {4291-4313},
          doi = {10.1093/mnras/staa1017},
archivePrefix = {arXiv},
       eprint = {1911.04557},
 primaryClass = {astro-ph.GA},
       adsurl = {https://ui.adsabs.harvard.edu/abs/2020MNRAS.494.4291C}
}

@ARTICLE{2023arXiv230312838O,
       author = {{Ou}, Xiaowei and {Eilers}, Anna-Christina and {Necib}, Lina and {Frebel}, Anna},
        title = "{The dark matter profile of the Milky Way inferred from its circular velocity curve}",
      journal = {arXiv e-prints},
         year = 2023,
        month = mar,
          eid = {arXiv:2303.12838},
        pages = {arXiv:2303.12838},
          doi = {10.48550/arXiv.2303.12838},
archivePrefix = {arXiv},
       eprint = {2303.12838},
 primaryClass = {astro-ph.GA},
       adsurl = {https://ui.adsabs.harvard.edu/abs/2023arXiv230312838O}
}

@ARTICLE{2022ApJ...936..103G,
       author = {{Guo}, Rui and {Shen}, Juntai and {Li}, Zhao-Yu and {Liu}, Chao and {Mao}, Shude},
        title = "{The North/South Asymmetry of the Galaxy: Possible Connection to the Vertical Phase-space Snail}",
      journal = {\apj},
         year = 2022,
        month = sep,
       volume = {936},
       number = {2},
          eid = {103},
        pages = {103},
          doi = {10.3847/1538-4357/ac86cd},
archivePrefix = {arXiv},
       eprint = {2208.03667},
 primaryClass = {astro-ph.GA},
       adsurl = {https://ui.adsabs.harvard.edu/abs/2022ApJ...936..103G}
}

@ARTICLE{2022MNRAS.510.4779S,
       author = {{Steigerwald}, Heinrich and {Rodrigues}, Davi and {Profumo}, Stefano and {Marra}, Valerio},
        title = "{Type Ia supernova magnitude step from the local dark matter environment}",
      journal = {\mnras},
         year = 2022,
        month = mar,
       volume = {510},
       number = {4},
        pages = {4779-4795},
          doi = {10.1093/mnras/stab3747},
archivePrefix = {arXiv},
       eprint = {2112.09739},
 primaryClass = {astro-ph.CO},
       adsurl = {https://ui.adsabs.harvard.edu/abs/2022MNRAS.510.4779S}
}

@ARTICLE{2020MNRAS.496.2107C,
       author = {{Crosta}, Mariateresa and {Giammaria}, Marco and {Lattanzi}, Mario G. and {Poggio}, Eloisa},
        title = "{On testing CDM and geometry-driven Milky Way rotation curve models with Gaia DR2}",
      journal = {\mnras},
         year = 2020,
        month = aug,
       volume = {496},
       number = {2},
        pages = {2107-2122},
          doi = {10.1093/mnras/staa1511},
archivePrefix = {arXiv},
       eprint = {1810.04445},
 primaryClass = {astro-ph.GA},
       adsurl = {https://ui.adsabs.harvard.edu/abs/2020MNRAS.496.2107C}
}

@ARTICLE{2020ApJ...896...26C,
       author = {{Casagrande}, Luca},
        title = "{Connecting the Local Stellar Halo and Its Dark Matter Density to Dwarf Galaxies via Blue Stragglers}",
      journal = {\apj},
         year = 2020,
        month = jun,
       volume = {896},
       number = {1},
          eid = {26},
        pages = {26},
          doi = {10.3847/1538-4357/ab929f},
archivePrefix = {arXiv},
       eprint = {2005.09131},
 primaryClass = {astro-ph.GA},
       adsurl = {https://ui.adsabs.harvard.edu/abs/2020ApJ...896...26C}
}

@ARTICLE{2020ApJ...895L..12A,
doi = {10.3847/2041-8213/ab8d45},
url = {https://dx.doi.org/10.3847/2041-8213/ab8d45},
year = {2020},
month = {may},
publisher = {The American Astronomical Society},
volume = {895},
number = {1},
pages = {L12},
author = {Iminhaji Ablimit and Gang Zhao and Chris Flynn and Sarah A. Bird},
title = {The Rotation Curve, Mass Distribution, and Dark Matter Content of the Milky Way from Classical Cepheids},
journal = {The Astrophysical Journal Letters}
}

@ARTICLE{2016MNRAS.458.3839X,
       author = {{Xia}, Qiran and {Liu}, Chao and {Mao}, Shude and {Song}, Yingyi and {Zhang}, Lan and {Long}, R.~J. and {Zhang}, Yong and {Hou}, Yonghui and {Wang}, Yuefei and {Wu}, Yue},
        title = "{Determining the local dark matter density with LAMOST data}",
      journal = {\mnras},
         year = 2016,
        month = jun,
       volume = {458},
       number = {4},
        pages = {3839-3850},
          doi = {10.1093/mnras/stw565},
archivePrefix = {arXiv},
       eprint = {1510.06810},
 primaryClass = {astro-ph.GA},
       adsurl = {https://ui.adsabs.harvard.edu/abs/2016MNRAS.458.3839X}
}

@ARTICLE{2015JCAP...12..001P,
       author = {{Pato}, Miguel and {Iocco}, Fabio and {Bertone}, Gianfranco},
        title = "{Dynamical constraints on the dark matter distribution in the Milky Way}",
      journal = {\jcap},
         year = 2015,
        month = dec,
       volume = {2015},
       number = {12},
        pages = {001-001},
          doi = {10.1088/1475-7516/2015/12/001},
archivePrefix = {arXiv},
       eprint = {1504.06324},
 primaryClass = {astro-ph.GA},
       adsurl = {https://ui.adsabs.harvard.edu/abs/2015JCAP...12..001P}
}

@ARTICLE{2019JCAP...03..033B,
       author = {{Benito}, Maria and {Cuoco}, Alessandro and {Iocco}, Fabio},
        title = "{Handling the uncertainties in the Galactic Dark Matter distribution for particle Dark Matter searches}",
      journal = {\jcap},
         year = 2019,
        month = mar,
       volume = {2019},
       number = {3},
          eid = {033},
        pages = {033},
          doi = {10.1088/1475-7516/2019/03/033},
archivePrefix = {arXiv},
       eprint = {1901.02460},
 primaryClass = {astro-ph.GA},
       adsurl = {https://ui.adsabs.harvard.edu/abs/2019JCAP...03..033B}
}

@ARTICLE{2021PDU....3200826B,
       author = {{Benito}, Mar{\'\i}a and {Iocco}, Fabio and {Cuoco}, Alessandro},
        title = "{Uncertainties in the Galactic Dark Matter distribution: An update}",
      journal = {Physics of the Dark Universe},
         year = 2021,
        month = may,
       volume = {32},
          eid = {100826},
        pages = {100826},
          doi = {10.1016/j.dark.2021.100826},
archivePrefix = {arXiv},
       eprint = {2009.13523},
 primaryClass = {astro-ph.GA},
       adsurl = {https://ui.adsabs.harvard.edu/abs/2021PDU....3200826B}
}

@ARTICLE{2019JCAP...09..046K,
       author = {{Karukes}, E.~V. and {Benito}, M. and {Iocco}, F. and {Trotta}, R. and {Geringer-Sameth}, A.},
        title = "{Bayesian reconstruction of the Milky Way dark matter distribution}",
      journal = {\jcap},
         year = 2019,
        month = sep,
       volume = {2019},
       number = {9},
          eid = {046},
        pages = {046},
          doi = {10.1088/1475-7516/2019/09/046},
archivePrefix = {arXiv},
       eprint = {1901.02463},
 primaryClass = {astro-ph.GA},
       adsurl = {https://ui.adsabs.harvard.edu/abs/2019JCAP...09..046K}
}

@ARTICLE{2016MNRAS.463.2623H,
       author = {{Huang}, Y. and {Liu}, X. -W. and {Yuan}, H. -B. and {Xiang}, M. -S. and {Zhang}, H. -W. et al.},
        title = "{The Milky Way's rotation curve out to 100 kpc and its constraint on the Galactic mass distribution}",
      journal = {\mnras},
         year = 2016,
        month = dec,
       volume = {463},
       number = {3},
        pages = {2623-2639},
          doi = {10.1093/mnras/stw2096},
archivePrefix = {arXiv},
       eprint = {1604.01216},
 primaryClass = {astro-ph.GA},
       adsurl = {https://ui.adsabs.harvard.edu/abs/2016MNRAS.463.2623H}
}

@ARTICLE{2019MNRAS.487.5679L,
       author = {{Lin}, Hai-Nan and {Li}, Xin},
        title = "{The dark matter profiles in the Milky Way}",
      journal = {\mnras},
         year = 2019,
        month = aug,
       volume = {487},
       number = {4},
        pages = {5679-5684},
          doi = {10.1093/mnras/stz1698},
archivePrefix = {arXiv},
       eprint = {1906.08419},
 primaryClass = {astro-ph.GA},
       adsurl = {https://ui.adsabs.harvard.edu/abs/2019MNRAS.487.5679L}
}

@ARTICLE{2019JCAP...10..037D,
       author = {{de Salas}, P.~F. and {Malhan}, K. and {Freese}, K. and {Hattori}, K. and {Valluri}, M.},
        title = "{On the estimation of the local dark matter density using the rotation curve of the Milky Way}",
      journal = {\jcap},
         year = 2019,
        month = oct,
       volume = {2019},
       number = {10},
          eid = {037},
        pages = {037},
          doi = {10.1088/1475-7516/2019/10/037},
archivePrefix = {arXiv},
       eprint = {1906.06133},
 primaryClass = {astro-ph.GA},
       adsurl = {https://ui.adsabs.harvard.edu/abs/2019JCAP...10..037D}
}

@ARTICLE{2020Galax...8...37S,
       author = {{Sofue}, Yoshiaki},
        title = "{Rotation Curve of the Milky Way and the Dark Matter Density}",
      journal = {Galaxies},
         year = 2020,
        month = apr,
       volume = {8},
       number = {2},
        pages = {37},
          doi = {10.3390/galaxies8020037},
archivePrefix = {arXiv},
       eprint = {2004.11688},
 primaryClass = {astro-ph.GA},
       adsurl = {https://ui.adsabs.harvard.edu/abs/2020Galax...8...37S}
}

@ARTICLE{2014A&A...571A..92B,
       author = {{Bienaym{\'e}}, O. and {Famaey}, B. and {Siebert}, A. and {Freeman}, K.~C. and {Gibson}, B.~K. et al.},
        title = "{Weighing the local dark matter with RAVE red clump stars}",
      journal = {\aap},
         year = 2014,
        month = nov,
       volume = {571},
          eid = {A92},
        pages = {A92},
          doi = {10.1051/0004-6361/201424478},
archivePrefix = {arXiv},
       eprint = {1406.6896},
 primaryClass = {astro-ph.GA},
       adsurl = {https://ui.adsabs.harvard.edu/abs/2014A&A...571A..92B}
}

@ARTICLE{2014MNRAS.445.3133P,
       author = {{Piffl}, T. and {Binney}, J. and {McMillan}, P.~J. and {Steinmetz}, M. and {Helmi}, A. and {Wyse}, R.~F.~G. et al.},
        title = "{Constraining the Galaxy's dark halo with RAVE stars}",
      journal = {\mnras},
         year = 2014,
        month = dec,
       volume = {445},
       number = {3},
        pages = {3133-3151},
          doi = {10.1093/mnras/stu1948},
archivePrefix = {arXiv},
       eprint = {1406.4130},
 primaryClass = {astro-ph.GA},
       adsurl = {https://ui.adsabs.harvard.edu/abs/2014MNRAS.445.3133P}
}

@ARTICLE{2023MNRAS.521.5100B,
       author = {{Buckley}, Matthew R. and {Lim}, Sung Hak and {Putney}, Eric and {Shih}, David},
        title = "{Measuring Galactic dark matter through unsupervised machine learning}",
      journal = {\mnras},
         year = 2023,
        month = jun,
       volume = {521},
       number = {4},
        pages = {5100-5119},
          doi = {10.1093/mnras/stad843},
archivePrefix = {arXiv},
       eprint = {2205.01129},
 primaryClass = {astro-ph.GA},
       adsurl = {https://ui.adsabs.harvard.edu/abs/2023MNRAS.521.5100B}
}

@ARTICLE{2023ApJ...942...26G,
       author = {{Green}, Gregory M. and {Ting}, Yuan-Sen and {Kamdar}, Harshil},
        title = "{Deep Potential: Recovering the Gravitational Potential from a Snapshot of Phase Space}",
      journal = {\apj},
         year = 2023,
        month = jan,
       volume = {942},
       number = {1},
          eid = {26},
        pages = {26},
          doi = {10.3847/1538-4357/aca3a7},
archivePrefix = {arXiv},
       eprint = {2205.02244},
 primaryClass = {astro-ph.GA},
       adsurl = {https://ui.adsabs.harvard.edu/abs/2023ApJ...942...26G}
}

@ARTICLE{galpy,
       author = {{Bovy}, Jo},
        title = "{galpy: A python Library for Galactic Dynamics}",
      journal = {\apjs},
         year = 2015,
        month = feb,
       volume = {216},
       number = {2},
          eid = {29},
        pages = {29},
          doi = {10.1088/0067-0049/216/2/29},
archivePrefix = {arXiv},
       eprint = {1412.3451},
 primaryClass = {astro-ph.GA},
       adsurl = {https://ui.adsabs.harvard.edu/abs/2015ApJS..216...29B}
}

@ARTICLE{2024arXiv240608158B,
       author = {{Bienaym{\'e}}, O. and {Robin}, A.~C. and {Salomon}, J. -B. and {Reyl{\'e}}, C.},
        title = "{Dark matter in the Milky Way: Measurements up to 3 kpc from the Galactic plane above the Sun}",
      journal = {arXiv e-prints},
         year = 2024,
        month = jun,
          eid = {arXiv:2406.08158},
        pages = {arXiv:2406.08158},
          doi = {10.48550/arXiv.2406.08158},
archivePrefix = {arXiv},
       eprint = {2406.08158},
 primaryClass = {astro-ph.GA},
       adsurl = {https://ui.adsabs.harvard.edu/abs/2024arXiv240608158B}
}

@ARTICLE{2024ApJ...960..133G,
       author = {{Guo}, Rui and {Li}, Zhao-Yu and {Shen}, Juntai and {Mao}, Shude and {Liu}, Chao},
        title = "{Measuring the Milky Way Vertical Potential with the Phase Snail in a Model-independent Way}",
      journal = {\apj},
         year = 2024,
        month = jan,
       volume = {960},
       number = {2},
          eid = {133},
        pages = {133},
          doi = {10.3847/1538-4357/ad037b},
archivePrefix = {arXiv},
       eprint = {2310.10225},
 primaryClass = {astro-ph.GA},
       adsurl = {https://ui.adsabs.harvard.edu/abs/2024ApJ...960..133G}
}

@ARTICLE{2023MNRAS.520.3329L,
       author = {{Li}, Haochuan and {Widrow}, Lawrence M.},
        title = "{Residuals of an equilibrium model for the galaxy reveal a state of disequilibrium in the Solar Neighbourhood}",
      journal = {\mnras},
         year = 2023,
        month = apr,
       volume = {520},
       number = {3},
        pages = {3329-3344},
          doi = {10.1093/mnras/stad244},
archivePrefix = {arXiv},
       eprint = {2207.03516},
 primaryClass = {astro-ph.GA},
       adsurl = {https://ui.adsabs.harvard.edu/abs/2023MNRAS.520.3329L}
}

@ARTICLE{2000ApJ...534L.143B,
       author = {{Burkert}, Andreas},
        title = "{The Structure and Evolution of Weakly Self-interacting Cold Dark Matter Halos}",
      journal = {\apjl},
         year = 2000,
        month = may,
       volume = {534},
       number = {2},
        pages = {L143-L146},
          doi = {10.1086/312674},
archivePrefix = {arXiv},
       eprint = {astro-ph/0002409},
 primaryClass = {astro-ph},
       adsurl = {https://ui.adsabs.harvard.edu/abs/2000ApJ...534L.143B}
}

@ARTICLE{2000ApJ...543..514K,
       author = {{Kochanek}, C.~S. and {White}, Martin},
        title = "{A Quantitative Study of Interacting Dark Matter in Halos}",
      journal = {\apj},
         year = 2000,
        month = nov,
       volume = {543},
       number = {2},
        pages = {514-520},
          doi = {10.1086/317149},
archivePrefix = {arXiv},
       eprint = {astro-ph/0003483},
 primaryClass = {astro-ph},
       adsurl = {https://ui.adsabs.harvard.edu/abs/2000ApJ...543..514K}
}

@ARTICLE{2000ApJ...544L..87Y,
       author = {{Yoshida}, Naoki and {Springel}, Volker and {White}, Simon D.~M. and {Tormen}, Giuseppe},
        title = "{Weakly Self-interacting Dark Matter and the Structure of Dark Halos}",
      journal = {\apjl},
         year = 2000,
        month = dec,
       volume = {544},
       number = {2},
        pages = {L87-L90},
          doi = {10.1086/317306},
archivePrefix = {arXiv},
       eprint = {astro-ph/0006134},
 primaryClass = {astro-ph},
       adsurl = {https://ui.adsabs.harvard.edu/abs/2000ApJ...544L..87Y}
}

@ARTICLE{2001ApJ...547..574D,
       author = {{Dav{\'e}}, Romeel and {Spergel}, David N. and {Steinhardt}, Paul J. and {Wandelt}, Benjamin D.},
        title = "{Halo Properties in Cosmological Simulations of Self-interacting Cold Dark Matter}",
      journal = {\apj},
         year = 2001,
        month = feb,
       volume = {547},
       number = {2},
        pages = {574-589},
          doi = {10.1086/318417},
archivePrefix = {arXiv},
       eprint = {astro-ph/0006218},
 primaryClass = {astro-ph},
       adsurl = {https://ui.adsabs.harvard.edu/abs/2001ApJ...547..574D}
}

@ARTICLE{2002ApJ...581..777C,
       author = {{Col{\'\i}n}, Pedro and {Avila-Reese}, Vladimir and {Valenzuela}, Octavio and {Firmani}, Claudio},
        title = "{Structure and Subhalo Population of Halos in a Self-interacting Dark Matter Cosmology}",
      journal = {\apj},
         year = 2002,
        month = dec,
       volume = {581},
       number = {2},
        pages = {777-793},
          doi = {10.1086/344259},
archivePrefix = {arXiv},
       eprint = {astro-ph/0205322},
 primaryClass = {astro-ph},
       adsurl = {https://ui.adsabs.harvard.edu/abs/2002ApJ...581..777C}
}

@ARTICLE{2011MNRAS.415.1125K,
       author = {{Koda}, Jun and {Shapiro}, Paul R.},
        title = "{Gravothermal collapse of isolated self-interacting dark matter haloes: N-body simulation versus the fluid model}",
      journal = {\mnras},
         year = 2011,
        month = aug,
       volume = {415},
       number = {2},
        pages = {1125-1137},
          doi = {10.1111/j.1365-2966.2011.18684.x},
archivePrefix = {arXiv},
       eprint = {1101.3097},
 primaryClass = {astro-ph.CO},
       adsurl = {https://ui.adsabs.harvard.edu/abs/2011MNRAS.415.1125K}
}

@ARTICLE{2012MNRAS.423.3740V,
       author = {{Vogelsberger}, Mark and {Zavala}, Jesus and {Loeb}, Abraham},
        title = "{Subhaloes in self-interacting galactic dark matter haloes}",
      journal = {\mnras},
         year = 2012,
        month = jul,
       volume = {423},
       number = {4},
        pages = {3740-3752},
          doi = {10.1111/j.1365-2966.2012.21182.x10.1002/asna.19141991009},
archivePrefix = {arXiv},
       eprint = {1201.5892},
 primaryClass = {astro-ph.CO},
       adsurl = {https://ui.adsabs.harvard.edu/abs/2012MNRAS.423.3740V}
}

@ARTICLE{2013MNRAS.430...81R,
       author = {{Rocha}, Miguel and {Peter}, Annika H.~G. and {Bullock}, James S. and {Kaplinghat}, Manoj and {Garrison-Kimmel}, Shea and {O{\~n}orbe}, Jose and {Moustakas}, Leonidas A.},
        title = "{Cosmological simulations with self-interacting dark matter - I. Constant-density cores and substructure}",
      journal = {\mnras},
         year = 2013,
        month = mar,
       volume = {430},
       number = {1},
        pages = {81-104},
          doi = {10.1093/mnras/sts514},
archivePrefix = {arXiv},
       eprint = {1208.3025},
 primaryClass = {astro-ph.CO},
       adsurl = {https://ui.adsabs.harvard.edu/abs/2013MNRAS.430...81R}
}

@ARTICLE{2013MNRAS.430..105P,
       author = {{Peter}, Annika H.~G. and {Rocha}, Miguel and {Bullock}, James S. and {Kaplinghat}, Manoj},
        title = "{Cosmological simulations with self-interacting dark matter - II. Halo shapes versus observations}",
      journal = {\mnras},
         year = 2013,
        month = mar,
       volume = {430},
       number = {1},
        pages = {105-120},
          doi = {10.1093/mnras/sts535},
archivePrefix = {arXiv},
       eprint = {1208.3026},
 primaryClass = {astro-ph.CO},
       adsurl = {https://ui.adsabs.harvard.edu/abs/2013MNRAS.430..105P}
}

@ARTICLE{2013MNRAS.431L..20Z,
       author = {{Zavala}, J. and {Vogelsberger}, M. and {Walker}, M.~G.},
        title = "{Constraining self-interacting dark matter with the Milky way's dwarf  spheroidals.}",
      journal = {\mnras},
         year = 2013,
        month = apr,
       volume = {431},
        pages = {L20-L24},
          doi = {10.1093/mnrasl/sls053},
archivePrefix = {arXiv},
       eprint = {1211.6426},
 primaryClass = {astro-ph.CO},
       adsurl = {https://ui.adsabs.harvard.edu/abs/2013MNRAS.431L..20Z}
}

@ARTICLE{2018ApJ...853..109E,
       author = {{Elbert}, Oliver D. and {Bullock}, James S. and {Kaplinghat}, Manoj and {Garrison-Kimmel}, Shea and {Graus}, Andrew S. and {Rocha}, Miguel},
        title = "{A Testable Conspiracy: Simulating Baryonic Effects on Self-interacting Dark Matter Halos}",
      journal = {\apj},
         year = 2018,
        month = feb,
       volume = {853},
       number = {2},
          eid = {109},
        pages = {109},
          doi = {10.3847/1538-4357/aa9710},
archivePrefix = {arXiv},
       eprint = {1609.08626},
 primaryClass = {astro-ph.GA},
       adsurl = {https://ui.adsabs.harvard.edu/abs/2018ApJ...853..109E}
}

@ARTICLE{dustpaperI,
       author = {{Putney}, Eric and {Shih}, David and {Lim}, Sung Hak and {Buckley}, Matthew R.},
        title = "{Mapping Dark Matter Through the Dust of the Milky Way Part I: Dust Correction and Phase Space Density}",
      journal = {arXiv e-prints},
         year = 2024,
        month = dec,
          eid = {arXiv:2412.14236},
        pages = {arXiv:2412.14236},
archivePrefix = {arXiv},
       eprint = {2412.14236},
 primaryClass = {astro-ph.GA},
       adsurl = {https://ui.adsabs.harvard.edu/abs/2024arXiv241214236P}
}

@ARTICLE{1991SvA....35...21K,
       author = {{Khlopov}, M.~Y. and {Beskin}, G.~M. and {Bochkarev}, N.~G. and {Pustilnik}, L.~A. and {Pustilnik}, S.~A.},
        title = "{Observational Physics of the Mirror World}",
      journal = {\sovast},
         year = 1991,
        month = feb,
       volume = {35},
        pages = {21},
       adsurl = {https://ui.adsabs.harvard.edu/abs/1991SvA....35...21K}
}

@ARTICLE{1996PhLB..375...26B,
       author = {{Berezhiani}, Z.~G. and {Dolgov}, A.~D. and {Mohapatra}, R.~N.},
        title = "{Asymmetric inflationary reheating and the nature of mirror universe}",
      journal = {Physics Letters B},
         year = 1996,
        month = feb,
       volume = {375},
        pages = {26-36},
          doi = {10.1016/0370-2693(96)00219-5},
archivePrefix = {arXiv},
       eprint = {hep-ph/9511221},
 primaryClass = {hep-ph},
       adsurl = {https://ui.adsabs.harvard.edu/abs/1996PhLB..375...26B}
}

@ARTICLE{2000PhRvD..62f3506M,
       author = {{Mohapatra}, Rabindra N. and {Teplitz}, Vigdor L.},
        title = "{Mirror dark matter and galaxy core densities}",
      journal = {\prd},
         year = 2000,
        month = sep,
       volume = {62},
       number = {6},
          eid = {063506},
        pages = {063506},
          doi = {10.1103/PhysRevD.62.063506},
archivePrefix = {arXiv},
       eprint = {astro-ph/0001362},
 primaryClass = {astro-ph},
       adsurl = {https://ui.adsabs.harvard.edu/abs/2000PhRvD..62f3506M}
}

@ARTICLE{2002PhRvD..66f3002M,
       author = {{Mohapatra}, R.~N. and {Nussinov}, S. and {Teplitz}, V.~L.},
        title = "{Mirror matter as self-interacting dark matter}",
      journal = {\prd},
         year = 2002,
        month = sep,
       volume = {66},
       number = {6},
          eid = {063002},
        pages = {063002},
          doi = {10.1103/PhysRevD.66.063002},
archivePrefix = {arXiv},
       eprint = {hep-ph/0111381},
 primaryClass = {hep-ph},
       adsurl = {https://ui.adsabs.harvard.edu/abs/2002PhRvD..66f3002M}
}

@ARTICLE{2004IJMPD..13.2161F,
       author = {{Foot}, R.},
        title = "{Mirror Matter-Type Dark Matter}",
      journal = {International Journal of Modern Physics D},
         year = 2004,
        month = jan,
       volume = {13},
       number = {10},
        pages = {2161-2192},
          doi = {10.1142/S0218271804006449},
archivePrefix = {arXiv},
       eprint = {astro-ph/0407623},
 primaryClass = {astro-ph},
       adsurl = {https://ui.adsabs.harvard.edu/abs/2004IJMPD..13.2161F}
}

@ARTICLE{2008arXiv0808.2595S,
       author = {{Silagadze}, Z.~K.},
        title = "{Mirror dark matter discovered?}",
      journal = {arXiv e-prints},
         year = 2008,
        month = aug,
          eid = {arXiv:0808.2595},
        pages = {arXiv:0808.2595},
          doi = {10.48550/arXiv.0808.2595},
archivePrefix = {arXiv},
       eprint = {0808.2595},
 primaryClass = {astro-ph},
       adsurl = {https://ui.adsabs.harvard.edu/abs/2008arXiv0808.2595S}
}

@ARTICLE{2013JCAP...08..031H,
       author = {{Higaki}, Tetsutaro and {Jeong}, Kwang Sik and {Takahashi}, Fuminobu},
        title = "{A parallel world in the dark}",
      journal = {\jcap},
         year = 2013,
        month = aug,
       volume = {2013},
       number = {8},
          eid = {031},
        pages = {031},
          doi = {10.1088/1475-7516/2013/08/031},
archivePrefix = {arXiv},
       eprint = {1302.2516},
 primaryClass = {hep-ph},
       adsurl = {https://ui.adsabs.harvard.edu/abs/2013JCAP...08..031H}
}

@ARTICLE{2010JCAP...05..021K,
       author = {{Kaplan}, David E. and {Krnjaic}, Gordan Z. and {Rehermann}, Keith R. and {Wells}, Christopher M.},
        title = "{Atomic dark matter}",
      journal = {\jcap},
         year = 2010,
        month = may,
       volume = {2010},
       number = {5},
          eid = {021},
        pages = {021},
          doi = {10.1088/1475-7516/2010/05/021},
archivePrefix = {arXiv},
       eprint = {0909.0753},
 primaryClass = {hep-ph},
       adsurl = {https://ui.adsabs.harvard.edu/abs/2010JCAP...05..021K}
}

@ARTICLE{2011JCAP...10..011K,
       author = {{Kaplan}, David E. and {Krnjaic}, Gordan Z. and {Rehermann}, Keith R. and {Wells}, Christopher M.},
        title = "{Dark atoms: asymmetry and direct detection}",
      journal = {\jcap},
         year = 2011,
        month = oct,
       volume = {2011},
       number = {10},
          eid = {011},
        pages = {011},
          doi = {10.1088/1475-7516/2011/10/011},
archivePrefix = {arXiv},
       eprint = {1105.2073},
 primaryClass = {hep-ph},
       adsurl = {https://ui.adsabs.harvard.edu/abs/2011JCAP...10..011K}
}

@ARTICLE{2012PhRvD..85j1302C,
       author = {{Cline}, James M. and {Liu}, Zuowei and {Xue}, Wei},
        title = "{Millicharged atomic dark matter}",
      journal = {\prd},
         year = 2012,
        month = may,
       volume = {85},
       number = {10},
          eid = {101302},
        pages = {101302},
          doi = {10.1103/PhysRevD.85.101302},
archivePrefix = {arXiv},
       eprint = {1201.4858},
 primaryClass = {hep-ph},
       adsurl = {https://ui.adsabs.harvard.edu/abs/2012PhRvD..85j1302C}
}

@article{gala,
  doi = {10.21105/joss.00388},
  url = {https://doi.org/10.21105%2Fjoss.00388},
  year = 2017,
  month = {oct},
  publisher = {The Open Journal},
  volume = {2},
  number = {18},
  author = {Adrian M. Price-Whelan},
  title = {Gala: A Python package for galactic dynamics},
  journal = {The Journal of Open Source Software}}

@ARTICLE{donlon,
       author = {{Donlon}, Thomas and {Chakrabarti}, Sukanya and {Widrow}, Lawrence M. and {Lam}, Michael T. and {Chang}, Philip and {Quillen}, Alice C.},
        title = "{Galactic structure from binary pulsar accelerations: Beyond smooth models}",
      journal = {\prd},
         year = 2024,
        month = jul,
       volume = {110},
       number = {2},
          eid = {023026},
        pages = {023026},
          doi = {10.1103/PhysRevD.110.023026},
archivePrefix = {arXiv},
       eprint = {2401.15808},
 primaryClass = {astro-ph.GA},
       adsurl = {https://ui.adsabs.harvard.edu/abs/2024PhRvD.110b3026D}
}

@ARTICLE{moran,
       author = {{Moran}, Abigail and {Mingarelli}, Chiara M.~F. and {Van Tilburg}, Ken and {Good}, Deborah},
        title = "{Pulsar-based map of galactic acceleration}",
      journal = {\prd},
         year = 2024,
        month = jun,
       volume = {109},
       number = {12},
          eid = {123015},
        pages = {123015},
          doi = {10.1103/PhysRevD.109.123015},
archivePrefix = {arXiv},
       eprint = {2306.13137},
 primaryClass = {astro-ph.GA},
       adsurl = {https://ui.adsabs.harvard.edu/abs/2024PhRvD.109l3015M}
}

@ARTICLE{2025NewAR.10001721H,
       author = {{Hunt}, Jason A.~S. and {Vasiliev}, Eugene},
        title = "{Milky Way dynamics in light of Gaia}",
      journal = {\nar},
         year = 2025,
        month = jun,
       volume = {100},
          eid = {101721},
        pages = {101721},
          doi = {10.1016/j.newar.2024.101721},
archivePrefix = {arXiv},
       eprint = {2501.04075},
 primaryClass = {astro-ph.GA},
       adsurl = {https://ui.adsabs.harvard.edu/abs/2025NewAR.10001721H}
}

@ARTICLE{lim2023mapping_journal,
       author = {{Lim}, Sung Hak and {Putney}, Eric and {Buckley}, Matthew R. and {Shih}, David},
        title = "{Mapping dark matter in the Milky Way using normalizing flows and Gaia DR3}",
      journal = {\jcap},
         year = 2025,
        month = jan,
       volume = {2025},
       number = {1},
          eid = {021},
        pages = {021},
          doi = {10.1088/1475-7516/2025/01/021},
archivePrefix = {arXiv},
       eprint = {2305.13358},
 primaryClass = {astro-ph.GA},
       adsurl = {https://ui.adsabs.harvard.edu/abs/2025JCAP...01..021L}
}

@ARTICLE{2025ApJ...985L..22N,
       author = {{Nibauer}, Jacob and {Bonaca}, Ana},
        title = "{Galactic Accelerations from the GD-1 Stream Suggest a Tilted Dark Matter Halo}",
      journal = {\apjl},
         year = 2025,
        month = may,
       volume = {985},
       number = {1},
          eid = {L22},
        pages = {L22},
          doi = {10.3847/2041-8213/add0a9},
archivePrefix = {arXiv},
       eprint = {2504.07187},
 primaryClass = {astro-ph.GA},
       adsurl = {https://ui.adsabs.harvard.edu/abs/2025ApJ...985L..22N}
}

@ARTICLE{2006MNRAS.367.1781A,
       author = {{Allgood}, Brandon and {Flores}, Ricardo A. and {Primack}, Joel R. and {Kravtsov}, Andrey V. and {Wechsler}, Risa H. and {Faltenbacher}, Andreas and {Bullock}, James S.},
        title = "{The shape of dark matter haloes: dependence on mass, redshift, radius and formation}",
      journal = {\mnras},
         year = 2006,
        month = apr,
       volume = {367},
       number = {4},
        pages = {1781-1796},
          doi = {10.1111/j.1365-2966.2006.10094.x},
archivePrefix = {arXiv},
       eprint = {astro-ph/0508497},
 primaryClass = {astro-ph},
       adsurl = {https://ui.adsabs.harvard.edu/abs/2006MNRAS.367.1781A}
}

@ARTICLE{2012JCAP...05..030S,
       author = {{Schneider}, Michael D. and {Frenk}, Carlos S. and {Cole}, Shaun},
        title = "{The shapes and alignments of dark matter halos}",
      journal = {\jcap},
         year = 2012,
        month = may,
       volume = {2012},
       number = {5},
          eid = {030},
        pages = {030},
          doi = {10.1088/1475-7516/2012/05/030},
archivePrefix = {arXiv},
       eprint = {1111.5616},
 primaryClass = {astro-ph.CO},
       adsurl = {https://ui.adsabs.harvard.edu/abs/2012JCAP...05..030S}
}

@ARTICLE{emcee,
       author = {{Foreman-Mackey}, Daniel and {Hogg}, David W. and {Lang}, Dustin and {Goodman}, Jonathan},
        title = "{emcee: The MCMC Hammer}",
      journal = {\pasp},
         year = 2013,
        month = mar,
       volume = {125},
       number = {925},
        pages = {306},
          doi = {10.1086/670067},
archivePrefix = {arXiv},
       eprint = {1202.3665},
 primaryClass = {astro-ph.IM},
       adsurl = {https://ui.adsabs.harvard.edu/abs/2013PASP..125..306F}
}

@ARTICLE{2003ApJ...585..151L,
       author = {{Lee}, Jounghun and {Suto}, Yasushi},
        title = "{Modeling Intracluster Gas in Triaxial Dark Halos: An Analytic Approach}",
      journal = {\apj},
         year = 2003,
        month = mar,
       volume = {585},
       number = {1},
        pages = {151-160},
          doi = {10.1086/345931},
archivePrefix = {arXiv},
       eprint = {astro-ph/0211007},
 primaryClass = {astro-ph},
       adsurl = {https://ui.adsabs.harvard.edu/abs/2003ApJ...585..151L}
}

@ARTICLE{1970ApJ...159..379R,
       author = {{Rubin}, Vera C. and {Ford}, Jr., W. Kent},
        title = "{Rotation of the Andromeda Nebula from a Spectroscopic Survey of Emission Regions}",
      journal = {\apj},
         year = 1970,
        month = feb,
       volume = {159},
        pages = {379},
          doi = {10.1086/150317},
       adsurl = {https://ui.adsabs.harvard.edu/abs/1970ApJ...159..379R}
}

@ARTICLE{1979ARA&A..17..135F,
       author = {{Faber}, S.~M. and {Gallagher}, J.~S.},
        title = "{Masses and mass-to-light ratios of galaxies.}",
      journal = {\araa},
         year = 1979,
        month = jan,
       volume = {17},
        pages = {135-187},
          doi = {10.1146/annurev.aa.17.090179.001031},
       adsurl = {https://ui.adsabs.harvard.edu/abs/1979ARA&A..17..135F}
}

@ARTICLE{2025arXiv250703742K,
       author = {{Kalda}, Taavet and {Green}, Gregory M.},
        title = "{Deep Potential: Recovering the gravitational potential and local pattern speed in the solar neighborhood with GDR3 using normalizing flows}",
      journal = {arXiv e-prints},
         year = 2025,
        month = jul,
          eid = {arXiv:2507.03742},
        pages = {arXiv:2507.03742},
          doi = {10.48550/arXiv.2507.03742},
archivePrefix = {arXiv},
       eprint = {2507.03742},
 primaryClass = {astro-ph.GA},
       adsurl = {https://ui.adsabs.harvard.edu/abs/2025arXiv250703742K}
}

@ARTICLE{2009MNRAS.399L.118C,
       author = {{Chakrabarti}, Sukanya and {Blitz}, Leo},
        title = "{Tidal imprints of a dark subhalo on the outskirts of the Milky Way}",
      journal = {\mnras},
         year = 2009,
        month = oct,
       volume = {399},
       number = {1},
        pages = {L118-L122},
          doi = {10.1111/j.1745-3933.2009.00735.x},
archivePrefix = {arXiv},
       eprint = {0812.0821},
 primaryClass = {astro-ph},
       adsurl = {https://ui.adsabs.harvard.edu/abs/2009MNRAS.399L.118C}
}

@ARTICLE{2006ApJ...643..881L,
       author = {{Levine}, E.~S. and {Blitz}, Leo and {Heiles}, Carl},
        title = "{The Vertical Structure of the Outer Milky Way H I Disk}",
      journal = {\apj},
         year = 2006,
        month = jun,
       volume = {643},
       number = {2},
        pages = {881-896},
          doi = {10.1086/503091},
archivePrefix = {arXiv},
       eprint = {astro-ph/0601697},
 primaryClass = {astro-ph},
       adsurl = {https://ui.adsabs.harvard.edu/abs/2006ApJ...643..881L}
}

@ARTICLE{2011Natur.477..301P,
       author = {{Purcell}, Chris W. and {Bullock}, James S. and {Tollerud}, Erik J. and {Rocha}, Miguel and {Chakrabarti}, Sukanya},
        title = "{The Sagittarius impact as an architect of spirality and outer rings in the Milky Way}",
      journal = {\nat},
         year = 2011,
        month = sep,
       volume = {477},
       number = {7364},
        pages = {301-303},
          doi = {10.1038/nature10417},
archivePrefix = {arXiv},
       eprint = {1109.2918},
 primaryClass = {astro-ph.GA},
       adsurl = {https://ui.adsabs.harvard.edu/abs/2011Natur.477..301P}
}

@ARTICLE{2018Natur.563...85H,
       author = {{Helmi}, Amina and {Babusiaux}, Carine and {Koppelman}, Helmer H. and {Massari}, Davide and {Veljanoski}, Jovan and {Brown}, Anthony G.~A.},
        title = "{The merger that led to the formation of the Milky Way's inner stellar halo and thick disk}",
      journal = {\nat},
         year = 2018,
        month = oct,
       volume = {563},
       number = {7729},
        pages = {85-88},
          doi = {10.1038/s41586-018-0625-x},
archivePrefix = {arXiv},
       eprint = {1806.06038},
 primaryClass = {astro-ph.GA},
       adsurl = {https://ui.adsabs.harvard.edu/abs/2018Natur.563...85H}
}

@ARTICLE{2019ApJ...886...67C,
       author = {{Chakrabarti}, Sukanya and {Chang}, Philip and {Price-Whelan}, Adrian M. and {Read}, Justin and {Blitz}, Leo and {Hernquist}, Lars},
        title = "{Antlia 2{\textquoteright}s Role in Driving the Ripples in the Outer Gas Disk of the Galaxy}",
      journal = {\apj},
         year = 2019,
        month = nov,
       volume = {886},
       number = {1},
          eid = {67},
        pages = {67},
          doi = {10.3847/1538-4357/ab4659},
archivePrefix = {arXiv},
       eprint = {1906.04203},
 primaryClass = {astro-ph.GA},
       adsurl = {https://ui.adsabs.harvard.edu/abs/2019ApJ...886...67C}
}

@ARTICLE{2025arXiv250716932C,
       author = {{Chakrabarti}, Sukanya and {Chang}, Philip and {Profumo}, Stefano and {Craig}, Peter},
        title = "{Detection of a dark matter sub-halo near the Sun from pulsar timing}",
      journal = {arXiv e-prints},
         year = 2025,
        month = jul,
          eid = {arXiv:2507.16932},
        pages = {arXiv:2507.16932},
          doi = {10.48550/arXiv.2507.16932},
archivePrefix = {arXiv},
       eprint = {2507.16932},
 primaryClass = {astro-ph.GA},
       adsurl = {https://ui.adsabs.harvard.edu/abs/2025arXiv250716932C}
}

@ARTICLE{2021A&A...649A...9G,
       author = {{Gaia Collaboration} and {Klioner}, S.~A. and {Mignard}, F. and {Lindegren}, L. and {Bastian}, U. and {McMillan}, P.~J. and {Hern{\'a}ndez}, J. and {Hobbs}, D. and {Ramos-Lerate}, M. and {Biermann}, M. and {Bombrun}, A. and {de Torres}, A. and {Gerlach}, E. and {Geyer}, R. and {Hilger}, T. and {Lammers}, U. and {Steidelm{\"u}ller}, H. and {Stephenson}, C.~A. and {Brown}, A.~G.~A. and {Vallenari}, A. and {Prusti}, T. and {de Bruijne}, J.~H.~J. and {Babusiaux}, C. and {Creevey}, O.~L. and {Evans}, D.~W. and {Eyer}, L. and {Hutton}, A. and {Jansen}, F. and {Jordi}, C. and {Luri}, X. and {Panem}, C. and {Pourbaix}, D. and {Randich}, S. and {Sartoretti}, P. and {Soubiran}, C. and {Walton}, N.~A. and {Arenou}, F. and {Bailer-Jones}, C.~A.~L. and {Cropper}, M. and {Drimmel}, R. and {Katz}, D. and {Lattanzi}, M.~G. and {van Leeuwen}, F. and {Bakker}, J. and {Casta{\~n}eda}, J. and {De Angeli}, F. and {Ducourant}, C. and {Fabricius}, C. and {Fouesneau}, M. and {Fr{\'e}mat}, Y. and {Guerra}, R. and {Guerrier}, A. and {Guiraud}, J. and {Jean-Antoine Piccolo}, A. and {Masana}, E. and {Messineo}, R. and {Mowlavi}, N. and {Nicolas}, C. and {Nienartowicz}, K. and {Pailler}, F. and {Panuzzo}, P. and {Riclet}, F. and {Roux}, W. and {Seabroke}, G.~M. and {Sordo}, R. and {Tanga}, P. and {Th{\'e}venin}, F. and {Gracia-Abril}, G. and {Portell}, J. and {Teyssier}, D. and {Altmann}, M. and {Andrae}, R. and {Bellas-Velidis}, I. and {Benson}, K. and {Berthier}, J. and {Blomme}, R. and {Brugaletta}, E. and {Burgess}, P.~W. and {Busso}, G. and {Carry}, B. and {Cellino}, A. and {Cheek}, N. and {Clementini}, G. and {Damerdji}, Y. and {Davidson}, M. and {Delchambre}, L. and {Dell'Oro}, A. and {Fern{\'a}ndez-Hern{\'a}ndez}, J. and {Galluccio}, L. and {Garc{\'\i}a-Lario}, P. and {Garcia-Reinaldos}, M. and {Gonz{\'a}lez-N{\'u}{\~n}ez}, J. and {Gosset}, E. and {Haigron}, R. and {Halbwachs}, J. -L. and {Hambly}, N.~C. and {Harrison}, D.~L. and {Hatzidimitriou}, D. and {Heiter}, U. and {Hestroffer}, D. and {Hodgkin}, S.~T. and {Holl}, B. and {Jan{\ss}en}, K. and {Jevardat de Fombelle}, G. and {Jordan}, S. and {Krone-Martins}, A. and {Lanzafame}, A.~C. and {L{\"o}ffler}, W. and {Lorca}, A. and {Manteiga}, M. and {Marchal}, O. and {Marrese}, P.~M. and {Moitinho}, A. and {Mora}, A. and {Muinonen}, K. and {Osborne}, P. and {Pancino}, E. and {Pauwels}, T. and {Recio-Blanco}, A. and {Richards}, P.~J. and {Riello}, M. and {Rimoldini}, L. and {Robin}, A.~C. and {Roegiers}, T. and {Rybizki}, J. and {Sarro}, L.~M. and {Siopis}, C. and {Smith}, M. and {Sozzetti}, A. and {Ulla}, A. and {Utrilla}, E. and {van Leeuwen}, M. and {van Reeven}, W. and {Abbas}, U. and {Abreu Aramburu}, A. and {Accart}, S. and {Aerts}, C. and {Aguado}, J.~J. and {Ajaj}, M. and {Altavilla}, G. and {{\'A}lvarez}, M.~A. and {{\'A}lvarez Cid-Fuentes}, J. and {Alves}, J. and {Anderson}, R.~I. and {Anglada Varela}, E. and {Antoja}, T. and {Audard}, M. and {Baines}, D. and {Baker}, S.~G. and {Balaguer-N{\'u}{\~n}ez}, L. and {Balbinot}, E. and {Balog}, Z. and {Barache}, C. and {Barbato}, D. and {Barros}, M. and {Barstow}, M.~A. and {Bartolom{\'e}}, S. and {Bassilana}, J. -L. and {Bauchet}, N. and {Baudesson-Stella}, A. and {Becciani}, U. and {Bellazzini}, M. and {Bernet}, M. and {Bertone}, S. and {Bianchi}, L. and {Blanco-Cuaresma}, S. and {Boch}, T. and {Bossini}, D. and {Bouquillon}, S. and {Bramante}, L. and {Breedt}, E. and {Bressan}, A. and {Brouillet}, N. and {Bucciarelli}, B. and {Burlacu}, A. and {Busonero}, D. and {Butkevich}, A.~G. and {Buzzi}, R. and {Caffau}, E. and {Cancelliere}, R. and {C{\'a}novas}, H. and {Cantat-Gaudin}, T. and {Carballo}, R. and {Carlucci}, T. and {Carnerero}, M.~I. and {Carrasco}, J.~M. and {Casamiquela}, L. and {Castellani}, M. and {Castro-Ginard}, A. and {Castro Sampol}, P. and {Chaoul}, L. and {Charlot}, P. and {Chemin}, L. and {Chiavassa}, A. and {Comoretto}, G. and {Cooper}, W.~J. and {Cornez}, T. and {Cowell}, S. and {Crifo}, F. and {Crosta}, M.},
        title = "{Gaia Early Data Release 3. Acceleration of the Solar System from Gaia astrometry}",
      journal = {\aap},
         year = 2021,
        month = may,
       volume = {649},
          eid = {A9},
        pages = {A9},
          doi = {10.1051/0004-6361/202039734},
archivePrefix = {arXiv},
       eprint = {2012.02036},
 primaryClass = {astro-ph.GA},
       adsurl = {https://ui.adsabs.harvard.edu/abs/2021A&A...649A...9G}
}

@ARTICLE{2024MNRAS.52712284K,
       author = {{Kalda}, Taavet and {Green}, Gregory M. and {Ghosh}, Soumavo},
        title = "{Recovering the gravitational potential in a rotating frame: Deep Potential applied to a simulated barred galaxy}",
      journal = {\mnras},
         year = 2024,
        month = feb,
       volume = {527},
       number = {4},
        pages = {12284-12297},
          doi = {10.1093/mnras/stae011},
archivePrefix = {arXiv},
       eprint = {2310.00040},
 primaryClass = {astro-ph.GA},
       adsurl = {https://ui.adsabs.harvard.edu/abs/2024MNRAS.52712284K}
}

@article{papamakarios2021normalizing,
  title={Normalizing flows for probabilistic modeling and inference},
  author={Papamakarios, George and Nalisnick, Eric and Rezende, Danilo Jimenez and Mohamed, Shakir and Lakshminarayanan, Balaji},
  journal={Journal of Machine Learning Research},
  volume={22},
  number={57},
  pages={1--64},
  year={2021}
}

@ARTICLE{2020ApJ...900..186E,
       author = {{Eilers}, Anna-Christina and {Hogg}, David W. and {Rix}, Hans-Walter and {Frankel}, Neige and {Hunt}, Jason A.~S. and {Fouvry}, Jean-Baptiste and {Buck}, Tobias},
        title = "{The Strength of the Dynamical Spiral Perturbation in the Galactic Disk}",
      journal = {\apj},
         year = 2020,
        month = sep,
       volume = {900},
       number = {2},
          eid = {186},
        pages = {186},
          doi = {10.3847/1538-4357/abac0b},
archivePrefix = {arXiv},
       eprint = {2003.01132},
 primaryClass = {astro-ph.GA},
       adsurl = {https://ui.adsabs.harvard.edu/abs/2020ApJ...900..186E}
}

@ARTICLE{2016ARA&A..54..529B,
       author = {{Bland-Hawthorn}, Joss and {Gerhard}, Ortwin},
        title = "{The Galaxy in Context: Structural, Kinematic, and Integrated Properties}",
      journal = {\araa},
         year = 2016,
        month = sep,
       volume = {54},
        pages = {529-596},
          doi = {10.1146/annurev-astro-081915-023441},
archivePrefix = {arXiv},
       eprint = {1602.07702},
 primaryClass = {astro-ph.GA},
       adsurl = {https://ui.adsabs.harvard.edu/abs/2016ARA&A..54..529B}
}

@ARTICLE{2020RAA....20..159S,
       author = {{Shen}, Juntai and {Zheng}, Xing-Wu},
        title = "{The bar and spiral arms in the Milky Way: structure and kinematics}",
      journal = {Research in Astronomy and Astrophysics},
         year = 2020,
        month = oct,
       volume = {20},
       number = {10},
          eid = {159},
        pages = {159},
          doi = {10.1088/1674-4527/20/10/159},
archivePrefix = {arXiv},
       eprint = {2012.10130},
 primaryClass = {astro-ph.GA},
       adsurl = {https://ui.adsabs.harvard.edu/abs/2020RAA....20..159S}
}

@ARTICLE{2017AstRv..13..113V,
       author = {{Vall{\'e}e}, Jacques P.},
        title = "{A guided map to the spiral arms in the galactic disk of the Milky Way}",
      journal = {The Astronomical Review},
         year = 2017,
        month = oct,
       volume = {13},
       number = {3-4},
        pages = {113-146},
          doi = {10.1080/21672857.2017.1379459},
archivePrefix = {arXiv},
       eprint = {1711.05228},
 primaryClass = {astro-ph.GA},
       adsurl = {https://ui.adsabs.harvard.edu/abs/2017AstRv..13..113V}
}

@ARTICLE{2012ApJ...750L..41W,
       author = {{Widrow}, Lawrence M. and {Gardner}, Susan and {Yanny}, Brian and {Dodelson}, Scott and {Chen}, Hsin-Yu},
        title = "{Galactoseismology: Discovery of Vertical Waves in the Galactic Disk}",
      journal = {\apjl},
         year = 2012,
        month = may,
       volume = {750},
       number = {2},
          eid = {L41},
        pages = {L41},
          doi = {10.1088/2041-8205/750/2/L41},
archivePrefix = {arXiv},
       eprint = {1203.6861},
 primaryClass = {astro-ph.GA},
       adsurl = {https://ui.adsabs.harvard.edu/abs/2012ApJ...750L..41W}
}

@ARTICLE{2009MNRAS.396L..56M,
       author = {{Minchev}, I. and {Quillen}, A.~C. and {Williams}, M. and {Freeman}, K.~C. and {Nordhaus}, J. and {Siebert}, A. and {Bienaym{\'e}}, O.},
        title = "{Is the Milky Way ringing? The hunt for high-velocity streams}",
      journal = {\mnras},
         year = 2009,
        month = jun,
       volume = {396},
       number = {1},
        pages = {L56-L60},
          doi = {10.1111/j.1745-3933.2009.00661.x},
archivePrefix = {arXiv},
       eprint = {0902.1531},
 primaryClass = {astro-ph.GA},
       adsurl = {https://ui.adsabs.harvard.edu/abs/2009MNRAS.396L..56M}
}

@ARTICLE{2008MNRAS.389.1041R,
       author = {{Read}, J.~I. and {Lake}, G. and {Agertz}, O. and {Debattista}, Victor P.},
        title = "{Thin, thick and dark discs in {\ensuremath{\Lambda}}CDM}",
      journal = {\mnras},
         year = 2008,
        month = sep,
       volume = {389},
       number = {3},
        pages = {1041-1057},
          doi = {10.1111/j.1365-2966.2008.13643.x},
archivePrefix = {arXiv},
       eprint = {0803.2714},
 primaryClass = {astro-ph},
       adsurl = {https://ui.adsabs.harvard.edu/abs/2008MNRAS.389.1041R}
}

@ARTICLE{2009MNRAS.397...44R,
       author = {{Read}, J.~I. and {Mayer}, L. and {Brooks}, A.~M. and {Governato}, F. and {Lake}, G.},
        title = "{A dark matter disc in three cosmological simulations of Milky Way mass galaxies}",
      journal = {\mnras},
         year = 2009,
        month = jul,
       volume = {397},
       number = {1},
        pages = {44-51},
          doi = {10.1111/j.1365-2966.2009.14757.x},
archivePrefix = {arXiv},
       eprint = {0902.0009},
 primaryClass = {astro-ph.GA},
       adsurl = {https://ui.adsabs.harvard.edu/abs/2009MNRAS.397...44R}
}

@ARTICLE{1989AJ.....98.1554L,
       author = {{Lake}, George},
        title = "{Must the Disk and Halo Dark Matter Be Different?}",
      journal = {\aj},
         year = 1989,
        month = nov,
       volume = {98},
        pages = {1554},
          doi = {10.1086/115238},
       adsurl = {https://ui.adsabs.harvard.edu/abs/1989AJ.....98.1554L}
}

@ARTICLE{2017MNRAS.472.3722G,
       author = {{G{\'o}mez}, Facundo A. and {Grand}, Robert J.~J. and {Monachesi}, Antonela and {White}, Simon D.~M. and {Bustamante}, Sebastian and {Marinacci}, Federico and {Pakmor}, R{\"u}diger and {Simpson}, Christine M. and {Springel}, Volker and {Frenk}, Carlos S.},
        title = "{Lessons from the Auriga discs: the hunt for the Milky Way's ex situ disc is not yet over}",
      journal = {\mnras},
         year = 2017,
        month = dec,
       volume = {472},
       number = {3},
        pages = {3722-3733},
          doi = {10.1093/mnras/stx2149},
archivePrefix = {arXiv},
       eprint = {1704.08261},
 primaryClass = {astro-ph.GA},
       adsurl = {https://ui.adsabs.harvard.edu/abs/2017MNRAS.472.3722G}
}

@ARTICLE{2015MNRAS.450.2874R,
       author = {{Ruchti}, G.~R. and {Read}, J.~I. and {Feltzing}, S. and {Serenelli}, A.~M. and {McMillan}, P. and {Lind}, K. and {Bensby}, T. and {Bergemann}, M. and {Asplund}, M. and {Vallenari}, A. and {Flaccomio}, E. and {Pancino}, E. and {Korn}, A.~J. and {Recio-Blanco}, A. and {Bayo}, A. and {Carraro}, G. and {Costado}, M.~T. and {Damiani}, F. and {Heiter}, U. and {Hourihane}, A. and {Jofr{\'e}}, P. and {Kordopatis}, G. and {Lardo}, C. and {de Laverny}, P. and {Monaco}, L. and {Morbidelli}, L. and {Sbordone}, L. and {Worley}, C.~C. and {Zaggia}, S.},
        title = "{The Gaia-ESO Survey: a quiescent Milky Way with no significant dark/stellar accreted disc}",
      journal = {\mnras},
         year = 2015,
        month = jul,
       volume = {450},
       number = {3},
        pages = {2874-2887},
          doi = {10.1093/mnras/stv807},
archivePrefix = {arXiv},
       eprint = {1504.02481},
 primaryClass = {astro-ph.GA},
       adsurl = {https://ui.adsabs.harvard.edu/abs/2015MNRAS.450.2874R}
}

@ARTICLE{2020NatAs...4.1078N,
       author = {{Necib}, Lina and {Ostdiek}, Bryan and {Lisanti}, Mariangela and {Cohen}, Timothy and {Freytsis}, Marat and {Garrison-Kimmel}, Shea and {Hopkins}, Philip F. and {Wetzel}, Andrew and {Sanderson}, Robyn},
        title = "{Evidence for a vast prograde stellar stream in the solar vicinity}",
      journal = {Nature Astronomy},
         year = 2020,
        month = jul,
       volume = {4},
        pages = {1078-1083},
          doi = {10.1038/s41550-020-1131-2},
archivePrefix = {arXiv},
       eprint = {1907.07190},
 primaryClass = {astro-ph.GA},
       adsurl = {https://ui.adsabs.harvard.edu/abs/2020NatAs...4.1078N}
}

@ARTICLE{1996MNRAS.278..488Z,
       author = {{Zhao}, Hongsheng},
        title = "{Analytical models for galactic nuclei}",
      journal = {\mnras},
         year = 1996,
        month = jan,
       volume = {278},
       number = {2},
        pages = {488-496},
          doi = {10.1093/mnras/278.2.488},
archivePrefix = {arXiv},
       eprint = {astro-ph/9509122},
 primaryClass = {astro-ph},
       adsurl = {https://ui.adsabs.harvard.edu/abs/1996MNRAS.278..488Z}
}

@ARTICLE{2015ApJS..216...29B,
       author = {{Bovy}, Jo},
        title = "{galpy: A python Library for Galactic Dynamics}",
      journal = {\apjs},
         year = 2015,
        month = feb,
       volume = {216},
       number = {2},
          eid = {29},
        pages = {29},
          doi = {10.1088/0067-0049/216/2/29},
archivePrefix = {arXiv},
       eprint = {1412.3451},
 primaryClass = {astro-ph.GA},
       adsurl = {https://ui.adsabs.harvard.edu/abs/2015ApJS..216...29B}
}

@ARTICLE{2025arXiv251115865D,
       author = {{Donlon}, II, Thomas and {Widrow}, Lawrence M. and {Chakrabarti}, Sukanya},
        title = "{Mean Mass Density near the Sun from the Divergence Theorem and Pulsar Accelerations}",
      journal = {arXiv e-prints},
         year = 2025,
        month = nov,
          eid = {arXiv:2511.15865},
        pages = {arXiv:2511.15865},
          doi = {10.48550/arXiv.2511.15865},
archivePrefix = {arXiv},
       eprint = {2511.15865},
 primaryClass = {astro-ph.GA},
       adsurl = {https://ui.adsabs.harvard.edu/abs/2025arXiv251115865D}
}

@ARTICLE{2025MNRAS.542.2987S,
       author = {{S{\"o}ding}, Laurin and {Bartel}, Ruben L. and {Mertsch}, Philipp},
        title = "{Local dark matter density from Gaia DR3 K-dwarfs using Gaussian processes}",
      journal = {\mnras},
         year = 2025,
        month = oct,
       volume = {542},
       number = {4},
        pages = {2987-2997},
          doi = {10.1093/mnras/staf1391},
archivePrefix = {arXiv},
       eprint = {2506.02956},
 primaryClass = {astro-ph.GA},
       adsurl = {https://ui.adsabs.harvard.edu/abs/2025MNRAS.542.2987S}
}

@ARTICLE{2025ApJ...990L..37L,
       author = {{Lian}, Jianhui and {Wang}, Tao and {Feng}, Qikang and {Huang}, Yang and {Guo}, Helong},
        title = "{The Milky Way Is a Less Massive Galaxy{\textemdash}New Estimates of the Milky Way's Local and Global Stellar Masses}",
      journal = {\apjl},
         year = 2025,
        month = sep,
       volume = {990},
       number = {2},
          eid = {L37},
        pages = {L37},
          doi = {10.3847/2041-8213/adfc73},
archivePrefix = {arXiv},
       eprint = {2508.13665},
 primaryClass = {astro-ph.GA},
       adsurl = {https://ui.adsabs.harvard.edu/abs/2025ApJ...990L..37L}
}

@ARTICLE{2025PhRvD.111j3036D,
       author = {{Donlon}, Thomas and {Chakrabarti}, Sukanya and {Vanderwaal}, Sophia and {Widrow}, Lawrence M. and {Ransom}, Scott and {Ramirez-Ruiz}, Enrico},
        title = "{Empirical modeling of magnetic braking in millisecond pulsars to measure the local dark matter density and effects of orbiting satellite galaxies}",
      journal = {\prd},
         year = 2025,
        month = may,
       volume = {111},
       number = {10},
          eid = {103036},
        pages = {103036},
          doi = {10.1103/PhysRevD.111.103036},
archivePrefix = {arXiv},
       eprint = {2501.03409},
 primaryClass = {astro-ph.HE},
       adsurl = {https://ui.adsabs.harvard.edu/abs/2025PhRvD.111j3036D}
}

@ARTICLE{2025ApJ...978...45L,
       author = {{L{\'o}pez-Corredoira}, Mart{\'\i}n},
        title = "{Milky Way Dark Matter Distribution or MOND Test from Vertical Stellar Kinematics with Gaia DR3}",
      journal = {\apj},
         year = 2025,
        month = jan,
       volume = {978},
       number = {1},
          eid = {45},
        pages = {45},
          doi = {10.3847/1538-4357/ad94f5},
archivePrefix = {arXiv},
       eprint = {2412.09665},
 primaryClass = {astro-ph.GA},
       adsurl = {https://ui.adsabs.harvard.edu/abs/2025ApJ...978...45L}
}

@ARTICLE{2024A&A...689A.280B,
       author = {{Bienaym{\'e}}, O. and {Robin}, A.~C. and {Salomon}, J.-B. and {Reyl{\'e}}, C.},
        title = "{Dark matter in the Milky Way: Measurements up to 3 kpc from the Galactic plane above the Sun}",
      journal = {\aap},
         year = 2024,
        month = sep,
       volume = {689},
          eid = {A280},
        pages = {A280},
          doi = {10.1051/0004-6361/202450327},
archivePrefix = {arXiv},
       eprint = {2406.08158},
 primaryClass = {astro-ph.GA},
       adsurl = {https://ui.adsabs.harvard.edu/abs/2024A&A...689A.280B}
}

@ARTICLE{2024PhRvD.110b3026D,
       author = {{Donlon}, Thomas and {Chakrabarti}, Sukanya and {Widrow}, Lawrence M. and {Lam}, Michael T. and {Chang}, Philip and {Quillen}, Alice C.},
        title = "{Galactic structure from binary pulsar accelerations: Beyond smooth models}",
      journal = {\prd},
         year = 2024,
        month = jul,
       volume = {110},
       number = {2},
          eid = {023026},
        pages = {023026},
          doi = {10.1103/PhysRevD.110.023026},
archivePrefix = {arXiv},
       eprint = {2401.15808},
 primaryClass = {astro-ph.GA},
       adsurl = {https://ui.adsabs.harvard.edu/abs/2024PhRvD.110b3026D}
}

@ARTICLE{2024MNRAS.529.4681B,
       author = {{Beordo}, William and {Crosta}, Mariateresa and {Lattanzi}, Mario G. and {Re Fiorentin}, Paola and {Spagna}, Alessandro},
        title = "{Geometry-driven and dark-matter-sustained Milky Way rotation curves with Gaia DR3}",
      journal = {\mnras},
         year = 2024,
        month = apr,
       volume = {529},
       number = {4},
        pages = {4681-4698},
          doi = {10.1093/mnras/stae855},
       adsurl = {https://ui.adsabs.harvard.edu/abs/2024MNRAS.529.4681B}
}

@ARTICLE{2024ApJ...962..165H,
       author = {{Horta}, Danny and {Price-Whelan}, Adrian M. and {Hogg}, David W. and {Johnston}, Kathryn V. and {Widrow}, Lawrence and {Dalcanton}, Julianne J. and {Ness}, Melissa K. and {Hunt}, Jason A.~S.},
        title = "{Orbital Torus Imaging: Acceleration, Density, and Dark Matter in the Galactic Disk Measured with Element Abundance Gradients}",
      journal = {\apj},
         year = 2024,
        month = feb,
       volume = {962},
       number = {2},
          eid = {165},
        pages = {165},
          doi = {10.3847/1538-4357/ad16e8},
archivePrefix = {arXiv},
       eprint = {2312.07664},
 primaryClass = {astro-ph.GA},
       adsurl = {https://ui.adsabs.harvard.edu/abs/2024ApJ...962..165H}
}

@ARTICLE{2024MNRAS.534.3387S,
       author = {{Syaifudin}, M.~A. and {Arifyanto}, M.~I. and {Wulandari}, H.~R.~T. and {Mulki}, F.~A.~M.},
        title = "{Comparing dark matter and MOND hyphotheses from the distribution function of A, F, early-G stars in the solar neighbourhood}",
      journal = {\mnras},
         year = 2024,
        month = nov,
       volume = {534},
       number = {4},
        pages = {3387-3399},
          doi = {10.1093/mnras/stae2316},
archivePrefix = {arXiv},
       eprint = {2401.11534},
 primaryClass = {astro-ph.GA},
       adsurl = {https://ui.adsabs.harvard.edu/abs/2024MNRAS.534.3387S}
}

@ARTICLE{2024JCAP...08..022S,
       author = {{Staudt}, Patrick G. and {Bullock}, James S. and {Boylan-Kolchin}, Michael and {Kirkby}, David and {Wetzel}, Andrew and {Ou}, Xiaowei},
        title = "{Sliding into DM: determining the local dark matter density and speed distribution using only the local circular speed of the galaxy}",
      journal = {\jcap},
         year = 2024,
        month = aug,
       volume = {2024},
       number = {8},
          eid = {022},
        pages = {022},
          doi = {10.1088/1475-7516/2024/08/022},
archivePrefix = {arXiv},
       eprint = {2403.04122},
 primaryClass = {astro-ph.GA},
       adsurl = {https://ui.adsabs.harvard.edu/abs/2024JCAP...08..022S}
}

@ARTICLE{2025ApJ...992...84K,
       author = {{Kalda}, Taavet and {Green}, Gregory M.},
        title = "{Deep Potential: Recovering the Gravitational Potential and Local Pattern Speed in the Solar Neighborhood with GDR3 Using Normalizing Flows}",
      journal = {\apj},
         year = 2025,
        month = oct,
       volume = {992},
       number = {1},
          eid = {84},
        pages = {84},
          doi = {10.3847/1538-4357/adf8ea},
archivePrefix = {arXiv},
       eprint = {2507.03742},
 primaryClass = {astro-ph.GA},
       adsurl = {https://ui.adsabs.harvard.edu/abs/2025ApJ...992...84K}
}

@ARTICLE{2024MNRAS.528..693O,
       author = {{Ou}, Xiaowei and {Eilers}, Anna-Christina and {Necib}, Lina and {Frebel}, Anna},
        title = "{The dark matter profile of the Milky Way inferred from its circular velocity curve}",
      journal = {\mnras},
         year = 2024,
        month = feb,
       volume = {528},
       number = {1},
        pages = {693-710},
          doi = {10.1093/mnras/stae034},
archivePrefix = {arXiv},
       eprint = {2303.12838},
 primaryClass = {astro-ph.GA},
       adsurl = {https://ui.adsabs.harvard.edu/abs/2024MNRAS.528..693O}
}

@ARTICLE{2023ApJ...946...73Z,
       author = {{Zhou}, Yuan and {Li}, Xinyi and {Huang}, Yang and {Zhang}, Huawei},
        title = "{The Circular Velocity Curve of the Milky Way from 5-25 kpc Using Luminous Red Giant Branch Stars}",
      journal = {\apj},
         year = 2023,
        month = apr,
       volume = {946},
       number = {2},
          eid = {73},
        pages = {73},
          doi = {10.3847/1538-4357/acadd9},
archivePrefix = {arXiv},
       eprint = {2212.10393},
 primaryClass = {astro-ph.GA},
       adsurl = {https://ui.adsabs.harvard.edu/abs/2023ApJ...946...73Z}
}

@ARTICLE{2022AJ....164..249H,
       author = {{Han}, Jiwon Jesse and {Conroy}, Charlie and {Johnson}, Benjamin D. and {Speagle}, Joshua S. and {Bonaca}, Ana and {Chandra}, Vedant and {Naidu}, Rohan P. and {Ting}, Yuan-Sen and {Woody}, Turner and {Zaritsky}, Dennis},
        title = "{The Stellar Halo of the Galaxy is Tilted and Doubly Broken}",
      journal = {\aj},
         year = 2022,
        month = dec,
       volume = {164},
       number = {6},
          eid = {249},
        pages = {249},
          doi = {10.3847/1538-3881/ac97e9},
archivePrefix = {arXiv},
       eprint = {2208.04327},
 primaryClass = {astro-ph.GA},
       adsurl = {https://ui.adsabs.harvard.edu/abs/2022AJ....164..249H}
}

@ARTICLE{2022ApJ...934...14H,
       author = {{Han}, Jiwon Jesse and {Naidu}, Rohan P. and {Conroy}, Charlie and {Bonaca}, Ana and {Zaritsky}, Dennis and {Caldwell}, Nelson and {Cargile}, Phillip and {Johnson}, Benjamin D. and {Chandra}, Vedant and {Speagle}, Joshua S. and {Ting}, Yuan-Sen and {Woody}, Turner},
        title = "{A Tilt in the Dark Matter Halo of the Galaxy}",
      journal = {\apj},
         year = 2022,
        month = jul,
       volume = {934},
       number = {1},
          eid = {14},
        pages = {14},
          doi = {10.3847/1538-4357/ac795f},
archivePrefix = {arXiv},
       eprint = {2202.07662},
 primaryClass = {astro-ph.GA},
       adsurl = {https://ui.adsabs.harvard.edu/abs/2022ApJ...934...14H}
}

@ARTICLE{2024arXiv240612969H,
       author = {{Han}, Jiwon Jesse and {Conroy}, Charlie and {Zaritsky}, Dennis and {Bonaca}, Ana and {Caldwell}, Nelson and {Chandra}, Vedant and {Ting}, Yuan-Sen},
        title = "{Our Halo of Ice and Fire: Strong Kinematic Asymmetries in the Galactic Halo}",
      journal = {arXiv e-prints},
         year = 2024,
        month = jun,
          eid = {arXiv:2406.12969},
        pages = {arXiv:2406.12969},
          doi = {10.48550/arXiv.2406.12969},
archivePrefix = {arXiv},
       eprint = {2406.12969},
 primaryClass = {astro-ph.GA},
       adsurl = {https://ui.adsabs.harvard.edu/abs/2024arXiv240612969H}
}

@ARTICLE{2023ApJ...957L..24H,
       author = {{Han}, Jiwon Jesse and {Semenov}, Vadim and {Conroy}, Charlie and {Hernquist}, Lars},
        title = "{Tilted Dark Halos Are Common and Long-lived, and Can Warp Galactic Disks}",
      journal = {\apjl},
         year = 2023,
        month = nov,
       volume = {957},
       number = {2},
          eid = {L24},
        pages = {L24},
          doi = {10.3847/2041-8213/ad0641},
archivePrefix = {arXiv},
       eprint = {2309.07208},
 primaryClass = {astro-ph.GA},
       adsurl = {https://ui.adsabs.harvard.edu/abs/2023ApJ...957L..24H}
}

@ARTICLE{2025ApJ...983...62D,
       author = {{Donlon}, Thomas and {Chakrabarti}, Sukanya and {Lam}, Michael T. and {Huber}, Daniel and {Hey}, Daniel and {Ramirez-Ruiz}, Enrico and {Shappee}, Benjamin and {Kaplan}, David L. and {Agazie}, Gabriella and {Anumarlapudi}, Akash and {Archibald}, Anne M. and {Arzoumanian}, Zaven and {Baker}, Paul T. and {Brook}, Paul R. and {Cromartie}, H. Thankful and {Crowter}, Kathryn and {DeCesar}, Megan E. and {Demorest}, Paul B. and {Dolch}, Timothy and {Ferrara}, Elizabeth C. and {Fiore}, William and {Fonseca}, Emmanuel and {Freedman}, Gabriel E. and {Garver-Daniels}, Nate and {Gentile}, Peter A. and {Glaser}, Joseph and {Good}, Deborah C. and {Hazboun}, Jeffrey S. and {Huber}, Mark and {Jennings}, Ross J. and {Jones}, Megan L. and {Kerr}, Matthew and {Lorimer}, Duncan R. and {Luo}, Jing and {Lynch}, Ryan S. and {McEwen}, Alexander and {McLaughlin}, Maura A. and {McMann}, Natasha and {Meyers}, Bradley W. and {Ng}, Cherry and {Nice}, David J. and {Pennucci}, Timothy T. and {Perera}, Benetge B.~P. and {Pol}, Nihan S. and {Radovan}, Henri A. and {Ransom}, Scott M. and {Ray}, Paul S. and {Schmiedekamp}, Ann and {Schmiedekamp}, Carl and {Shapiro-Albert}, Brent J. and {Stairs}, Ingrid H. and {Stovall}, Kevin and {Susobhanan}, Abhimanyu and {Swiggum}, Joseph K. and {Tucker}, Michael A. and {Wahl}, Haley M.},
        title = "{The Anomalous Acceleration of PSR J2043+1711: Long-period Orbital Companion or Stellar Flyby?}",
      journal = {\apj},
         year = 2025,
        month = apr,
       volume = {983},
       number = {1},
          eid = {62},
        pages = {62},
          doi = {10.3847/1538-4357/adbf90},
archivePrefix = {arXiv},
       eprint = {2407.06482},
 primaryClass = {astro-ph.SR},
       adsurl = {https://ui.adsabs.harvard.edu/abs/2025ApJ...983...62D}
}
\end{document}